\documentclass[acmsmall,screen]{acmart}

\usepackage{colortbl}
\usepackage{xcolor}
\usepackage{multirow}
\usepackage{amsmath}
\usepackage{subcaption}
\usepackage{listings}
\usepackage{xspace}

\newcommand{\colorrect}[2]{\raisebox{2pt}{\colorbox[HTML]{#2}{\rule{0.25cm}{0cm}}} \space {\footnotesize #1}}
\newcommand{\colorrectRight}[2]{{\footnotesize #1} \space \raisebox{2pt}{\colorbox[HTML]{#2}{\rule{0.25cm}{0cm}}}}

\definecolor{cstring}{rgb}{0, 0.3, 0}

\AtBeginDocument{%
  }

\lstdefinelanguage{JavaScript}{
  keywords={typeof, new, true, false, catch, function, return, null, catch, switch, var, if, in, while, do, else, case, break},
  keywordstyle=\color{blue}\bfseries,
  ndkeywords={class, export, boolean, throw, implements, import, this},
  ndkeywordstyle=\color{darkgray}\bfseries,
  identifierstyle=\color{black},
  sensitive=false,
  comment=[l]{//},
  morecomment=[s]{/*}{*/},
  morestring=[b]',
  morestring=[b]"
}

\lstdefinelanguage
[x86gnuatt]{Assembler}
[x86masm]{Assembler}
{
  morekeywords={rax, xmm0, rbx},
  comment=[l]{\#},
  morecomment=[l]{\#},
  escapechar=\%,
  mathescape=true
}

\lstdefinelanguage
[arm]{Assembler}
{
  morekeywords={bmi, cmp, lslv, w0, w1, w2, w3},
  comment=[l]{\#},
  morecomment=[l]{\#},
  escapechar=\%,
  mathescape=true,
}

\lstdefinestyle{customc}{
  language=C,
  morekeywords={inline, bool, int64_t, uint32_t, uint16_t, uint64_t},
  commentstyle=\rmfamily\footnotesize\color{darkgray},
  stringstyle=\color{cstring},
  showstringspaces=false
}

\setcopyright{acmlicensed}
\copyrightyear{2025}
\acmYear{2025}
\acmDOI{XXXXXXX.XXXXXXX}

\acmConference[OOPSLA '25]{Object-Oriented Programming, Systems, Languages \& Applications}{October 12--18,
  2025}{Singapore}
\acmISBN{978-1-4503-XXXX-X/18/06}

\newcommand\redrock{AMD Ryzen 7955WX, 62GB, Linux 6.12.31 x86\_64, Debian 10.13, gcc 14.2.0}
\newcommand\redrockTT{Intel(R) Xeon(R) W-2245, 32GB, Linux 6.12.31 x86\_64, Debian 10.13, gcc 14.2.0}
\newcommand\starfive{riscv sifive, u74-mc, 8GB, Linux starfive 6.1.31-starfive riscv64, gcc 14.2.0}
\newcommand\sizo{Apple M2 Max, 64GB, macOS 15.4.1, gcc 14.3.0}

\begin{document}

\newcommand{\ms}[1]{\textcolor{red}{[MS: #1]}}
\newcommand{\om}[1]{\textcolor{blue}{[OM: #1]}}

\makeatletter
\newcommand\citeartifact{\if@ACM@anonymous\text{[\textit{hidden for review}]}\else\cite{artifact}\fi}
\makeatother

\title{Float Self-Tagging}

\author{Olivier Melan\c{c}on}
\email{olivier.melancon.1@umontreal.ca}
\orcid{0009-0007-7688-3208}
\affiliation{%
  \institution{Universit{\'e} de Montr{\'e}al}
  \city{Montr{\'e}al}
  \country{Canada}
}

\author{Manuel Serrano}
\email{Manuel.Serrano@inria.fr}
\orcid{0000-0002-5240-1610}
\affiliation{%
  \institution{Inria/UCA}
  \city{Sophia Antipolis}
  \country{France}
}

\author{Marc Feeley}
\email{feeley@iro.umontreal.ca}
\orcid{0009-0005-5237-8712}
\affiliation{%
  \institution{Universit{\'e} de Montr{\'e}al}
  \city{Montr{\'e}al}
  \country{Canada}
}


\begin{abstract}
  Dynamic and polymorphic languages attach information, such as types, to run
  time objects, and therefore adapt the memory layout of values to include space
  for this information. This makes it difficult to efficiently implement IEEE754
  floating-point numbers as this format does not leave an easily accessible
  space to store type information. The three main floating-point number
  encodings in use today, tagged pointers, NaN-boxing, and NuN-boxing,
  have drawbacks. Tagged pointers entail a heap allocation of all float objects,
  and NaN/NuN-boxing puts additional run time costs on type checks and the handling
  of other objects.
  
  This paper introduces self-tagging, a new approach to object tagging that uses
  an invertible bitwise transformation to map floating-point numbers to tagged
  values that contain the correct type information at the correct position in
  their bit pattern, superimposing both their value and type information in a
  single machine word. Such a transformation can only map a subset of all floats
  to correctly typed tagged values, hence self-tagging takes advantage of the
  non-uniform distribution of floating point numbers used in practice to avoid
  heap allocation of the most frequently encountered floats.
  
  Variants of self-tagging were implemented in two distinct Scheme compilers and
  evaluated on four microarchitectures to assess their performance and compare
  them to tagged pointers, NaN-boxing, and NuN-boxing. Experiments demonstrate
  that, in practice, the approach eliminates heap allocation of nearly all
  floating-point numbers and provides good execution speed of float-intensive
  benchmarks in Scheme with a negligible performance impact on other benchmarks,
  making it an attractive alternative to tagged pointers, alongside NaN-boxing
  and NuN-boxing.
\end{abstract}



\received{25 March 2025}
\received[revised]{-}
\received[accepted]{-}

\maketitle

\newcommand{\htmlColorOneTag}{FF00FF}
\newcommand{\htmlColorTwoTag}{ffb6f3}
\newcommand{\htmlColorThreeTag}{61D836}
\newcommand{\htmlColorFourTag}{909090}
\newcommand{\htmlColorTwoTagNZ}{00A2FF}
\newcommand{\htmlColorNan}{edd20b}
\newcommand{\htmlColorAlloc}{aa0000}

\definecolor{tag0}{HTML}{61D836}
\definecolor{tag3}{HTML}{00A2FF}
\definecolor{tag4}{HTML}{82BEFA}
\definecolor{tag7}{HTML}{909090}
\definecolor{1tag}{HTML}{FF00FF}
\definecolor{2tag}{HTML}{ffb6f3}
\definecolor{alloc}{HTML}{aa0000}
\newcommand{\tagzerocolorname}{green\xspace}
\newcommand{\tagthreecolorname}{dark blue\xspace}
\newcommand{\tagfourcolorname}{light blue\xspace}
\newcommand{\tagsevencolorname}{gray\xspace}
\newcommand{\onetagcolorname}{bold pink\xspace}
\newcommand{\twotagcolorname}{light pink\xspace}
\newcommand{\alloccolorname}{aa0000\xspace}

\newcommand{\bigloocommit}{5b1118\xspace}
\newcommand{\gambitcommit}{768900\xspace}
\newcommand{\legendbiglootimenun}{
\begin{center}
\begin{tabular}{llllllll}
\colorrect{self-tagging (1-tag)}{FF00FF}     &
\colorrect{self-tagging (2-tag w/ prealloc. zero)}{00A2FF}    &
\colorrect{self-tagging (3-tag)}{61D836}   &
\colorrect{NaN-boxing}{edd20b}    \\
\end{tabular}
\end{center}
}
\newcommand{\legendgambittimenun}{
\begin{center}
\begin{tabular}{llllllll}
\colorrect{self-tagging (1-tag)}{FF00FF}     &
\colorrect{self-tagging (2-tag)}{ffb6f3}    &
\colorrect{self-tagging (3-tag)}{61D836}   &
\colorrect{self-tagging (4-tag)}{909090}  \\
\end{tabular}
\end{center}
}
\newcommand{\legendbiglootimealloc}{
\begin{center}
\begin{tabular}{l}
\colorrect{self-tagging (1-tag)}{FF00FF}
\end{tabular}
\end{center}
}
\newcommand{\legendgambittimealloc}{
\begin{center}
\begin{tabular}{l}
\colorrect{self-tagging (4-tag)}{909090}
\end{tabular}
\end{center}
}
\newcommand{\legendbiglootimemantissa}{
\begin{center}
\colorrect{self-tagging (2-tag, mantissa low bits)}{ac2db7}
\end{center}
}
\newcommand{\legendbigloomem}{
\begin{center}
\begin{tabular}{lll}
\colorrect{self-tagging (1-tag)}{FF00FF}     &
\colorrect{self-tagging (2-tag w/ prealloc. zero)}{00A2FF}    &
\colorrect{self-tagging (3-tag)}{61D836}   \\
\colorrect{NaN-boxing}{edd20b}    &
\colorrect{NuN-boxing}{cf0034}    \\
\end{tabular}
\end{center}
}
\newcommand{\legendbigloobranch}{
\begin{center}
\begin{tabular}{lll}
\colorrect{self-tagging (2-tag, mantissa low-bits)}{ac2db7}     &
\colorrect{self-tagging (1-tag)}{FF00FF} & \\
\colorrect{self-tagging (2-tag w/ prealloc. zero)}{00A2FF}    &
\colorrect{NaN-boxing}{edd20b}    &
\colorrect{NuN-boxing}{cf0034}    \\
\end{tabular}
\end{center}
}

\newcommand{\experimentsrepetitions}{10}

\newcommand{\biglooredrockFltOneNunRatioFloats}{1.20}
\newcommand{\biglooredrockFltOneNunRatioNonFloats}{1.01}
\newcommand{\biglooredrockFltOneNunRatioAll}{1.09}

\newcommand{\biglooredrockFltnzNunRatioFloats}{1.22}
\newcommand{\biglooredrockFltnzNunRatioNonFloats}{1.01}
\newcommand{\biglooredrockFltnzNunRatioAll}{1.09}

\newcommand{\biglooredrockFltThreeNunRatioFloats}{1.13}
\newcommand{\biglooredrockFltThreeNunRatioNonFloats}{1.00}
\newcommand{\biglooredrockFltThreeNunRatioAll}{1.05}

\newcommand{\biglooredrockNanNunRatioFloats}{.87}
\newcommand{\biglooredrockNanNunRatioNonFloats}{1.01}
\newcommand{\biglooredrockNanNunRatioAll}{.95}

\newcommand{\gambitredrockFltOneNunRatioFloats}{1.06}
\newcommand{\gambitredrockFltOneNunRatioNonFloats}{.95}
\newcommand{\gambitredrockFltOneNunRatioAll}{1.00}

\newcommand{\gambitredrockFltTwoNunRatioFloats}{1.11}
\newcommand{\gambitredrockFltTwoNunRatioNonFloats}{.96}
\newcommand{\gambitredrockFltTwoNunRatioAll}{1.02}

\newcommand{\gambitredrockFltThreeNunRatioFloats}{1.09}
\newcommand{\gambitredrockFltThreeNunRatioNonFloats}{.96}
\newcommand{\gambitredrockFltThreeNunRatioAll}{1.01}

\newcommand{\gambitredrockFltFourNunRatioFloats}{.97}
\newcommand{\gambitredrockFltFourNunRatioNonFloats}{.97}
\newcommand{\gambitredrockFltFourNunRatioAll}{.97}

\newcommand{\biglooredrockTTFltOneNunRatioFloats}{1.20}
\newcommand{\biglooredrockTTFltOneNunRatioNonFloats}{.97}
\newcommand{\biglooredrockTTFltOneNunRatioAll}{1.06}

\newcommand{\biglooredrockTTFltnzNunRatioFloats}{1.34}
\newcommand{\biglooredrockTTFltnzNunRatioNonFloats}{.96}
\newcommand{\biglooredrockTTFltnzNunRatioAll}{1.11}

\newcommand{\biglooredrockTTFltThreeNunRatioFloats}{1.09}
\newcommand{\biglooredrockTTFltThreeNunRatioNonFloats}{1.00}
\newcommand{\biglooredrockTTFltThreeNunRatioAll}{1.04}

\newcommand{\biglooredrockTTNanNunRatioFloats}{.84}
\newcommand{\biglooredrockTTNanNunRatioNonFloats}{1.02}
\newcommand{\biglooredrockTTNanNunRatioAll}{.94}

\newcommand{\gambitredrockTTFltOneNunRatioFloats}{.94}
\newcommand{\gambitredrockTTFltOneNunRatioNonFloats}{.87}
\newcommand{\gambitredrockTTFltOneNunRatioAll}{.90}

\newcommand{\gambitredrockTTFltTwoNunRatioFloats}{1.00}
\newcommand{\gambitredrockTTFltTwoNunRatioNonFloats}{.88}
\newcommand{\gambitredrockTTFltTwoNunRatioAll}{.93}

\newcommand{\gambitredrockTTFltThreeNunRatioFloats}{1.06}
\newcommand{\gambitredrockTTFltThreeNunRatioNonFloats}{.90}
\newcommand{\gambitredrockTTFltThreeNunRatioAll}{.97}

\newcommand{\gambitredrockTTFltFourNunRatioFloats}{.90}
\newcommand{\gambitredrockTTFltFourNunRatioNonFloats}{.89}
\newcommand{\gambitredrockTTFltFourNunRatioAll}{.89}

\newcommand{\bigloosizoFltOneNunRatioFloats}{1.25}
\newcommand{\bigloosizoFltOneNunRatioNonFloats}{.97}
\newcommand{\bigloosizoFltOneNunRatioAll}{1.08}

\newcommand{\bigloosizoFltnzNunRatioFloats}{1.33}
\newcommand{\bigloosizoFltnzNunRatioNonFloats}{.98}
\newcommand{\bigloosizoFltnzNunRatioAll}{1.12}

\newcommand{\bigloosizoFltThreeNunRatioFloats}{1.10}
\newcommand{\bigloosizoFltThreeNunRatioNonFloats}{1.00}
\newcommand{\bigloosizoFltThreeNunRatioAll}{1.04}

\newcommand{\bigloosizoNanNunRatioFloats}{.90}
\newcommand{\bigloosizoNanNunRatioNonFloats}{1.00}
\newcommand{\bigloosizoNanNunRatioAll}{.95}

\newcommand{\gambitsizoFltOneNunRatioFloats}{1.05}
\newcommand{\gambitsizoFltOneNunRatioNonFloats}{.91}
\newcommand{\gambitsizoFltOneNunRatioAll}{.97}

\newcommand{\gambitsizoFltTwoNunRatioFloats}{1.16}
\newcommand{\gambitsizoFltTwoNunRatioNonFloats}{.95}
\newcommand{\gambitsizoFltTwoNunRatioAll}{1.04}

\newcommand{\gambitsizoFltThreeNunRatioFloats}{1.14}
\newcommand{\gambitsizoFltThreeNunRatioNonFloats}{.98}
\newcommand{\gambitsizoFltThreeNunRatioAll}{1.05}

\newcommand{\gambitsizoFltFourNunRatioFloats}{1.04}
\newcommand{\gambitsizoFltFourNunRatioNonFloats}{.96}
\newcommand{\gambitsizoFltFourNunRatioAll}{.99}

\newcommand{\bigloostarfiveFltOneNunRatioFloats}{1.28}
\newcommand{\bigloostarfiveFltOneNunRatioNonFloats}{.97}
\newcommand{\bigloostarfiveFltOneNunRatioAll}{1.10}

\newcommand{\bigloostarfiveFltnzNunRatioFloats}{1.29}
\newcommand{\bigloostarfiveFltnzNunRatioNonFloats}{.97}
\newcommand{\bigloostarfiveFltnzNunRatioAll}{1.10}

\newcommand{\bigloostarfiveFltThreeNunRatioFloats}{1.19}
\newcommand{\bigloostarfiveFltThreeNunRatioNonFloats}{1.02}
\newcommand{\bigloostarfiveFltThreeNunRatioAll}{1.09}

\newcommand{\bigloostarfiveNanNunRatioFloats}{.90}
\newcommand{\bigloostarfiveNanNunRatioNonFloats}{1.06}
\newcommand{\bigloostarfiveNanNunRatioAll}{.99}

\newcommand{\gambitstarfiveFltOneNunRatioFloats}{1.03}
\newcommand{\gambitstarfiveFltOneNunRatioNonFloats}{.88}
\newcommand{\gambitstarfiveFltOneNunRatioAll}{.94}

\newcommand{\gambitstarfiveFltTwoNunRatioFloats}{1.09}
\newcommand{\gambitstarfiveFltTwoNunRatioNonFloats}{.89}
\newcommand{\gambitstarfiveFltTwoNunRatioAll}{.97}

\newcommand{\gambitstarfiveFltThreeNunRatioFloats}{1.06}
\newcommand{\gambitstarfiveFltThreeNunRatioNonFloats}{.88}
\newcommand{\gambitstarfiveFltThreeNunRatioAll}{.95}

\newcommand{\gambitstarfiveFltFourNunRatioFloats}{.94}
\newcommand{\gambitstarfiveFltFourNunRatioNonFloats}{.89}
\newcommand{\gambitstarfiveFltFourNunRatioAll}{.91}

\section{Introduction}

In dynamic and other polymorphic languages, efficiently supporting
floating-point numbers (floats) remains a challenging issue. These languages
require attaching type information to values at run time, including numeric
types like integers and floats. This requirement conflicts with the IEEE754
standard~\cite{ieee754} for encoding floats, which uses the entire 64-bit word
to represent a double-precision float, leaving no space for additional type
information. A straightforward but suboptimal solution is to represent floats as
heap allocated objects. To reduce the overhead of heap allocating floats,
several alternative representations have been proposed, each offering different
trade-offs.

This paper presents a novel technique for encoding floats, called
\emph{self-tagging}, that avoids most float allocations without adversely
affecting the cost of checking types and accessing objects, which are frequent
operations of dynamic languages.

Self-tagging exploits the fact that any invertible bitwise transformation on an
IEEE754 float will map some floats to a bit arrangement that contains the
correct type information at the correct location, thus requiring no memory
allocation for these floats. Using the fact that some ranges of floats appear
more frequently in practical applications~\cite{Lindstrom06, Brown07,
Afroozeh23, Zurstrassen23}, it is possible to define a transformation that
avoids heap allocating almost all floats encountered in practice.

This technique does not require any program static analysis and could be
integrated in most runtime systems of languages that need run time types. It is
developed in the context of 64-bit runtime systems but is also applicable to 32-bit
systems. The rest of this section presents the main encodings in-use as well as
their trade-offs.

\subsection{Tagged Objects}
\label{section:tagged-objects}

The classical and popular solution to preserve type information is to
attach a tag to all objects~\cite{Gudeman1993RepresentingTI,
Leijen2022, gchandbooktags, moderncompilerinctagging}.
An N-bit machine word is used to encode an \emph{object reference} (either a value or a
pointer to a heap allocated value) and a small bit field is reserved in
the word to store type information. For the rest of the paper the term \emph{object}
is used interchangeably with object reference. Tagging allows low-cost type
checks and object access at the cost of losing a few bits for data.
Aligning all heap allocated values to 64-bit machine words conveniently
frees the low bits of pointers to store a 3-bit tag. On many architectures, accessing the fields of the
object can be done at no additional cost by using an offset in the memory dereference instruction to
take into account the tag for that type of object.
Objects can thus be encoded using one of the three
following representations.

\begin{itemize}
\item \textbf{Tagged values} store type information with a tag, and the
  object's value in the remaining bits. This representation does not
  require heap allocation.
\item \textbf{Tagged pointers} also store type information with a tag, but
  values are stored in the heap. The objects are represented by
  pointers with a tag in their low bits.
\item \textbf{Generic pointers} represent all remaining objects as pointers
  with a generic tag. The generic tag is associated with multiple types, and
  type information is stored in a header, in the heap.
\end{itemize}

\begin{figure}
  \includegraphics[width=0.95\textwidth]{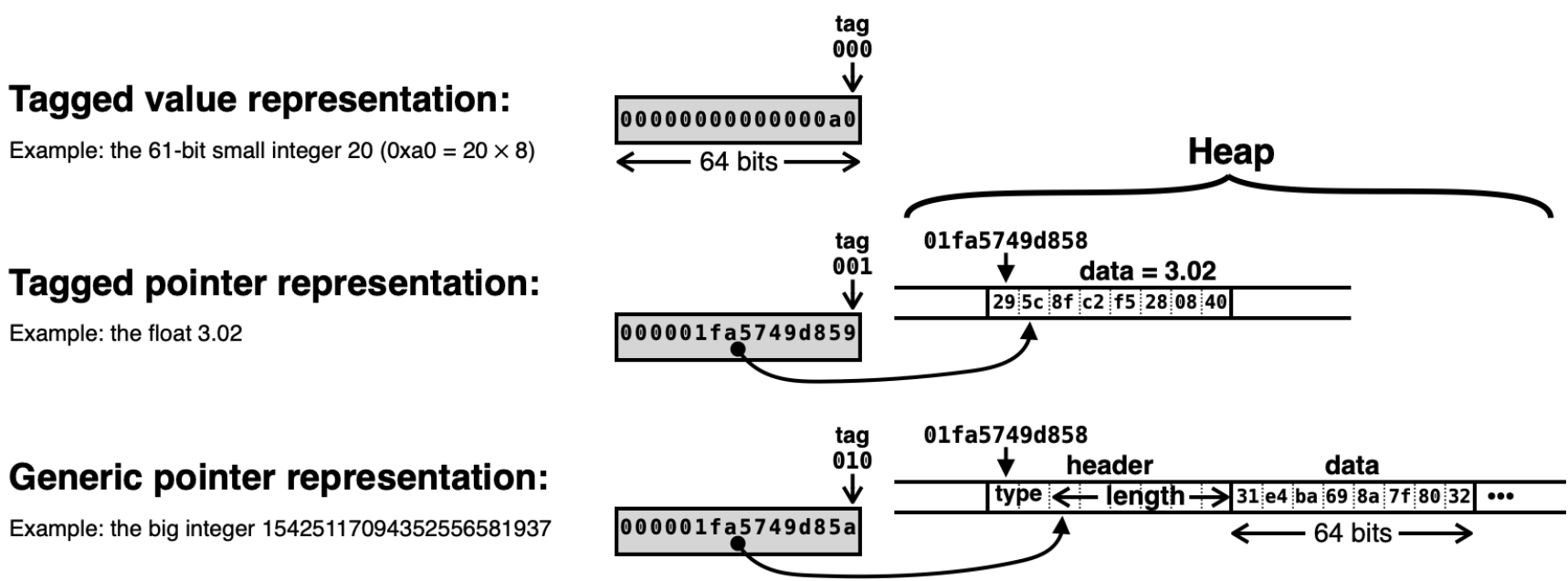}
  \caption{The three representations in a tagged object system (here shown on
  a little-endian machine).}
  \label{figure:tagged-objects}
\end{figure}

Hence, a tag indicates a type, but also a memory representation. A
tagged value representation is the most efficient as it needs no heap
allocation and no memory read.  A tagged pointer still allows
efficient type checks, but requires heap allocation. This introduces
the overhead of dereferencing the pointer, and adds a strain on the
garbage collector. A generic pointer is the least efficient
representation as it involves storing type information in the heap, which
increases space usage and the cost of type checks and memory management. Consequently,
specific tags are a precious resource that must be carefully assigned to the most frequent types
observed in programs, such as small integers or
floats. Figure~\ref{figure:tagged-objects} shows a typical memory
layout of each representation.

As an example, consider an implementation using 3-bit tags. Due to the frequent
use of small integers, a tagged value representation is appropriate for small
integers and the choice of the tag \texttt{000} allows direct
addition/subtraction. Since 3 bits are occupied by the tag, small integers are
limited to 61 bits. This is generally considered an acceptable trade-off since a
two's complement representation of signed integers can accommodate any number of
bits, and 61 bits still offers a wide range of values. For larger values,
generic pointers to heap allocated big integers can be used~\cite{Leijen2022}.
This provides a hybrid representation where the most common integers appear as
efficient tagged values and less-frequent big integers are represented with more
costly generic pointers.

Unfortunately, when it comes to the representation of floats, none of the
aforementioned options are well-suited. Double-precision floats
cannot be represented as tagged values since the IEEE754 standard
enforces the use of 1 sign bit, 11 exponent bits and 52 mantissa bits
(see Figure~\ref{figure:ieee754-encoding}), totaling
64 bits and leaving no space for a tag~\cite{ieee754}.
Parting from this standard is impracticable due to the set of instructions offered by
x86~\cite{intel_instruction_set} and ARM~\cite{arm_instruction_set} architectures
being specific to 32 or 64-bit floats.

Yet, tagged pointers introduce two major inefficiencies for floats.
First, all uses of the value require fetching it from memory.
Second, the heap allocation of all floats increases memory usage
and the cost of computing float results. This is especially disconcerting considering that
allocated floats are often short-lived intermediate results that add a
strain on garbage collectors~\cite{DBLP:conf/icfp/SerranoF96,
gchandbookgenerations}. This has costly implications for programs
performing extensive numerical computations or languages such as
JavaScript where floats are the only available primitive number
type~\cite{ecmascript6}.

Still, tagged objects are straightforward to implement and are thus found in numerous
compilers such as V8~\cite{v8_pointer_compression}, QuickJS~\cite{quickjs}, and Hopc~\cite{DBLP:journals/pacmpl/Serrano21}
(JavaScript), the
Lua interpreter~\cite{luainterpreter} (Lua),
CRuby~\cite{ruby_under_microscope} (Ruby), SBCL~\cite{sbcl} (Lisp), and Gambit~\cite{gambit} and Bigloo~\cite{bigloo} (Scheme).
Recently, CPython also moved toward tagged pointers in the process
of removing its global interpreter lock~\cite{cpythonpep703}.

\subsection{N\MakeLowercase{a}N-boxing}

NaN-boxing circumvents the drawbacks of tagged pointer floats by
reclaiming unused bits in the encodings of floating-point NaN values
to store data~\cite{Gudeman1993RepresentingTI, Leijen2022}.
As per the IEEE754 standard, a NaN value is represented by setting all
11 bits of the exponent to \texttt{1} and a non-zero mantissa (zero is used to
encode \texttt{Infinity}). The mantissa's highest bit distinguishes between a quiet
and signaling NaN, which determines whether the NaN should
signal an exception or fall through operations.

\begin{figure}
  \begin{subfigure}{0.94\textwidth}
    \includegraphics[width=0.94\textwidth]{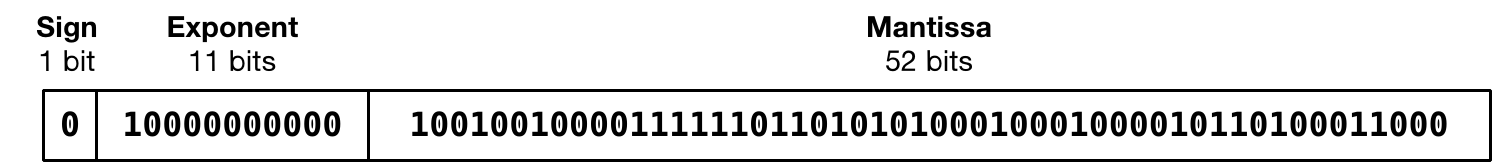}
    \caption{
        IEEE754 floating-point representation of $\pi$.
    }
    \label{figure:ieee754-encoding}
  \end{subfigure}

  \vspace{0.2cm}

  \begin{subfigure}{0.94\textwidth}
    \includegraphics[width=0.94\textwidth]{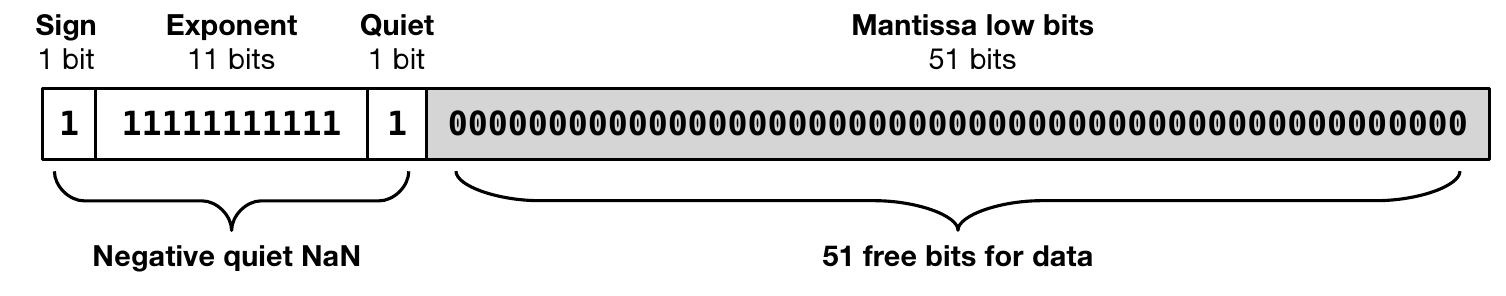}
    \caption{
        NaN-boxing uses the 51 free bits on negative, quiet NaN values
        to store data.
    }
    \label{figure:nan-boxing-encoding}
  \end{subfigure}

  \caption{Floating-point representations of numbers and NaN.}
  
\end{figure}

A subset of NaNs can thus be reserved to encode non-float objects.
A convenient subset is that of negative, quiet NaNs, illustrated in
Figure~\ref{figure:nan-boxing-encoding}, which correspond to the interval from
\texttt{0xfff8\_0000\_0000\_0000} to \texttt{0xffff\_ffff\_ffff\_ffff}. This partitions
floating-point numbers in two intervals. Bit patterns above \texttt{0xfff8\_0000\_0000\_0000}
are reserved for non-float objects. The bit patterns below \texttt{0xfff8\_0000\_0000\_0000}
are reserved for floats (including signaling NaN but not quiet NaN)
and negative quiet NaN are mapped to the bit pattern \texttt{0xfff8\_0000\_0000\_0000}.

On current hardware, memory addresses typically fit in 48 bits. The 51 bits
uncovered by NaN-boxing are thus sufficient for storing tagged pointers with
3-bit tags, and 32-bit small integers. It offers the advantage of unboxed
floats, but impacts the performance of other, non-float objects due to
higher-cost machine instructions to check the type and dereference NaN-boxed
pointers.

This reliance on hardware specific details also interferes with other optimizations and
portability.
Moreover, on 32-bit architectures, NaN-boxing would cripple memory management
because 32-bit float NaNs leave space for only 22-bit pointers, which only
allows addressing 4~MiB.

NaN-boxing is used in a few language implementations, including
SpiderMonkey~\cite{spidermonkey} (JavaScript), tinylisp~\cite{tinylisp} (Lisp),
LuaJIT~\cite{luajit} (Lua), and Zag~\cite{DBLP:conf/iwst/Mason22} (Smalltalk).

\subsection{Pointer-Biased N\MakeLowercase{a}N-boxing}

\textit{NuN}-boxing is a variant of NaN-boxing that alleviates the cost
of dereferencing NaN-boxed pointers~\cite{Leijen2022}. It relies on the
(currently valid) fact that common hardware never returns negative,
quiet NaNs greater than or equal to \texttt{0xfffe\_0000\_0000\_0000}.

\begin{figure}
  \includegraphics[width=0.63\textwidth]{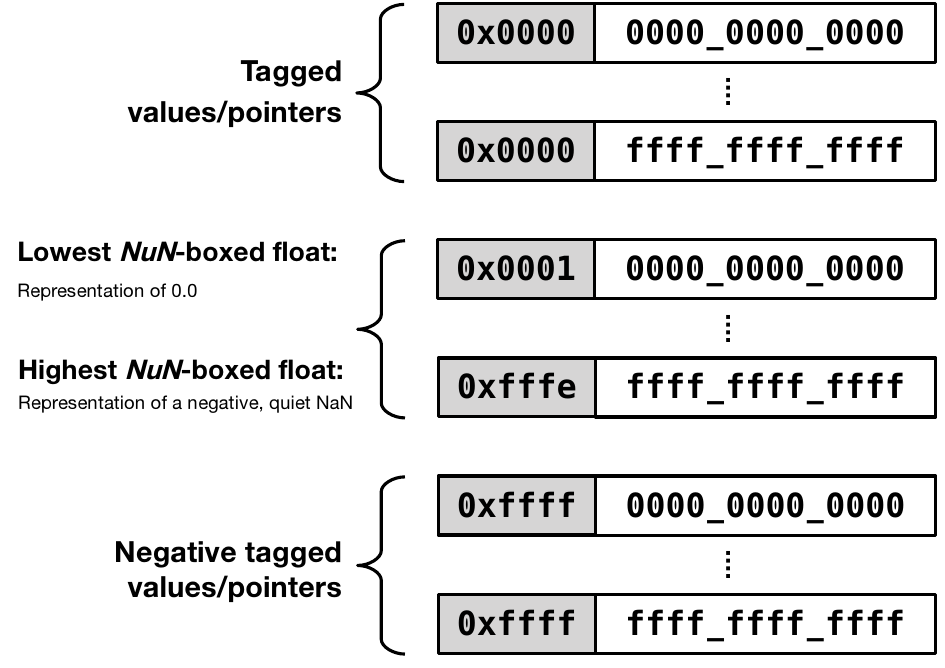}
  \caption{
    Encoding of NuN-boxed values, after having applied the
    \texttt{0x0001\_0000\_0000\_0000} bias to floats. The low and high ranges (whose 16
    highest bits are \texttt{0x0000} or \texttt{0xffff}) are
    used for non-float objects that can be represented as tagged values or
    pointers.
  }
  \label{figure:nun-boxed-encoding}
\end{figure}

Instead of taking advantage of the unused NaN value range
from \texttt{0xfffe\_0000\_0000\_0000} to \texttt{0xffff\_ffff\_ffff\_ffff},
which would result in an encoding similar to standard NaN-boxing,
NuN-boxing biases
all floats by adding \texttt{0x0001\_0000\_0000\_0000}. This frees up the lowest
and highest ranges for 48-bit tagged objects as shown in
Figure~\ref{figure:nun-boxed-encoding}. It however reintroduces a cost
for encoding and decoding floats (adding/subtracting the bias) to operate on their value.

NuN-boxing is used by JavaScriptCore~\cite{javascriptcore} (JavaScript).

\subsection{Contribution}

The contribution of this paper is a thorough analysis of a new approach to
object tagging, named \textit{self-tagging}, that allows attaching type
information to \emph{some} 64-bit objects by applying a bitwise transformation
that preserves their value without heap allocation. This approach is applied to
implement double-precision floats as tagged values instead of tagged pointers.
Contrarily to NaN-boxing, it does not impact the performance of encoding and
decoding other tagged objects and is applicable in the context of 32-bit
architectures. Self-tagging cannot represent all floats and thus requires an
hybrid representation for fast self-tagged and slow heap allocated floats. Yet,
the conducted experiments show that the set of numbers covered by self-tagging
is large enough to remove nearly all heap allocations of floats in
practice. Hence, it provides another alternative to allocated floats with
different trade-offs than NaN/NuN-boxing.

\subsection{Paper Structure}

This paper is structured as follows. Section~\ref{section:self-tagging}
describes the general idea of self-tagging with a focus on its application to
floats. Section~\ref{section:implementation} presents technical details for
portable implementations that could be used by compilers that generate either C
or assembly code. Section~\ref{section:experiment} shows experimental results in
the context of Scheme with the Bigloo and Gambit compilers.
Section~\ref{section:st-on-32-bits} discusses the adaptation of self-tagging to
32-bit architectures. Finally, Section~\ref{section:related-work} presents
related work.


\section{Self-Tagging}
\label{section:self-tagging}

\begin{figure}
  \includegraphics[width=0.95\textwidth]{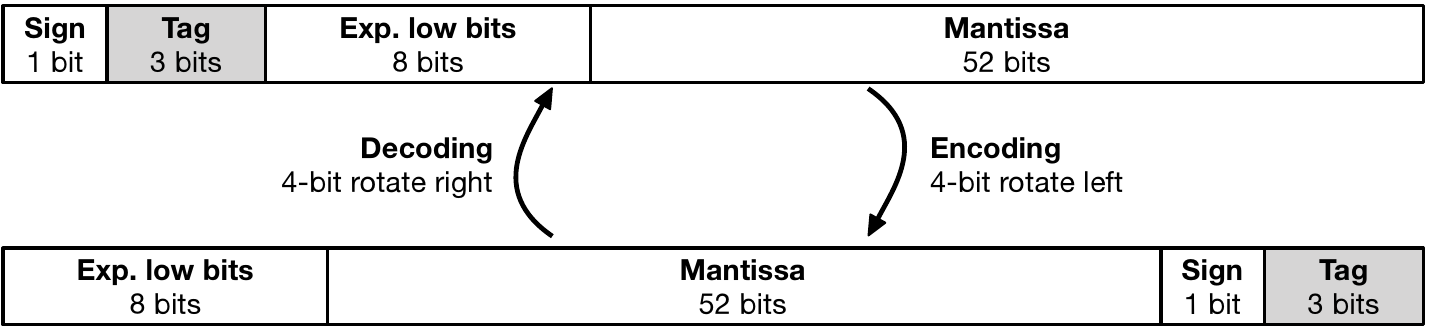}
  
  \caption{
      A float self-tagging representation where the tag corresponds to the
      high bits of the exponent.
      The top bit sequence is a float where the tag is superimposed with
      the exponent's three most significant bits.
      The bottom sequence is the tagged value representation of the float where a 4-bit left rotation
      is applied to place the tag on low bits.
  }
  \label{figure:tagged-float-exponent}
\end{figure}

This section describes \emph{self-tagging}, a tagging technique that
exploits the fact that some values can be encoded to a tagged value by
a bitwise transformation containing the tag corresponding to their
type. Such self-tagged objects avoid the cost of heap allocation.
In this paper,
self-tagging is applied to the specific case of IEEE754 floats. Unless
stated otherwise, double-precision floats and a 64-bit architecture
are implied. However, self-tagging also pertains to 32-bit architectures,
which is discussed in Section~\ref{section:st-on-32-bits}.

Consider the common case of tagged objects with 3-bit tags (8 available tags).
Given tag $T$ for floats, any invertible bitwise transformation will map $1 / 8$
of possible floats to a bit arrangement that contains $T$ on its low bits. These
floats can thus be encoded to tagged values with this transformation and decoded
back to a IEEE754 representation by applying the inverse transformation. The set
of floats that are mapped to a correctly tagged value by a given transformation
are said to be \emph{self-tagged} and require no memory allocation. Since only a
subset of all floats can be self-tagged, the remaining floats must be heap
allocated with tagged or generic pointers. Yet, the proportion of self-tagged
floats can be increased by assigning more than one tag to self-tagged floats. In
general, using $n$ tags for self-tagging allows avoiding heap allocation of $n /
8$ of all possible floats.

As an example, consider the case where the encoding transformation is a 4-bit
rotation to the left, which has the effect of using the 3 high bits of an
IEEE754 float's exponent as tag for self-tagged floats (Figure
\ref{figure:tagged-float-exponent}). This transformation has the interesting
property that each tag captures a contiguous range of floats. For instance,
assigning the tags \texttt{000}, \texttt{011} and \texttt{100} to self-tagged
floats, avoids heap allocating all floats in the ranges $0.0$ .. $7 \times 10^{-251}$
and $1.7 \times 10^{-77} $ .. $2.3 \times 10^{77}$ as well as the corresponding
negative ranges (detailed calculations of these ranges are provided in
Section~\ref{section:float-coverage}).

This suggests that, while many transformations can be chosen to implement
self-tagging, some choices are better than others. Three properties make for a
useful self-tagging scheme. It must:

\begin{itemize}
  \item \textbf{Capture intervals of common floats}: most values computed in
  practice are either $\pm 0.0$ or centered around $\pm 1.0$~\cite{Lindstrom06,
  Brown07, Afroozeh23, Zurstrassen23}. These should be prioritized to reduce
  memory allocation.
  \item \textbf{Have an efficient encoding/decoding}: the cost of the
  transformation must not outweigh the benefit of decreased memory allocations
  and strain on the garbage collector.
  \item \textbf{Use few tags}: it should reserve as few tags as possible
  to avoid having to defer to generic pointers for other important types.
  Ideally, a single tag should be used for all common floats, and uncommon floats
  should be represented either with tagged or generic pointers.
\end{itemize}

Section~\ref{section:float-coverage} presents self-tagging variants that offer
different tradeoffs between captured intervals and number of reserved tags.
Section~\ref{section:implementation} describes efficient implementations for
each variant.


\begin{figure}
    \centering
    \setlength\extrarowheight{2pt}
    \setlength{\tabcolsep}{3.7pt}
    \begin{tabular}{|c|c|c|r|r|r|r|r|r|r|r|r|r|}
      \multicolumn{1}{c}{\textbf{5 most}} & \multicolumn{1}{c}{\textbf{5 most}} & \multicolumn{1}{c}{} &
      \multicolumn{9}{c}{}
        \\[-1ex]

      \multicolumn{1}{c}{\textbf{signif.}} & \multicolumn{1}{c}{\textbf{signif.}} & \multicolumn{1}{c}{} &
      \multicolumn{9}{c}{}
        \\[-1ex]

      \multicolumn{1}{c}{\textbf{expo.}} & \multicolumn{1}{c}{\textbf{expo.}} & \multicolumn{1}{c}{\textbf{Value}} &
      \multicolumn{9}{c}{}
        \\[-1ex]

        \multicolumn{1}{c}{\textbf{bits}} & \multicolumn{1}{c}{\textbf{bits + 1}} & \multicolumn{1}{c}{\textbf{range}} &
        \multicolumn{1}{l}{\raisebox{-2 pt}[0 pt][0 pt]{\makebox[1px][l]{\rotatebox[origin=l]{40}{\small \textbf{fft}}}}} &
        \multicolumn{1}{l}{\raisebox{-2 pt}[0 pt][0 pt]{\makebox[1px][l]{\rotatebox[origin=l]{40}{\small \textbf{fibfp}}}}} &
        \multicolumn{1}{l}{\raisebox{-2 pt}[0 pt][0 pt]{\makebox[1px][l]{\rotatebox[origin=l]{40}{\small \textbf{mbrot}}}}} &
        \multicolumn{1}{l}{\raisebox{-2 pt}[0 pt][0 pt]{\makebox[1px][l]{\rotatebox[origin=l]{40}{\small \textbf{nucleic}}}}} &
        \multicolumn{1}{l}{\raisebox{-2 pt}[0 pt][0 pt]{\makebox[1px][l]{\rotatebox[origin=l]{40}{\small \textbf{pnpoly}}}}} &
        \multicolumn{1}{l}{\raisebox{-2 pt}[0 pt][0 pt]{\makebox[1px][l]{\rotatebox[origin=l]{40}{\small \textbf{ray}}}}} &
        \multicolumn{1}{l}{\raisebox{-2 pt}[0 pt][0 pt]{\makebox[1px][l]{\rotatebox[origin=l]{40}{\small \textbf{simplex}}}}} &
        \multicolumn{1}{l}{\raisebox{-2 pt}[0 pt][0 pt]{\makebox[1px][l]{\rotatebox[origin=l]{40}{\small \textbf{sumfp}}}}} &
        \multicolumn{1}{l}{\raisebox{-2 pt}[0 pt][0 pt]{\makebox[1px][l]{\rotatebox[origin=l]{40}{\small \textbf{sum1}}}}} &
        \multicolumn{1}{l}{\raisebox{-2 pt}[0 pt][0 pt]{\makebox[1px][l]{\rotatebox[origin=l]{40}{\small \textbf{\cite{Zurstrassen23}}}}}}
        \\
        \hline

        \cellcolor{tag0} & \cellcolor{1tag} & {\texttt{0.0}}
        &  79\% &   9\% &   0\% &  1\% &   11\% &   7\% &  11\% &  13\% &  0\% & 0\%
        \\

        \cline{3-13}
        \raisebox{0.8 em}[0 pt]{\cellcolor{tag0}\underline{000}00} & \raisebox{0.8 em}[0 pt]{\cellcolor{1tag}0\underline{000}1} & {\footnotesize 5e-324 .. 2.1e-289}
        & \multicolumn{1}{c|}{-}    & \multicolumn{1}{c|}{-}    & \multicolumn{1}{c|}{-}    &  \multicolumn{1}{c|}{-}   &  \multicolumn{1}{c|}{-}   &  \multicolumn{1}{c|}{-}   &  \multicolumn{1}{c|}{-}   &  \multicolumn{1}{c|}{-}   &  \multicolumn{1}{c|}{-} &  0\%
        \\
        \hline

        {\cellcolor{tag0}\underline{000}01} & {\cellcolor{2tag}00010} & {\footnotesize 2.1e-289 .. 3.8e-270}
        & \multicolumn{1}{c|}{-}    & \multicolumn{1}{c|}{-}    & \multicolumn{1}{c|}{-}    &  \multicolumn{1}{c|}{-}   &  \multicolumn{1}{c|}{-}   &  \multicolumn{1}{c|}{-}   &  \multicolumn{1}{c|}{-}   &  \multicolumn{1}{c|}{-}   &  \multicolumn{1}{c|}{-} &  0\%
        \\
        \hline

        {\cellcolor{tag0}\underline{000}10} & {00011} & {\footnotesize 3.8e-270 .. 7e-251}
        & \multicolumn{1}{c|}{-}    & \multicolumn{1}{c|}{-}    & \multicolumn{1}{c|}{-}    &  \multicolumn{1}{c|}{-}   &  \multicolumn{1}{c|}{-}   &  \multicolumn{1}{c|}{-}   &  \multicolumn{1}{c|}{-}   &  \multicolumn{1}{c|}{-}   &  \multicolumn{1}{c|}{-} &  0\%
        \\
        \hline

        {\cellcolor{tag0}\underline{000}11} & {00100} & {\footnotesize 7e-251 .. 1.3e-231}
        & \multicolumn{1}{c|}{-}    & \multicolumn{1}{c|}{-}    & \multicolumn{1}{c|}{-}    &  \multicolumn{1}{c|}{-}   &  \multicolumn{1}{c|}{-}   &  \multicolumn{1}{c|}{-}   &  \multicolumn{1}{c|}{-}   &  \multicolumn{1}{c|}{-}   &  \multicolumn{1}{c|}{-} &  0\%
        \\
        \hline

        {00100} & {00101} & {\footnotesize 1.3e-231 .. 2.4e-212}
        & \multicolumn{1}{c|}{-}    & \multicolumn{1}{c|}{-}    & \multicolumn{1}{c|}{-}    &  \multicolumn{1}{c|}{-}   &  \multicolumn{1}{c|}{-}   &  \multicolumn{1}{c|}{-}   &  \multicolumn{1}{c|}{-}   &  \multicolumn{1}{c|}{-}   &  \multicolumn{1}{c|}{-} &  0\%
        \\
        \hline

        {00101} & {00110} & {\footnotesize 2.4e-212 .. 4.4e-193}
        & \multicolumn{1}{c|}{-}    & \multicolumn{1}{c|}{-}    & \multicolumn{1}{c|}{-}    &  \multicolumn{1}{c|}{-}   &  \multicolumn{1}{c|}{-}   &  \multicolumn{1}{c|}{-}   &  \multicolumn{1}{c|}{-}   &  \multicolumn{1}{c|}{-}   &  \multicolumn{1}{c|}{-} &  0\%
        \\
        \hline

        {00110} & {00111} & {\footnotesize 4.4e-193 .. 8.1e-174}
        & \multicolumn{1}{c|}{-}    & \multicolumn{1}{c|}{-}    & \multicolumn{1}{c|}{-}    &  \multicolumn{1}{c|}{-}   &  \multicolumn{1}{c|}{-}   &  \multicolumn{1}{c|}{-}   &  \multicolumn{1}{c|}{-}   &  \multicolumn{1}{c|}{-}   &  \multicolumn{1}{c|}{-} &  0\%
        \\
        \hline

        {00111} & {01000} & {\footnotesize 8.1e-174 .. 1.5e-154}
        & \multicolumn{1}{c|}{-}    & \multicolumn{1}{c|}{-}    & \multicolumn{1}{c|}{-}    &  \multicolumn{1}{c|}{-}   &  \multicolumn{1}{c|}{-}   &  \multicolumn{1}{c|}{-}   &  \multicolumn{1}{c|}{-}   &  \multicolumn{1}{c|}{-}   &  \multicolumn{1}{c|}{-} &  0\%
        \\
        \hline

        {01000} & {01001} & {\footnotesize 1.5e-154 .. 2.8e-135}
        & \multicolumn{1}{c|}{-}    & \multicolumn{1}{c|}{-}    & \multicolumn{1}{c|}{-}    &  \multicolumn{1}{c|}{-}   &  \multicolumn{1}{c|}{-}   &  \multicolumn{1}{c|}{-}   &  \multicolumn{1}{c|}{-}   &  \multicolumn{1}{c|}{-}   &  \multicolumn{1}{c|}{-} &  0\%
        \\
        \hline

        {01001} & {01010} & {\footnotesize 2.8e-135 .. 5.1e-116}
        & \multicolumn{1}{c|}{-}    & \multicolumn{1}{c|}{-}    & \multicolumn{1}{c|}{-}    &  \multicolumn{1}{c|}{-}   &  \multicolumn{1}{c|}{-}   &  \multicolumn{1}{c|}{-}   &  \multicolumn{1}{c|}{-}   &  \multicolumn{1}{c|}{-}   &  \multicolumn{1}{c|}{-} &  0\%
        \\
        \hline

        {01010} & {01011} & {\footnotesize 5.1e-116 .. 9.4e-97}
        & \multicolumn{1}{c|}{-}    & \multicolumn{1}{c|}{-}    & \multicolumn{1}{c|}{-}    &  \multicolumn{1}{c|}{-}   &  \multicolumn{1}{c|}{-}   &  \multicolumn{1}{c|}{-}   &  \multicolumn{1}{c|}{-}   &  \multicolumn{1}{c|}{-}   &  \multicolumn{1}{c|}{-} &  0\%
        \\
        \hline

        {01011} & {01100} & {\footnotesize 9.4e-97 .. 1.7e-77}
        & \multicolumn{1}{c|}{-}    & \multicolumn{1}{c|}{-}    & \multicolumn{1}{c|}{-}    &  \multicolumn{1}{c|}{-}   &  \multicolumn{1}{c|}{-}   &  \multicolumn{1}{c|}{-}   &  \multicolumn{1}{c|}{-}   &  \multicolumn{1}{c|}{-}   &  \multicolumn{1}{c|}{-} &  0\%
        \\
        \hline

        {\cellcolor{tag3}\underline{011}00} & {01101} & {\footnotesize 1.7e-77 .. 3.2e-58}
        & \multicolumn{1}{c|}{-}    & \multicolumn{1}{c|}{-}    & \multicolumn{1}{c|}{-}    &  \multicolumn{1}{c|}{-}   &  \multicolumn{1}{c|}{-}   &  \multicolumn{1}{c|}{-}   &  \multicolumn{1}{c|}{-}   &  \multicolumn{1}{c|}{-}   &  \multicolumn{1}{c|}{-} &  0\%
        \\
        \hline

        {\cellcolor{tag3}\underline{011}01} & {01110} & {\footnotesize 3.2e-58 .. 5.9e-39}
        & \multicolumn{1}{c|}{-}    & \multicolumn{1}{c|}{-}    & \multicolumn{1}{c|}{-}    &  \multicolumn{1}{c|}{-}   &  \multicolumn{1}{c|}{-}   &  \multicolumn{1}{c|}{-}   &  \multicolumn{1}{c|}{-}   &  \multicolumn{1}{c|}{-}   &  \multicolumn{1}{c|}{-} &  1\%
        \\
        \hline

        {\cellcolor{tag3}\underline{011}10} & {\cellcolor{2tag}01111} & {\footnotesize 5.9e-39 .. 1.1e-19}
        & 0\%    & \multicolumn{1}{c|}{-}    & \multicolumn{1}{c|}{-}    &  \multicolumn{1}{c|}{-}   &  \multicolumn{1}{c|}{-}   &  0\%   &  \multicolumn{1}{c|}{-}   &  \multicolumn{1}{c|}{-}   &  \multicolumn{1}{c|}{-} &  3\%
        \\
        \hline

        {\cellcolor{tag3}\underline{011}11} & {\cellcolor{1tag}1\underline{000}0} & {\footnotesize 1.1e-19 .. 2}
        &  21\% &  28\% &   89\% &  61\% &  63\% &  22\% &  64\% &  13\% &  0\% &  61\% 
        \\
        \hline

        {\cellcolor{tag4}\underline{100}00} & {\cellcolor{1tag}1\underline{000}1} & {\footnotesize 2 .. 3.7e19}
        &   0\% &  63\% & 11\% &  38\% &  26\% &  70\% &  25\% &  75\% &  100\% &  28\% 
        \\
        \hline

        {\cellcolor{tag4}\underline{100}01} & {\cellcolor{2tag}10010} & {\footnotesize 3.7e19 .. 6.8e38}
        & \multicolumn{1}{c|}{-}    & \multicolumn{1}{c|}{-}    & \multicolumn{1}{c|}{-}    &  \multicolumn{1}{c|}{-}   &  \multicolumn{1}{c|}{-}   &  \multicolumn{1}{c|}{-}   &  \multicolumn{1}{c|}{-}   &  \multicolumn{1}{c|}{-}   &  \multicolumn{1}{c|}{-} &  2\% 
        \\
        \hline

        {\cellcolor{tag4}\underline{100}10} & {10011} & {\footnotesize 6.8e38 .. 1.3e58}
        & \multicolumn{1}{c|}{-}    & \multicolumn{1}{c|}{-}    & \multicolumn{1}{c|}{-}    &  \multicolumn{1}{c|}{-}   &  \multicolumn{1}{c|}{-}   &  \multicolumn{1}{c|}{-}   &  \multicolumn{1}{c|}{-}   &  \multicolumn{1}{c|}{-}   &  \multicolumn{1}{c|}{-} &  1\% 
        \\
        \hline

        {\cellcolor{tag4}\underline{100}11} & {10100} & {\footnotesize 1.3e58 .. 2.3e77}
        & \multicolumn{1}{c|}{-}    & \multicolumn{1}{c|}{-}    & \multicolumn{1}{c|}{-}    &  \multicolumn{1}{c|}{-}   &  \multicolumn{1}{c|}{-}   &  \multicolumn{1}{c|}{-}   &  \multicolumn{1}{c|}{-}   &  \multicolumn{1}{c|}{-}   &  \multicolumn{1}{c|}{-} &  0\% 
        \\
        \hline

        {10100} & {10101} & {\footnotesize 2.3e77 .. 4.3e96}
        & \multicolumn{1}{c|}{-}    & \multicolumn{1}{c|}{-}    & \multicolumn{1}{c|}{-}    &  \multicolumn{1}{c|}{-}   &  \multicolumn{1}{c|}{-}   &  \multicolumn{1}{c|}{-}   &  \multicolumn{1}{c|}{-}   &  \multicolumn{1}{c|}{-}   &  \multicolumn{1}{c|}{-} &  0\% 
        \\
        \hline

        {10101} & {10110} & {\footnotesize 4.3e96 .. 7.9e115}
        & \multicolumn{1}{c|}{-}    & \multicolumn{1}{c|}{-}    & \multicolumn{1}{c|}{-}    &  \multicolumn{1}{c|}{-}   &  \multicolumn{1}{c|}{-}   &  \multicolumn{1}{c|}{-}   &  \multicolumn{1}{c|}{-}   &  \multicolumn{1}{c|}{-}   &  \multicolumn{1}{c|}{-} &  0\% 
        \\
        \hline

        {10110} & {10111} & {\footnotesize 7.9e115 .. 1.5e135}
        & \multicolumn{1}{c|}{-}    & \multicolumn{1}{c|}{-}    & \multicolumn{1}{c|}{-}    &  \multicolumn{1}{c|}{-}   &  \multicolumn{1}{c|}{-}   &  \multicolumn{1}{c|}{-}   &  \multicolumn{1}{c|}{-}   &  \multicolumn{1}{c|}{-}   &  \multicolumn{1}{c|}{-} &  0\% 
        \\
        \hline

        {10111} & {11000} & {\footnotesize 1.5e135 .. 2.7e154}
        & \multicolumn{1}{c|}{-}    & \multicolumn{1}{c|}{-}    & \multicolumn{1}{c|}{-}    &  \multicolumn{1}{c|}{-}   &  \multicolumn{1}{c|}{-}   &  \multicolumn{1}{c|}{-}   &  \multicolumn{1}{c|}{-}   &  \multicolumn{1}{c|}{-}   &  \multicolumn{1}{c|}{-} &  0\% 
        \\
        \hline

        {11000} & {11001} & {\footnotesize 2.7e154 .. 4.9e173}
        & \multicolumn{1}{c|}{-}    & \multicolumn{1}{c|}{-}    & \multicolumn{1}{c|}{-}    &  \multicolumn{1}{c|}{-}   &  \multicolumn{1}{c|}{-}   &  \multicolumn{1}{c|}{-}   &  \multicolumn{1}{c|}{-}   &  \multicolumn{1}{c|}{-}   &  \multicolumn{1}{c|}{-} &  0\% 
        \\
        \hline

        {11001} & {11010} & {\footnotesize 4.9e173 .. 9.1e192}
        & \multicolumn{1}{c|}{-}    & \multicolumn{1}{c|}{-}    & \multicolumn{1}{c|}{-}    &  \multicolumn{1}{c|}{-}   &  \multicolumn{1}{c|}{-}   &  \multicolumn{1}{c|}{-}   &  \multicolumn{1}{c|}{-}   &  \multicolumn{1}{c|}{-}   &  \multicolumn{1}{c|}{-} &  0\% 
        \\
        \hline

        {11010} & {11011} & {\footnotesize 9.1e192 .. 1.7e212}
        & \multicolumn{1}{c|}{-}    & \multicolumn{1}{c|}{-}    & \multicolumn{1}{c|}{-}    &  \multicolumn{1}{c|}{-}   &  \multicolumn{1}{c|}{-}   &  \multicolumn{1}{c|}{-}   &  \multicolumn{1}{c|}{-}   &  \multicolumn{1}{c|}{-}   &  \multicolumn{1}{c|}{-} &  0\% 
        \\
        \hline

        {11011} & {11100} & {\footnotesize 1.7e212 .. 3.1e231}
        & \multicolumn{1}{c|}{-}    & \multicolumn{1}{c|}{-}    & \multicolumn{1}{c|}{-}    &  \multicolumn{1}{c|}{-}   &  \multicolumn{1}{c|}{-}   &  \multicolumn{1}{c|}{-}   &  \multicolumn{1}{c|}{-}   &  \multicolumn{1}{c|}{-}   &  \multicolumn{1}{c|}{-} &  0\% 
        \\
        \hline

        {\cellcolor{tag7}\underline{111}00} & {11101} & {\footnotesize 3.1e231 .. 5.7e250}
        & \multicolumn{1}{c|}{-}    & \multicolumn{1}{c|}{-}    & \multicolumn{1}{c|}{-}    &  \multicolumn{1}{c|}{-}   &  \multicolumn{1}{c|}{-}   &  \multicolumn{1}{c|}{-}   &  \multicolumn{1}{c|}{-}   &  \multicolumn{1}{c|}{-}   &  \multicolumn{1}{c|}{-} &  0\% 
        \\
        \hline

        {\cellcolor{tag7}\underline{111}01} & {11110} & {\footnotesize 5.7e250 .. 1.1e270}
        & \multicolumn{1}{c|}{-}    & \multicolumn{1}{c|}{-}    & \multicolumn{1}{c|}{-}    &  \multicolumn{1}{c|}{-}   &  \multicolumn{1}{c|}{-}   &  \multicolumn{1}{c|}{-}   &  \multicolumn{1}{c|}{-}   &  \multicolumn{1}{c|}{-}   &  \multicolumn{1}{c|}{-} &  0\% 
        \\
        \hline

        {\cellcolor{tag7}\underline{111}10} & {\cellcolor{2tag}11111} & {\footnotesize 1.1e270 .. 1.9e289}
        & \multicolumn{1}{c|}{-}    & \multicolumn{1}{c|}{-}    & \multicolumn{1}{c|}{-}    &  \multicolumn{1}{c|}{-}   &  \multicolumn{1}{c|}{-}   &  \multicolumn{1}{c|}{-}   &  \multicolumn{1}{c|}{-}   &  \multicolumn{1}{c|}{-}   &  \multicolumn{1}{c|}{-} &  0\% 
        \\
        \hline

        \cellcolor{tag7} & \cellcolor{1tag} & {\footnotesize 1.9e289 .. 1.8e308}
        & \multicolumn{1}{c|}{-}    & \multicolumn{1}{c|}{-}    & \multicolumn{1}{c|}{-}    &  \multicolumn{1}{c|}{-}   &  \multicolumn{1}{c|}{-}   &  0\%   &  \multicolumn{1}{c|}{-}   &  \multicolumn{1}{c|}{-}   &  \multicolumn{1}{c|}{-} &  0\% 
        \\
        \cline{3-13}
        \raisebox{0.8 em}[0 pt]{\cellcolor{tag7}\underline{111}11} & \raisebox{0.8 em}[0 pt]{\cellcolor{1tag}0\underline{000}0} & {\textit{Infinity/NaN}}
        & \multicolumn{1}{c|}{-}    & \multicolumn{1}{c|}{-}    & \multicolumn{1}{c|}{-}    &  \multicolumn{1}{c|}{-}   &  \multicolumn{1}{c|}{-}   &  \multicolumn{1}{c|}{-}   &  \multicolumn{1}{c|}{-}   &  \multicolumn{1}{c|}{-}   &  \multicolumn{1}{c|}{-}  &  {\small N/A}
        \\
        \hline
    \end{tabular}

    \caption{ The absolute value ranges captured by each combination of the 5
      most significant exponent bits and the proportion of all floats computed
      by float-intensive R7RS benchmarks. Intervals of interest are highlighted.
      Empty entries indicate no float in this range, whereas 0\% means that very
      few floats were generated (less than 0.5\%). The last column shows the
      relative occurence of float for each interval reported
      by~\cite[Fig.~7]{Zurstrassen23}. }
\label{figure:tag-interval-table}

\end{figure}

\subsection{Float Coverage}
\label{section:float-coverage}

Self-tagging introduces the notion of tags that indicate which values of
a type are not heap allocated.
This contrasts with standard object tagging that invites assigning tags
with efficient handling of frequent types in mind, regardless of values.
Consequently, self-tagging should be used for the \emph{most common} floats to
maximize the likelihood that a float will be self-tagged. Floats whose magnitude
is concentrated near $1.0$ have been observed to be more common~\cite{Brown07,
Lindstrom06, Afroozeh23, Zurstrassen23}, with a relative occurrence that
decreases sharply away from $1.0$~\cite{Zurstrassen23}. Zeros are common, but
subnormal numbers are rarely used~\cite{Zurstrassen23}.

These observations from~\cite{Brown07, Lindstrom06, Afroozeh23, Zurstrassen23}
are reproduced by instrumenting the R7RS benchmarks, the standard benchmark
suite for Scheme~\cite{r7rs-benchmarks}, to create a profile of computed floats.
All benchmarks labelled as \emph{float} benchmarks were profiled (this includes
\texttt{fibfp}=recursive Fibonacci with floats, \texttt{fft}=1024 points Fast
Fourier Transform, \texttt{mbrot}=mandelbrot set, \texttt{nucleic}=3D structure
determination of a nucleic acid, \texttt{pnpoly}=determine if a 2D point is in a
20 sided polygon, \texttt{ray}=raytrace a scene of 33 spheres,
\texttt{simplex}=linear programming using simplex method, \texttt{sumfp}=add the
floats 0 to $10^6$, and \texttt{sum1}=add a file of floats).

The results are shown in Figure~\ref{figure:tag-interval-table}. The table
contains 32 rows, one for each combination of the 5 most significant bits of the
float's 11-bit exponent field, which is indicated in the first column (for now,
ignore the second column). Each of the right columns indicates what proportion
of computed floats have a specific combination of most significant exponent
bits. The computed floats are clearly clustered around $1.0$, specifically in
the range $1.1 \times 10^{-19}$ .. $3.7 \times 10^{19}$, and the value 0.0 also
occurs frequently for some programs. In fact, R7RS benchmarks have a
distribution of floats that appears even more tightly concentrated near $1.0$
than that reported by~\cite[Fig.~7]{Zurstrassen23}, shown in the rightmost
column. Hence, other scientific programs may exhibit profiles that use slightly
wider ranges of floats. It is apparent that many programs will operate with
floats that are predominantly those around $1.0$ on a logarithmic scale, and
also 0.0.

\subsection{Self-Tagging with 3 Tags and 4 Tags}
\label{section:3tags}

The float profile of Figure~\ref{figure:tag-interval-table} has the property
that almost all floats computed have the 3 most significant bits of the exponent
field that are either \texttt{000} (\tagzerocolorname rows), \texttt{011}
(\tagthreecolorname rows), or \texttt{100} (\tagfourcolorname rows). Moreover,
among the values captured by \texttt{000}, only $\pm$\texttt{0.0} are used.

These 3 exponent bits can be shifted using a bit rotation to align them with the
lower 3 tag bits as shown in Figure~\ref{figure:tagged-float-exponent}.  Thus a
float is encoded to a tagged value by a 4-bit rotation to the left (or 60-bit to
the right). A tagged value is decoded to a float with the inverse rotation. This
corresponds to the encoding discussed in the previous section
(Figure~\ref{figure:tagged-float-exponent}).

Test programs did not compute the float values $\pm$Infinity and NaN, but those
values would be assigned the tag \texttt{111} (\tagsevencolorname rows).
Assigning this fourth tag to self-tagged floats might be useful to avoid
heap allocated floats in programs frequently computing these values.

\subsection{Self-Tagging with 2 Tags and Preallocated Zeros}
\label{section:2tags}

As is shown in the float profile of Figure~\ref{figure:tag-interval-table} the
tag \texttt{000} is only used to capture the values $\pm$\texttt{0.0}.  It can
be freed for other uses by preallocating the values $\pm$\texttt{0.0} using tagged pointer or generic
pointer representation. When a float must be encoded as an object it is compared
to zero, in which case one of the preallocated zeroes is returned, thus avoiding
a new allocation. If this is done then only 2 tags are needed to cover the most
frequent cases: \texttt{011} and \texttt{100}. Unfortunately, this approach may
raise the number of failed branch predictions due to the test for
$\pm$\texttt{0.0} (this is further discussed in
Section~\ref{section:branch-prediction}). 

\subsection{Self-Tagging with 1 Tag}
\label{section:1tag}

Self-tagging can also be achieved with a single tag by a different
transformation of the most significant exponent bits. The 5 most significant
bits of the exponent are added to 1 and the middle 3 bits are used as the tag
(using a rotation 5 places to the left). This corresponds to the second column
of Figure~\ref{figure:tag-interval-table}. Note that the tag \texttt{000}
(\onetagcolorname rows) covers all the useful ranges, including $\pm$Infinity
and NaN that are not covered by the 2-tag and 3-tag variants.

It is noteworthy that there is nothing special about the tag
\texttt{000}.  If for some external reason some other tag is more
convenient it can easily be assigned to 1-tag self-tagged floats by
adding $1 + 2 \times tag$ rather than 1 to the 5 most significant
exponent bits.

Putting all of this together, the self-tagged encoding of a float
whose IEEE754 64-bit representation is the integer $n$ can be computed
as $(n \oplus ((1 + 2 \times tag) << 58)) <<_{rot} 5$, where $\oplus$
is addition modulo $2^{64}$ and $<<_{rot}$ is the left rotation
operator. The operations are reversed to perform the decoding of a
self-tagged float to its IEEE754 64-bit representation.

This variant has the advantages of covering all the useful ranges of floats
computed by R7RS benchmarks and about 90\% of floats reported
by~\cite{Zurstrassen23}, while using a single tag, which leaves more tags
available for other types.

\subsection{Self-Tagging with a Different Exponent Bias}
\label{section:2tags2}

A small change in the exponent bias of the 1-tag variant can be used to get
another self-tagging variant.  If $2 \times tag_1$ is added to the highest 5
bits and the middle 3 bits are examined, then when they are equal to $tag_1$ or
$tag_2=tag_1-1$ they are self-tagged floats. This effectively doubles coverage
by including both the \twotagcolorname and \onetagcolorname rows of the second
column of Figure~\ref{figure:tag-interval-table}. The following ranges will be
self-tagged: $0 .. 3.8\times 10^{-270} \cup 5.9\times 10^{-39} .. 6.8\times
10^{38} \cup 1.1\times 10^{270} .. Infinity/NaN$. To put this into perspective,
this is a superset of the IEEE754 32-bit floats, so any floating point value
that can be represented as a IEEE754 32-bit float can be self-tagged with this
variant which uses 2 of the 8 3-bit tags.  This is likely a better use of 2 tags
than the 2-tag variant of Section~\ref{section:2tags}, which does not cover the
ranges containing 0, Infinity, and NaN.

When compared to the 1-tag variant, there will be almost no difference in heap
allocations for programs with similar float profiles as
Figure~\ref{figure:tag-interval-table}.  However, as will be explained in the
next section, it yields more compact code, because the bias can be added after
the rotation, and the need for a bias disappears when $tag_1$=\texttt{000}.



\section{Implementation}
\label{section:implementation}

This section discusses implementation details for float-related operations in
dynamic languages. The types \texttt{f64} and \texttt{f32} will refer to IEEE754
64-bit and 32-bit doubles and the types \texttt{i64} and \texttt{i32} will refer
to 64-bit and 32-bit integers. Moreover, the type \texttt{object} will refer to
values in the language (the sum type of floats, integers, booleans, and other
objects). To illustrate the low-level implications of the implementation and
give a sense of the execution time performance, assembly code for the
Intel x86 architecture is provided. This section focusses on a 64-bit
implementation where \texttt{object} is an \texttt{i64} and floats are
\texttt{f64}, but it is straightforward to implement the operations similarly on
a 32-bit implementation where the type \texttt{object} is an \texttt{i32} and
floats are \texttt{f32}.

Testing the low tag bits of an \texttt{object} value is a basic need
for float-related operations to convert \texttt{f64} to and from
\texttt{object} and for dynamic type checks. In particular it is
possible to determine if a \texttt{f64} can be self-tagged by going
through the encoding process and checking that the resulting tag is
one of the tags (or \emph{the} tag) appropriate for the chosen
self-tagging variant.

\subsection{Single Tag Testing}

Checking that the \emph{N} low bits are equal to a specific tag can be
achieved by masking those bits followed by a comparison. Assuming the
\texttt{object} value is in the x86 64-bit register \texttt{rax}, then
the following sequence will branch to label \texttt{tag\_matches}
when the 3 low bits are equal to \textit{tag}\footnote{Intel
assembler syntax is used throughout.}:

\begin{minipage}{\textwidth}
\footnotesize
\begin{lstlisting}[language={[x86gnuatt]Assembler},mathescape]
  and al, 7
  cmp al, $tag$
  jz  tag_matches
\end{lstlisting}
\end{minipage}

\noindent
Note that by using \texttt{al} only the lower 8 bits of \texttt{rax} are
accessed, leading to smaller constants and instruction encodings. However, this
sequence modifies \texttt{rax} so if the value is needed after the tag test, as
often will be the case, then an additional instruction and register are required
to keep a copy. In the special case of \textit{tag}=0, the two first
instructions can be replaced with ``\texttt{test~al, 7}'', which is a
\emph{bitwise-and} that tests if all 3 lowest bits of \texttt{rax} are zero
without modifying \texttt{rax}. In the general case when \textit{tag}$\neq$0, the
\texttt{test} instruction can be combined with a \texttt{lea}
(\emph{Load Effective Address}) instruction to copy the source register, plus
a constant to cancel \textit{tag} (mod 8), to a temporary register:

\begin{minipage}{\textwidth}
\footnotesize
\begin{lstlisting}[language={[x86gnuatt]Assembler},mathescape]
  lea  ebx, [eax+(8-$tag$)]  # ebx = eax + (8-$tag$)
  test bl, 7
  jz   tag_matches
\end{lstlisting}
\end{minipage}

\subsection{Multiple Tag Testing}

When it is necessary to check if an object's tag is one of a set of tags
it is often possible to use a single branch instruction rather than a sequence
of single tag checks. This is particularly the case when the implementer has
leeway in the assignments of tags, as is often the case.

Checking that $tag \in \{tag_1,tag_2\}$ where $tag_2 = tag_1 + 2^n$ (mod 8) can
be done just as efficiently as the single tag test by using the \texttt{lea}
instruction to map $tag_1$ to 0 and $tag_2$ to $2^n$ and then using a mask of
$7-2^n$ in the \texttt{test} instruction to ignore the bit that may be 0 or 1.
For example, for 2-tag self-tagging (Section~\ref{section:2tags2}) with
the tags \verb|010| and \verb|110| the check is:

\begin{minipage}{\textwidth}
\footnotesize
\begin{lstlisting}[language={[x86gnuatt]Assembler},mathescape]
  lea  ebx, [eax+(8-2)]  # 010 -> 000 and 110 -> 100
  test bl, 3             # 3 = 7-4
  jz   tag_matches_010_or_110
\end{lstlisting}
\end{minipage}

4-tag self-tagging (Section~\ref{section:3tags}) with the tags \texttt{010},
\texttt{011}, \texttt{110}, and \texttt{111} is even simpler:

\begin{minipage}{\textwidth}
\footnotesize
\begin{lstlisting}[language={[x86gnuatt]Assembler},mathescape]
  test al, 2
  jnz  tag_matches_010_or_011_or_110_or_111
\end{lstlisting}
\end{minipage}

The general case of checking that $tag$ is in a set of tags can be done
efficiently with the x86 \emph{bit-test} instruction ``\texttt{bt}~\textit{n},
\textit{i}'' that reads the bit at index \textit{i} of \textit{n} and puts it in
the carry flag.  The \texttt{bt} instruction comes in 16, 32, and 64 bit
variants, but not 8 bit.  For example, for 3-tag self-tagging
(Section~\ref{section:3tags}) with the tags \verb|000|, \verb|011|, and \verb|100| the check is:

\begin{minipage}{\textwidth}
\footnotesize
\begin{lstlisting}[language={[x86gnuatt]Assembler}]
  mov bx, 0x1919  # set of tags
  bt  bx, ax      # test bit at index ax of bx
  jc  tag_matches_000_or_011_or_100
\end{lstlisting}
\end{minipage}

\noindent
The bit index in \texttt{ax} is obtained \textit{mod~16}, so the byte
\texttt{0b00011001}=\texttt{0x19} (all 0's except at bit indices 0, 3, and 4) is
repeated twice in register \texttt{bx}.  Using the \texttt{bt} instruction has
the advantage of neither modifying the \texttt{rax} register nor the register
holding the set of tags, amortizing the cost of \texttt{mov} over multiple
checks of that set of tags (a possibly near zero cost if that register is
globally reserved).

On architectures without a bit test instruction, a dynamic count shift of the
tag set register can achieve the bit indexing. If done by modifying the tag set
register, the initialization can't be shared by multiple tag tests. On most
3-address RISC architectures the shift can be done non-destructively. On ARM
A64, this instruction sequence tests the 3 low bits of 64-bit register
\texttt{x1} for a tags 0, 3, or 4:

\begin{minipage}{\textwidth}
\footnotesize
\begin{lstlisting}[language={[arm]Assembler}]
  lslv w3, w2, w1
  cmp  w3, 0
  bmi  matching_tag
\end{lstlisting}
\end{minipage}

\noindent
It assumes that the 32-bit register \texttt{w2} has been preloaded with
0x98989898, the bit reversed tag set that aligns the bit for tag 0 with the
sign bit.

The cost of the check does not depend on the number of tags
tested. This can be advantageous if a tagged pointer is used for heap allocated
floats, say with tag \textit{h}. In that case, a check for a float
(either self-tagged or heap allocated) can be done by adding \textit{h} to the tag
set. Once it is known that an object is a float, it is easy to check
for the single tag \textit{h} to discriminate between the self-tagged and heap allocated
representations.

\subsection{Boxing: conversion \texttt{f64}$\rightarrow$\texttt{object}}

To convert a \texttt{f64} to an \texttt{object}, it is converted to an
\texttt{i64}, added to a constant $bias$ modulo $2^{64}$, and bit-rotated left by
$r$. For the 1-tag variant, $bias=(1+2\times tag)\times 2^{58}$ and
$r=5$. For the 2-tag, 3-tag, and 4-tag variants, $bias=0$ (no bias) and $r=4$.
After this there is a check to verify if the resulting object's tag is in the
set of self-tagged tags. If this is the case the conversion is done, otherwise
an out of line routine can be called to heap allocate a float and return an
appropriately tagged pointer. As an example, if the \texttt{f64} value is in the
64-bit float register \texttt{xmm0}, then the following x86 code will set
register \texttt{rax} to the corresponding \texttt{object} when 3-tag
self-tagging is used:

\begin{minipage}{\textwidth}
\footnotesize
\begin{lstlisting}[language={[x86gnuatt]Assembler}]
  mov  rax, xmm0   # rax $\leftarrow$ xmm0
  rol  rax, 4      # rotate left 4 places
  bt   bx, ax      # assumes bx initialized to 0x1919 elsewhere
  jnc  heap_alloc_float # tag is not one of 000, 011, or 100?
done:
\end{lstlisting}
\end{minipage}

\noindent
Assuming the initialization of the tag mask register (\texttt{bx}) is amortized over
multiple float-related operations, the cost for the hot path is low: 3
register-to-register operations and an easily predictable conditional
jump.

The following x86 code is appropriate for 1-tag self-tagging
using \texttt{000} as the tag:

\begin{minipage}{\textwidth}
\footnotesize
\begin{lstlisting}[language={[x86gnuatt]Assembler}]
  mov  rax, xmm0   # rax $\leftarrow$ xmm0
  add  rax, rbx    # assumes rbx initialized to 1<<58 elsewhere
  rol  rax, 5      # rotate left 5 places
  test al, 7
  jnz  heap_alloc_float # tag is not 000?
done:
\end{lstlisting}
\end{minipage}

\noindent
Here also, the cost for the
hot path is just a bit more due to the bias: 4 register-to-register
operations and an easily predictable conditional jump.

For the 2-tag variant with no special handling of zeroes
(Section~\ref{section:2tags2}), the rotation can be done before
adding the bias to use a compact addition instruction (because then the bias is the small constant $tag_1$). In the
special case of $tag_1$=\texttt{000} the bias vanishes. For example, here is the
code when $tag_1$=\texttt{011}:

\begin{minipage}{\textwidth}
\footnotesize
\begin{lstlisting}[language={[x86gnuatt]Assembler}]
  mov  rax, xmm0   # rax $\leftarrow$ xmm0
  rol  rax, 5      # rotate left 5 places
  add  al, 3       # sufficient to modify the lower 8 bits of rax
  bt   bx, ax      # assumes bx initialized to 0x0c0c elsewhere
  jnc  heap_alloc_float # tag is not one of 010 or 011?
done:
\end{lstlisting}
\end{minipage}

\noindent
The \texttt{add} instruction uses a small constant as a bias, contrary
to the code for the 1-tag version (large constants typically need to
be setup in a register, which is additional work even if it is
amortized). Moreover the 1-tag version cannot eliminate the bias, but
here the \texttt{add} instruction can be removed when
$tag_1$=\texttt{000} which makes it slighly faster than the 1-tag
variant and as efficient as the 3-tag variant: 3 register-to-register
operations and an easily predictable conditional jump.

\subsection{Unboxing: conversion \texttt{object}$\rightarrow$\texttt{f64}}

To convert an \texttt{object} (that is known to be a float) to a
\texttt{f64}, the tag must be tested to see if it is a self-tagged
float. If it is, the boxing operations are done in reverse order and inverted, replacing the
\texttt{rol} by a \texttt{ror} (rotate-right), and if there is a bias,
replacing the \texttt{add} by a \texttt{sub} (subtract).
If it is not a self-tagged float, a memory read gets the
\texttt{f64} result out of the heap allocated float.  As
an example using the 3-tag variant, if the \texttt{object}
value is in the 64-bit register \texttt{rax}, then the following x86
code will set register \texttt{xmm0} to the corresponding
\texttt{f64}:

\begin{minipage}{\textwidth}
\footnotesize
\begin{lstlisting}[language={[x86gnuatt]Assembler}]
  bt  bx, ax             # assumes bx initialized to 0x1919 elsewhere
  jc  self_tagged
  mov xmm0,[rax+offset]  # xmm0 $\leftarrow$ value field of float
  jmp done
self_tagged:
  ror rax, 4             # rotate right 4 places
  mov xmm0, rax          # xmm0 $\leftarrow$ rax
done:
\end{lstlisting}
\end{minipage}

\noindent
The number and type of instructions on the hot path is the same
as for the boxing operation.

In summary, the self-tagging operations require few machine instructions
but there are small variations depending on the instruction set and also the
assignment of tags which can often be optimized for self-tagging without
impacting the speed of operations on objects or small integers.

\subsection{C Implementation}

This section explains how the above operations can be implemented in C, as
this is a common implementation language, in particular it is used by
the Bigloo and Gambit Scheme to C compilers used in the experiments of
Section~\ref{section:experiment}.

Testing for a specific tag or a pair of tags is fairly straightforward
to express in C and follows the same principles as the assembly code,
so details are skipped.  Testing for a set of tags is more challenging
in portable C code.  It can be implemented with the following pure C
function, which is easily inlined by a C compiler (here testing
for the tags \texttt{000}, \texttt{011}, and \texttt{100}):

\begin{minipage}{\textwidth}
\footnotesize
\begin{lstlisting}[style=customc]
#define TAG_SET ((1<<0)|(1<<3)|(1<<4)) /* 0x19 for tags 000, 011, and 100 */

inline bool has_tag_0_or_3_or_4(int64_t n) {
  return (((uint32_t)1 << (n & 31)) & (~(uint32_t)0/0xff * TAG_SET)) != 0;
}
\end{lstlisting}
\end{minipage}

The expression \verb|n & 31| computes a shift count \textit{mod~32} and the bitwise-and
therefore tests the bit at index \texttt{n}~\textit{mod~32} of its second operand. The code
uses the compile-time constant \verb|~(uint32_t)0/0xff * TAG_SET| that repeats the
8 bit mask \texttt{TAG\_SET} 4 times to fill a 32 bit word, giving 0x19191919.
This achieves the equivalent of
testing the bit at index \texttt{n}~\textit{mod~8} of \texttt{TAG\_SET}.

Using this pure C definition with GCC version 13.2.0 and clang version 18.1.3 on
x86-64 results in the non-destructive approach based on the bit test
instruction. The expression \texttt{n \& 31} is optimized out by those compilers
because the bit test instruction on x86 implicitly does this operation. On older
versions of those compilers that would implement this with a destructive shift
instruction, the bit test approach can be achieved using the following
definition that uses an \texttt{asm} statement:

\begin{minipage}{\textwidth}
\footnotesize
\begin{lstlisting}[style=customc]
inline bool has_tag_0_or_3_or_4(int64_t n) {
  bool carry;
  __asm__("bt %%ax, %2;" /* get one bit of mask into carry (ax is index) */
          : "=@ccc"(carry)
          : "a"((uint16_t)n),
            "r"(~(uint16_t)0/0xff * TAG_SET)); /* 0x1919 */
  return carry;
}
\end{lstlisting}
\end{minipage}

Boxing and unboxing operations need to convert between a float
value and its bit representation. This can be achieved portably with a
\texttt{union} type whose fields are of type \texttt{f64} and
\texttt{i64}. Moreover, boxing and unboxing need to rotate the bit
representation of the float. Although C does not provide an operator
for bit rotation, both GCC and clang recognize an equivalent pair of
shifts and generate a single machine rotate instruction.

\begin{minipage}{\textwidth}
\vspace{0.2em}
The implementation of boxing in C for the 3-tag variant is:
\begin{lstlisting}[style=customc]
union di { double d; int64_t i; };

#define ROTL(n,s) ((int64_t)(((uint64_t)n << s) | ((uint64_t)n >> (64 - s))))

inline int64_t f64_to_object(double f) {
  int64_t result = ROTL(((union di)f).i, 4);
  if (has_tag_0_or_3_or_4(result)) return result;
  return heap_allocate_float(f);
}
\end{lstlisting}
\end{minipage}

\begin{minipage}{\textwidth}
The implementation of unboxing for the 3-tag variant is:
\begin{lstlisting}[style=customc]
inline double object_to_f64(int64_t o) {
  int64_t result = ROTL(o, 60);
  if (has_tag_0_or_3_or_4(result)) return ((union di)result).d;
  return value_of_heap_allocated_float(o);
}
\end{lstlisting}
\end{minipage}


\section{Experiments}
\label{section:experiment}

This section evaluates float self-tagging and demonstrates that it presents an
interesting performance trade-off between programs that perform frequent
operations on objects but few floating-point operations, and float-intensive
programs where this balance is reversed. This evaluation is conducted through
experiments that measure performance metrics across different object
representations (both self-tagging and existing ones), multiple machine
architectures, and two state-of-the-art Scheme implementations: the Bigloo
(commit \texttt{\bigloocommit})~\cite{bigloo} and Gambit v4.9.7 (commit
\texttt{\gambitcommit})~\cite{gambit} Scheme compilers. Using two independently
developed compilers helps to demonstrate that self-tagging can be adapted to
preserve the different design philosophies and historical implementation choices
of these systems.

The following self-tagging variants are evaluated:

\begin{itemize}
\item \textbf{3-tag self-tagging} uses 3 tags for self-tagging floats
  corresponding to the \tagzerocolorname, \tagthreecolorname, and
  \tagfourcolorname ranges from Figure~\ref{figure:tag-interval-table}. The
  remaining floats are heap allocated. Bigloo uses the tags \texttt{000},
  \texttt{011} and \texttt{100}, and heap allocated floats are represented with
  generic pointers with a header type. Gambit offsets the tags by 3 (i.e.~it
  uses \texttt{011}, \texttt{110} and \texttt{111}) so that the tag \texttt{000}
  is kept for tagging small integers, and tagged pointers with the tag
  \texttt{010} are used for the heap allocated floats.
\item \textbf{4-tag self-tagging} (Gambit only) is like the 3-tag variant but
  uses the tag \texttt{010} to self-tag floats in the \tagsevencolorname range
  from Figure~\ref{figure:tag-interval-table}. Allocated floats are represented
  with generic pointers.
\item \textbf{2-tag self-tagging with preallocated zeros} (Bigloo only) tests
  the impact of expending the tag \texttt{000} for self-tagging. It only uses
  the tags \texttt{011} and \texttt{100} for float self-tagging the
  \tagthreecolorname and \tagfourcolorname ranges from
  Figure~\ref{figure:tag-interval-table} and reclaims the tag \texttt{001} to
  represent all heap allocated floats as tagged pointers. The floats
  $\pm$\texttt{0.0} are preallocated and the float boxing operation contains an
  explicit zero check to return the preallocated zeros.
\item \textbf{2-tag self-tagging} (Gambit only) tests the variant that uses the
  tags \texttt{010} and \texttt{110} for float self-tagging the \onetagcolorname
  and \twotagcolorname ranges from Figure~\ref{figure:tag-interval-table}.
\item \textbf{1-tag self-tagging} tests the variant that uses a single tag for
  self-tagging the \onetagcolorname ranges from
  Figure~\ref{figure:tag-interval-table}. Bigloo uses the tag \texttt{001} and
  Gambit uses the tag \texttt{110}. heap allocated floats are represented with
  tagged pointers.
\end{itemize}

Additionally both compilers can be configured to represent all floats
as heap allocated tagged objects or NuN-boxing, and Bigloo
also supports NaN-boxing. This provides a flexible
environment for comparing these techniques.

In this paper the C back end of the Bigloo and Gambit compilers are
used.  This way the resulting machine code benefits from the C
compiler's optimizations and in particular excellent architecture specific
instruction selection. In both cases the same version of the gcc C
compiler is used.

\subsection{Overview of the Bigloo Compiler}

The Bigloo compiler conforms to the R5RS Scheme specification and adds
several extensions, including optional type annotations.  Exact
rational numbers and complex numbers, which are optional in the R5RS
Scheme specification, are not supported.

The Bigloo compiler builds a typed abstract syntax
tree from the compiled Scheme program. Several analysis and optimizations
refine the initial optional type annotations contained in the source
program, the main ones being occurrence-typing~\cite{occurrence-typing}
and storage use analysis~\cite{DBLP:conf/icfp/SerranoF96}. During compilation,
primitive values have a dual representation: one for polymorphic
contexts and one for specific contexts. For instance, when the compiler can
establish that a function is always invoked with floats, the
compiler assigns the C \texttt{double} type to that function and, when
needed, it introduces cast operations from and to the generic
representation of floats and double.

The result of this
compilation technique is that many local variables and local functions are
precisely typed and avoid polymorphic representations. This contributes
to removing otherwise necessary heap allocations and casts for floats.
The compiler also tracks cases where only floats
are stored into vectors, which are then transformed into arrays of C
doubles.

Bigloo uses the Boehm-Demers-Weiser mark-and-sweep conservative garbage
collector~\cite{BoehmW88}. The complexity of collecting dead objects is
proportional to the number of allocated objects that are still live, contrary to
copying collectors. For such a collector, avoiding allocating short lived floats
is crucial for performance.

\subsection{Overview of the Gambit Compiler}

The Gambit compiler conforms to the R7RS Scheme language. It features
the full numerical tower including arbitrary-precision integers, exact
rational numbers and complex numbers. Arithmetic operators are generic
and the result type depends on the specific values, for example
multiplying two complex numbers may yield a number of any type. Like
Bigloo, the compiler inlines the small integer and float
cases of the dispatch and handles other cases in an out-of-line
function.

First-class continuations (\texttt{call/cc}) and tail-calls are fully
supported, and stack-overflows are gracefully handled. With the C
back end, this requires the use of a trampoline and the management of
the stack through an explicit array, stack pointer and stack-overflow
checks. This causes a small slowdown when compared to a machine code
back end where the trampoline would not be required.  Memory management
is based on a fast bump-allocator and a stop-and-copy garbage
collector.

The compiler front end performs several source-to-source optimizations, such as
constant folding, lambda lifting, and function inlining. It does not perform
type analysis or type inference. However, the back end uses a simple mechanism
to optimize float calculations by tracking the use of their results within basic
blocks. The result of a float operation is kept in raw form so that other
operations within the same basic block that consume it do not need to unbox the
value. A float value is boxed only if it remains live at the end of the basic
block.

\subsection{Experimental Setup}

Self-tagging variants are implemented in Bigloo and Gambit, and tested using a
subset of the R7RS benchmarks, which is the standard benchmark suite for
Scheme~\cite{r7rs-benchmarks}. The experiment includes all macro-benchmarks
(more than 500 lines of code) and benchmarks whose calculations mainly involve
floats (tagged as such in the R7RS benchmarks suite). For the purpose of
analyzing results, it is useful to distinguish between benchmarks that use few
or no floats and float-intensive benchmarks (\texttt{fibfp}, \texttt{fft},
\texttt{mbrot}, \texttt{nucleic}, \texttt{pnpoly}, \texttt{ray},
\texttt{simplex}, \texttt{sum1}, and \texttt{sumfp}).

Experiments were repeated on the following machines that operate on distinct
microarchitectures:

\begin{itemize}
  \item \textsc{Intel} - \redrockTT
  \item \textsc{Amd} - \redrock
  \item \textsc{M2} - \sizo
  \item \textsc{Risc-V} - \starfive
\end{itemize}

For each combination of machine, compiler and self-tagging variant, benchmarks
are executed {\experimentsrepetitions} times, and each repetition is configured
to last at least 5 seconds. Repetitions of each variant and baseline are paired
in order of execution to compute relative execution time for each pair. Upcoming
sections report geometric means of these relative times, with geometric standard
deviations.

To reduce variance on \textsc{Intel} and \textsc{Amd}, benchmarks are executed
on the same CPU core by using \texttt{taskset} and address space randomization
is disabled with \texttt{setarch}. On \textsc{M2}, no equivalent tools are
available and hence the variance is higher on some benchmarks. On
\textsc{Risc-V}, the variance remains very low despite not using
\texttt{taskset} and \texttt{setarch}.

For measurements that were similar across
all machines, a single microarchitectures is shown. Omitted figures can be found
in the artifact that comes with this paper~\citeartifact.

\subsection{Memory Profiling and Execution Time}
\label{section:experiment-results}

\newcommand\memlinewidth{0.98}
\begin{figure}
  \center
  {\large Heap Allocations Relative to Allocated Floats (Bigloo)}\\[0.5em]
  {\small \redrockTT}\\
  \includegraphics[width=\linewidth]{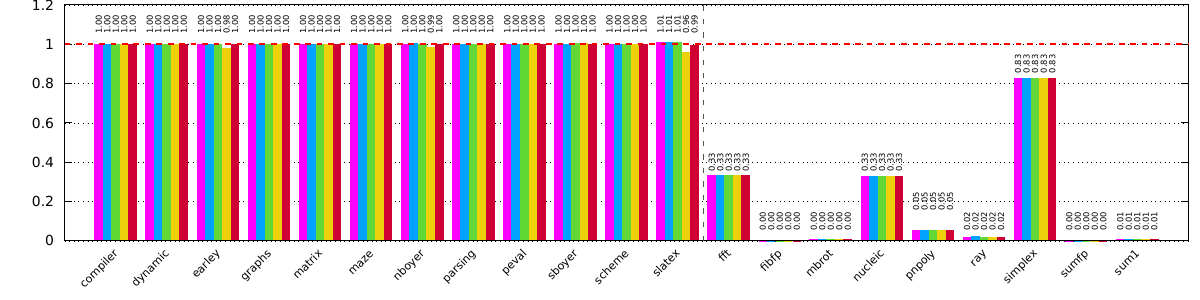}
  \legendbigloomem
  \caption{ Heap allocations of Bigloo with each variant of self-tagging,
    NaN-boxing, NuN-boxing, and allocated floats. The y-axis shows the number of
    allocated bytes relative to the version that heap allocates all
    floats (dashed horizontal red line). $1.00\times$ indicates no change, while
    $0.00\times$ indicates no heap allocation.  }
  \label{figure:bigloo-memory}
\end{figure}

\newcommand\timenunlinewidth{0.95}

\begin{figure}
  {\large Execution Time Relative to NuN-Boxing (Bigloo)}\\[0.5em]
  {\small \redrock}
  \begin{subfigure}{\timenunlinewidth\linewidth}
    \center
    \includegraphics[width=\linewidth]{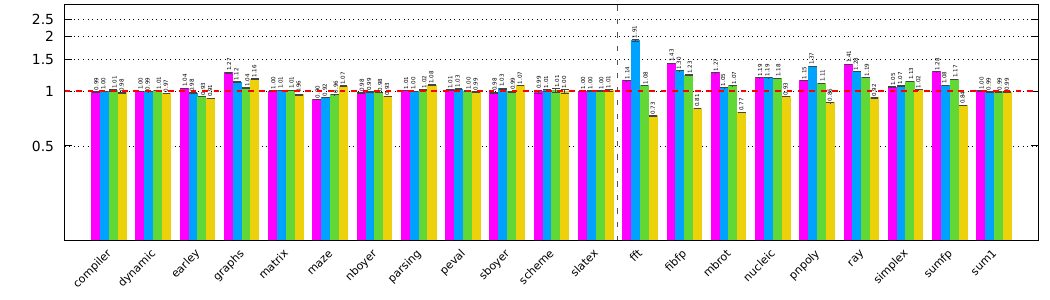}
  \end{subfigure}\\
  {\small \redrockTT}
  \begin{subfigure}{\timenunlinewidth\linewidth}
    \center
    \includegraphics[width=\linewidth]{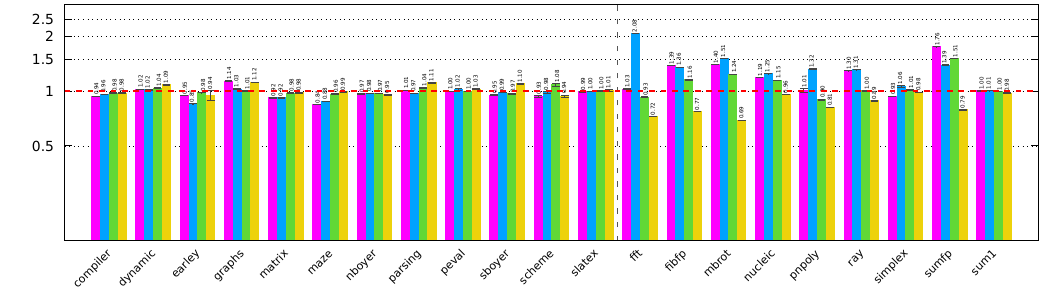}
  \end{subfigure}\\
  {\small \sizo}
  \begin{subfigure}{\timenunlinewidth\linewidth}
    \center
    \includegraphics[width=\linewidth]{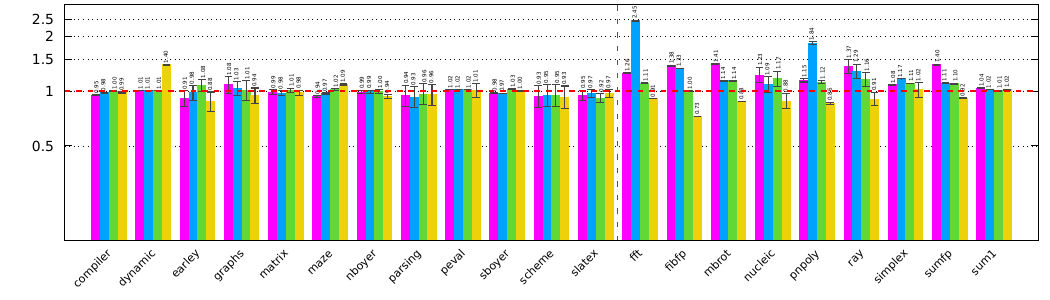}
  \end{subfigure}\\
  {\small \starfive}
  \begin{subfigure}{\timenunlinewidth\linewidth}
    \center
    \includegraphics[width=\linewidth]{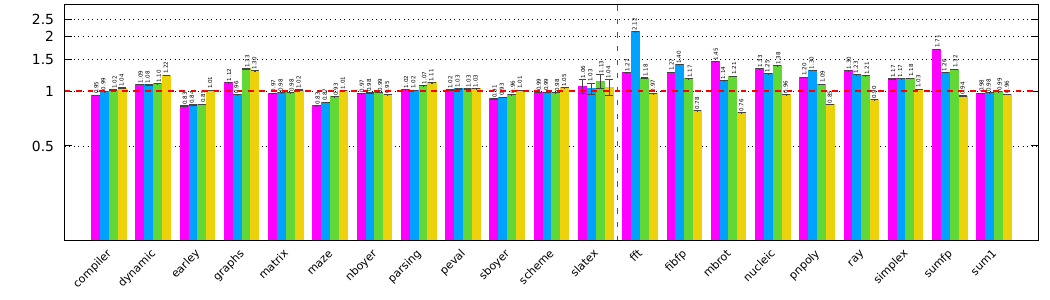}
  \end{subfigure}
  \legendbiglootimenun
  \caption{ Execution time of Bigloo with each variant of self-tagging and
    NaN-boxing relative to NuN-boxing on four distinct microarchitectures (from
    top to bottom: \textsc{Amd}, \textsc{Intel}, \textsc{M2}, and \textsc{Risc-V}).
    The y-axis shows execution time
    relative to NuN-boxing (dashed horizontal red line) on a logarithmic scale.
    $1.00\times$ indicates no change, lower means faster, and higher means
    slower execution than NuN-boxing. }
  \label{figure:bigloo-time-vs-nun}
\end{figure}

\begin{figure}
  {\large Execution Time Relative to NuN-Boxing (Gambit)}\\[0.5em]
  {\small \redrock}
  \begin{subfigure}{\timenunlinewidth\linewidth}
    \center
    \includegraphics[width=\linewidth]{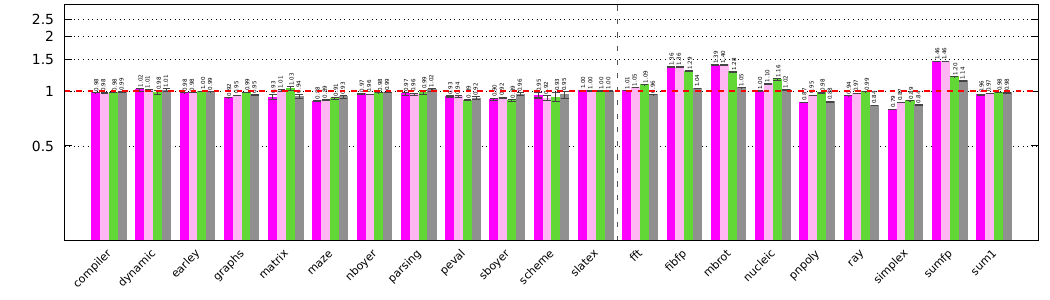}
  \end{subfigure}\\
  {\small \redrockTT}
  \begin{subfigure}{\timenunlinewidth\linewidth}
    \center
    \includegraphics[width=\linewidth]{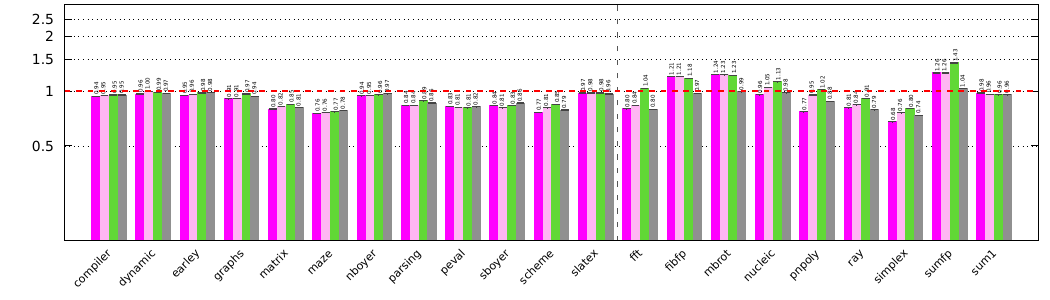}
  \end{subfigure}\\
  {\small \sizo}
  \begin{subfigure}{\timenunlinewidth\linewidth}
    \center
    \includegraphics[width=\linewidth]{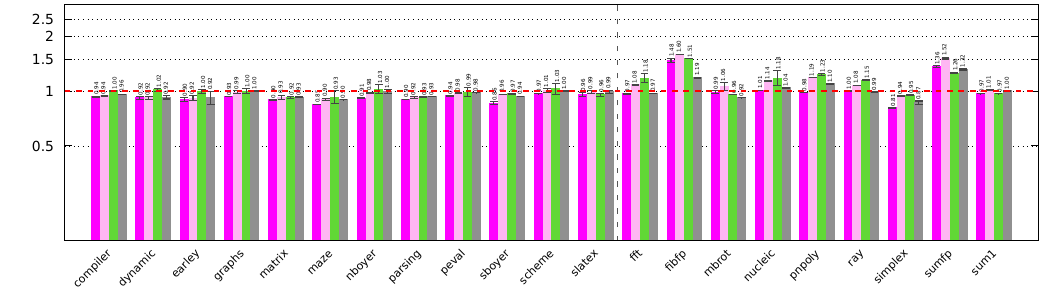}
  \end{subfigure}\\
  {\small \starfive}
  \begin{subfigure}{\timenunlinewidth\linewidth}
    \center
    \includegraphics[width=\linewidth]{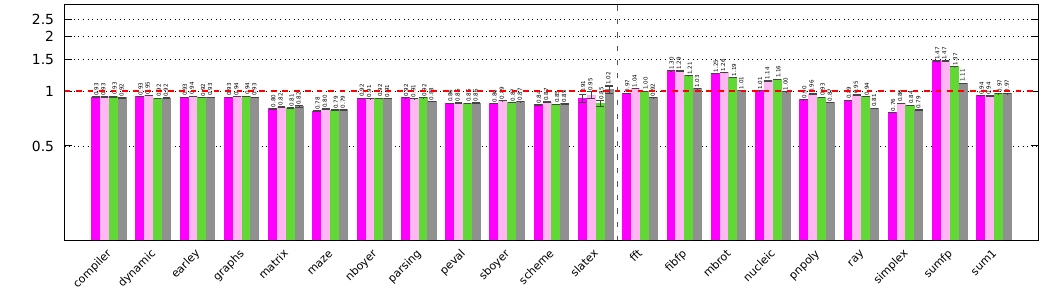}
  \end{subfigure}
  \legendgambittimenun
  \caption{ Execution time of Gambit with each variant of self-tagging relative
    to NuN-boxing on four distinct microarchitectures (from top to bottom: \textsc{Amd},
    \textsc{Intel}, \textsc{M2}, and \textsc{Risc-V}). The y-axis shows execution time relative to
    NuN-boxing (dashed horizontal red line) on a logarithmic scale. $1.00\times$
    indicates no change, lower means faster, and higher means slower execution
    than NuN-boxing. }
  \label{figure:gambit-time-vs-nun}
\end{figure}

Figure~\ref{figure:bigloo-memory} shows the memory allocations of Bigloo's
self-tagging variants, NaN-boxing, and NuN-boxing compared to allocated floats.
Since NaN/NuN-boxing are known to allocate no floats at all, these correspond to
executions where heap usage is for non-float objects only. A few self-tagging variants
allocate some floats on some benchmarks (see
Figure~\ref{figure:tag-interval-table}), but these allocations are rare enough
that memory profiling of all variants is nearly identical to that of
NaN/NuN-boxing.

Detailed execution time comparisons with NuN-boxing are shown in
Figures~\ref{figure:bigloo-time-vs-nun}~and~\ref{figure:gambit-time-vs-nun}
(Bigloo and Gambit respectively). Figure~\ref{figure:bigloo-time-vs-nun} also
shows time for NaN-boxing, which Bigloo implements. These results are summarized
in Figure~\ref{figure:time-vs-nun-summary}, which provides the geometric mean of
execution times relative to NuN-boxing for each float encoding, compiler, and
microarchitecture, for both float-intensive and non-float benchmarks.

\begin{figure}
\center
\setlength\extrarowheight{2.3pt}
\setlength{\tabcolsep}{3.7pt}
{\large Summary of Execution Time Relative to NuN-Boxing}\\[0.5em]

\small
\begin{tabular}{|cr|r|r|r|r|r|r|r|r|}
\cline{3-10}
\multicolumn{2}{c}{}  & \multicolumn{4}{|c|}{\large Non-float benchmarks} & \multicolumn{4}{c|}{\large Float benchmarks}  \\
\multicolumn{2}{c}{}  & \multicolumn{1}{|c}{\textsc{Amd}} & \multicolumn{1}{c}{\textsc{Intel}} & \multicolumn{1}{c}{\textsc{M2}} & \multicolumn{1}{c|}{\textsc{Risc-V}} & \multicolumn{1}{c}{\textsc{Amd}} & \multicolumn{1}{c}{\textsc{Intel}} & \multicolumn{1}{c}{\textsc{M2}} & \multicolumn{1}{c|}{\textsc{Risc-V}} \\
\cline{1-10}
\multirow{4}{*}{\rotatebox{90}{\large Bigloo}} & \colorrectRight{\normalsize 1-tag}{\htmlColorOneTag} & $\biglooredrockFltOneNunRatioNonFloats\times$ & $\biglooredrockTTFltOneNunRatioNonFloats\times$ & $\bigloosizoFltOneNunRatioNonFloats\times$ & $\bigloostarfiveFltOneNunRatioNonFloats\times$
      & $\biglooredrockFltOneNunRatioFloats\times$ & $\biglooredrockTTFltOneNunRatioFloats\times$ & $\bigloosizoFltOneNunRatioFloats\times$ & $\bigloostarfiveFltOneNunRatioFloats\times$ \\
\cline{3-10}
 & \colorrectRight{\normalsize 2-tag w/ prealloc. zero}{\htmlColorTwoTagNZ} & $\biglooredrockFltnzNunRatioNonFloats\times$ & $\biglooredrockTTFltnzNunRatioNonFloats\times$ & $\bigloosizoFltnzNunRatioNonFloats\times$ & $\bigloostarfiveFltnzNunRatioNonFloats\times$
      & $\biglooredrockFltnzNunRatioFloats\times$ & $\biglooredrockTTFltnzNunRatioFloats\times$ & $\bigloosizoFltnzNunRatioFloats\times$ & $\bigloostarfiveFltnzNunRatioFloats\times$ \\
\cline{3-10}
 & \colorrectRight{\normalsize 3-tag}{\htmlColorThreeTag} & $\biglooredrockFltThreeNunRatioNonFloats\times$ & $\biglooredrockTTFltThreeNunRatioNonFloats\times$ & $\bigloosizoFltThreeNunRatioNonFloats\times$ & $\bigloostarfiveFltThreeNunRatioNonFloats\times$
      & $\biglooredrockFltThreeNunRatioFloats\times$ & $\biglooredrockTTFltThreeNunRatioFloats\times$ & $\bigloosizoFltThreeNunRatioFloats\times$ & $\bigloostarfiveFltThreeNunRatioFloats\times$ \\
\cline{3-10}
 & \colorrectRight{\normalsize NaN-boxing}{\htmlColorNan} & $\biglooredrockNanNunRatioNonFloats\times$ & $\biglooredrockTTNanNunRatioNonFloats\times$ & $\bigloosizoNanNunRatioNonFloats\times$ & $\bigloostarfiveNanNunRatioNonFloats\times$
      & $\biglooredrockNanNunRatioFloats\times$ & $\biglooredrockTTNanNunRatioFloats\times$ & $\bigloosizoNanNunRatioFloats\times$ & $\bigloostarfiveNanNunRatioFloats\times$ \\
\cline{1-10}
\multirow{4}{*}{\rotatebox{90}{\large Gambit}} & \colorrectRight{\normalsize 1-tag}{\htmlColorOneTag} & $\gambitredrockFltOneNunRatioNonFloats\times$ & $\gambitredrockTTFltOneNunRatioNonFloats\times$ & $\gambitsizoFltOneNunRatioNonFloats\times$ & $\gambitstarfiveFltOneNunRatioNonFloats\times$
      & $\gambitredrockFltOneNunRatioFloats\times$ & $\gambitredrockTTFltOneNunRatioFloats\times$ & $\gambitsizoFltOneNunRatioFloats\times$ & $\gambitstarfiveFltOneNunRatioFloats\times$ \\
\cline{3-10}
 & \colorrectRight{\normalsize \phantom{w/ prealloc. zero} 2-tag}{\htmlColorTwoTag} & $\gambitredrockFltTwoNunRatioNonFloats\times$ & $\gambitredrockTTFltTwoNunRatioNonFloats\times$ & $\gambitsizoFltTwoNunRatioNonFloats\times$ & $\gambitstarfiveFltTwoNunRatioNonFloats\times$
      & $\gambitredrockFltTwoNunRatioFloats\times$ & $\gambitredrockTTFltTwoNunRatioFloats\times$ & $\gambitsizoFltTwoNunRatioFloats\times$ & $\gambitstarfiveFltTwoNunRatioFloats\times$ \\
\cline{3-10}
 & \colorrectRight{\normalsize 3-tag}{\htmlColorThreeTag} & $\gambitredrockFltThreeNunRatioNonFloats\times$ & $\gambitredrockTTFltThreeNunRatioNonFloats\times$ & $\gambitsizoFltThreeNunRatioNonFloats\times$ & $\gambitstarfiveFltThreeNunRatioNonFloats\times$
      & $\gambitredrockFltThreeNunRatioFloats\times$ & $\gambitredrockTTFltThreeNunRatioFloats\times$ & $\gambitsizoFltThreeNunRatioFloats\times$ & $\gambitstarfiveFltThreeNunRatioFloats\times$ \\
\cline{3-10}
 & \colorrectRight{\normalsize 4-tag}{\htmlColorFourTag} & $\gambitredrockFltFourNunRatioNonFloats\times$ & $\gambitredrockTTFltFourNunRatioNonFloats\times$ & $\gambitsizoFltFourNunRatioNonFloats\times$ & $\gambitstarfiveFltFourNunRatioNonFloats\times$
      & $\gambitredrockFltFourNunRatioFloats\times$ & $\gambitredrockTTFltFourNunRatioFloats\times$ & $\gambitsizoFltFourNunRatioFloats\times$ & $\gambitstarfiveFltFourNunRatioFloats\times$ \\
\cline{1-10}
\end{tabular}

\caption{Geometric mean of execution times relative to NuN-boxing of float and
 non-float benchmarks for each combination of float encoding, compiler, and
 microarchitecture. $1.00\times$ indicates no change, lower means faster, and
 higher means slower execution than NuN-boxing.}
\label{figure:time-vs-nun-summary}
\end{figure}

As a general observation, none of the evaluated encodings is a better
alternative overall; the best encoding depends on the nature of benchmarks,
microarchitecture, and implementation. For instance with Bigloo, NaN-boxing is
faster across microarchitectures on float-intensive benchmarks, but slowest on
non-float benchmarks. Conversely, self-tagging generally fares better than
NaN/NuN-boxing on non-float benchmarks since it introduces no overhead on
non-float objects.

Bigloo's self-tagging does not consistently match the performance of NuN-boxing
on float benchmarks across microarchitectures (but is faster than allocated
floats, as will be discussed in Section~\ref{section:gc-experiment}). The 1-tag
and 3-tag variants offer the highest performance overall for Bigloo, with the
3-tag variant being faster for float benchmarks and slightly slower for
non-float benchmarks. Bigloo's 2-tag with preallocated zeros has similar average
performance as 1-tag, however it shows more variability across benchmarks (see
\text{fft} and \text{pnpoly} in Figure~\ref{figure:bigloo-time-vs-nun}) since it
has to test for $\pm0.0$ out-of-line. Hence, Gambit's 2-tag variant
(Section~\ref{section:2tags2}) should be preferred to that of Bigloo.

Gambit's 4-tag variant provides consistently faster average execution times for
both float and non-float benchmarks on all architectures except \textsc{M2} where it is
nearly the same speed as NuN-boxing. Gambit's 1-tag variant also fares well on
all benchmarks on \textsc{Intel} and \textsc{Risc-V}.

Self-tagging and NaN/NuN-boxing have competitive performance and the best choice
depends on the specific setting. When floating point computation performance is
important but less so than operations on other objects, self-tagging is a good
option to consider. Additionally, a significant portion of this study was spent
adapting existing systems (Bigloo and Gambit) to new float encodings, and
self-tagging was found to be qualitatively simpler to implement as it only
affects the encoding of floats, while NaN/NuN-boxing has deeper object encoding
implications.

\subsection{Memory Management Considerations}
\label{section:gc-experiment}

\newcommand\timealloclinewidth{0.95}

\begin{figure}
  {\large Execution Time Relative to Allocated Floats (Bigloo)}\\[0.5em]
  {\small \redrock}
  \begin{subfigure}{\timealloclinewidth\linewidth}
    \center
    \includegraphics[width=\linewidth]{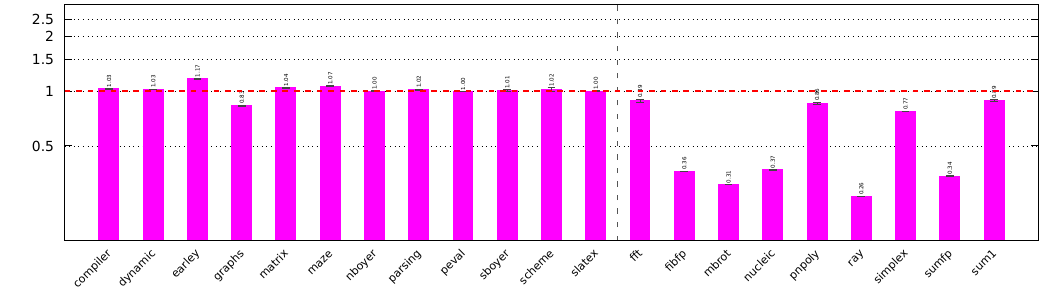}
  \end{subfigure}\\
  {\small \redrockTT}
  \begin{subfigure}{\timealloclinewidth\linewidth}
    \center
    \includegraphics[width=\linewidth]{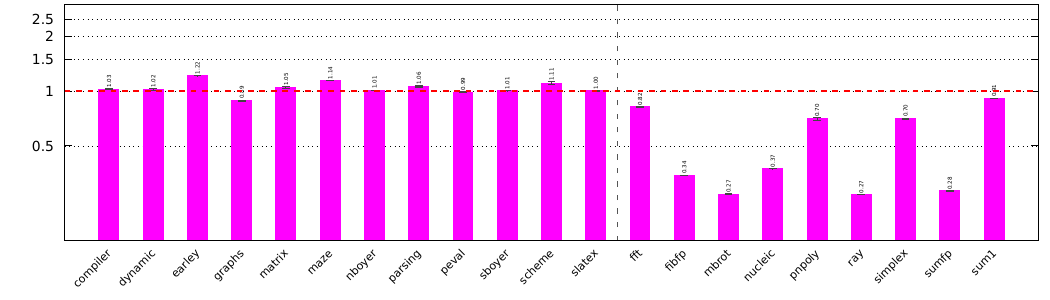}
  \end{subfigure}\\
  {\small \sizo}
  \begin{subfigure}{\timealloclinewidth\linewidth}
    \center
    \includegraphics[width=\linewidth]{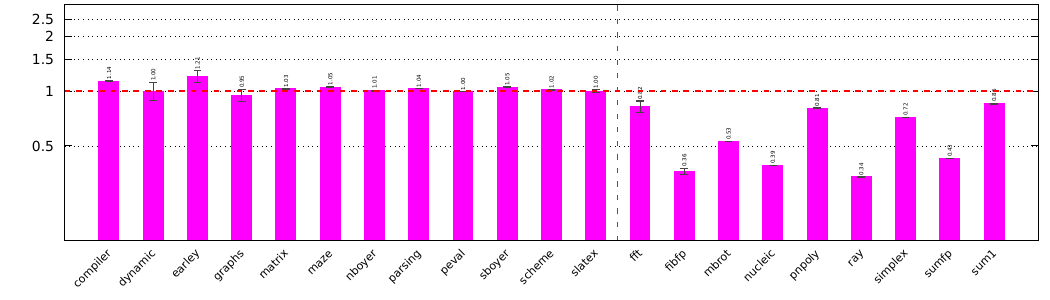}
  \end{subfigure}\\
  {\small \starfive}
  \begin{subfigure}{\timealloclinewidth\linewidth}
    \center
    \includegraphics[width=\linewidth]{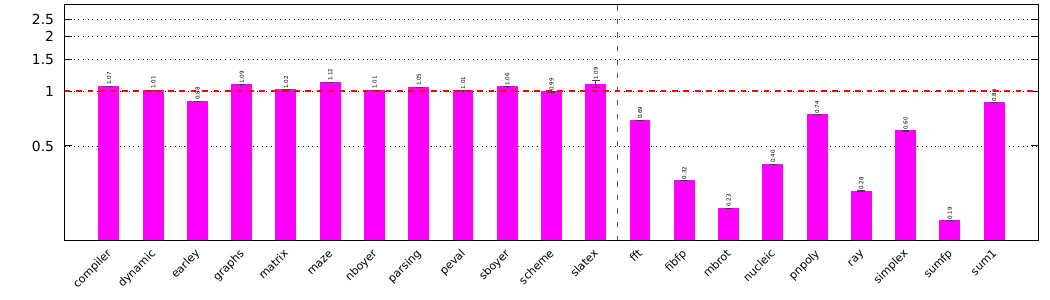}
  \end{subfigure}
  \legendbiglootimealloc
  \caption{ Execution time of Bigloo with self-tagging (1-tag) relative to
    allocated floats on four distinct microarchitectures (from top to bottom:
    \textsc{Amd}, \textsc{Intel}, \textsc{M2}, and \textsc{Risc-V}). The y-axis shows execution time relative to the
    version that heap allocates all floats (dashed horizontal red line) on a
    logarithmic scale. $1.00\times$ indicates no change, lower means faster
    execution than allocated floats. }
\label{figure:bigloo-time-vs-alloc}
\end{figure}

\begin{figure}
  {\large Execution Time Relative to Allocated Floats (Gambit)}\\[0.5em]
  {\small \redrock}
  \begin{subfigure}{\timealloclinewidth\linewidth}
    \center
    \includegraphics[width=\linewidth]{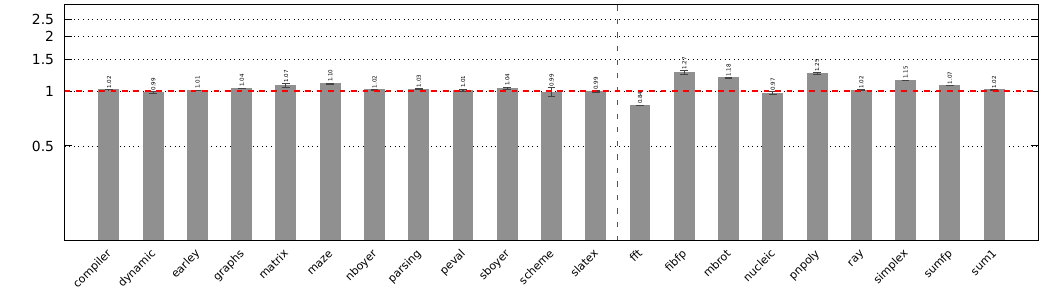}
  \end{subfigure}\\
  {\small \redrockTT}
  \begin{subfigure}{\timealloclinewidth\linewidth}
    \center
    \includegraphics[width=\linewidth]{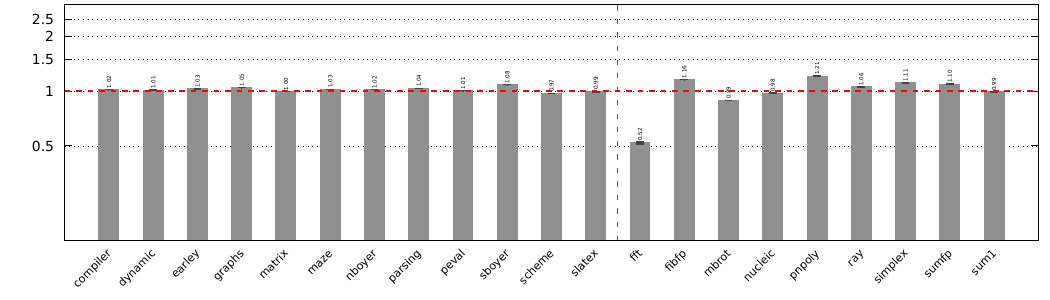}
  \end{subfigure}\\
  {\small \sizo}
  \begin{subfigure}{\timealloclinewidth\linewidth}
    \center
    \includegraphics[width=\linewidth]{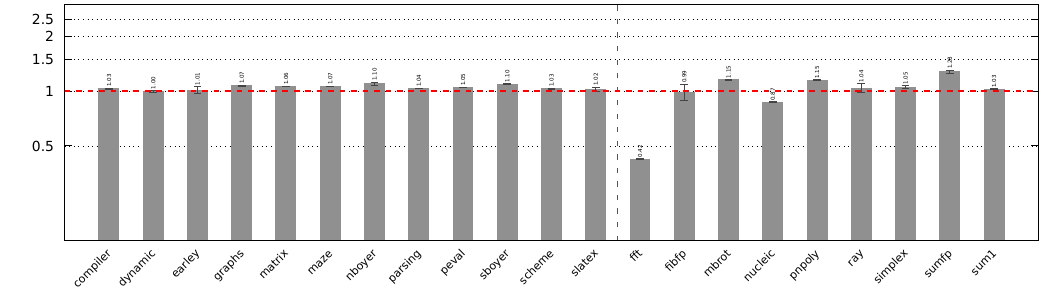}
  \end{subfigure}\\
  {\small \starfive}
  \begin{subfigure}{\timealloclinewidth\linewidth}
    \center
    \includegraphics[width=\linewidth]{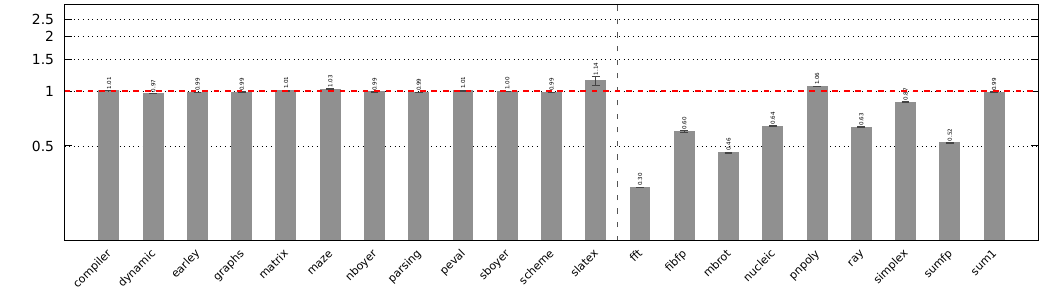}
  \end{subfigure}
  \legendgambittimealloc
  \caption{ Execution time of Gambit with self-tagging (4-tag) relative to
    allocated floats on four distinct microarchitectures (from top to bottom:
    \textsc{Amd}, \textsc{Intel}, \textsc{M2}, and \textsc{Risc-V}). The y-axis shows execution time relative to the
    version that heap allocates all floats (dashed horizontal red line) on a
    logarithmic scale. $1.00\times$ indicates no change, lower means faster
    execution than allocated floats. }
  \label{figure:gambit-time-vs-alloc}
\end{figure}

\begin{figure}
  \centering
  \text{{\Large Execution Time with Non-Empty Heap}}\\
  {\small \redrock}\\[0.2em]
  \parbox{0.04\textwidth}{
    \centering
    \raisebox{6em}{\rotatebox{90}{\text{{\large Execution time (s)}}}}
  }%
  \parbox{0.94\textwidth}{
  \centering
  \begin{subfigure}{0.3\columnwidth}
    \centering
    \texttt{fft}
    \includegraphics[width=\columnwidth]{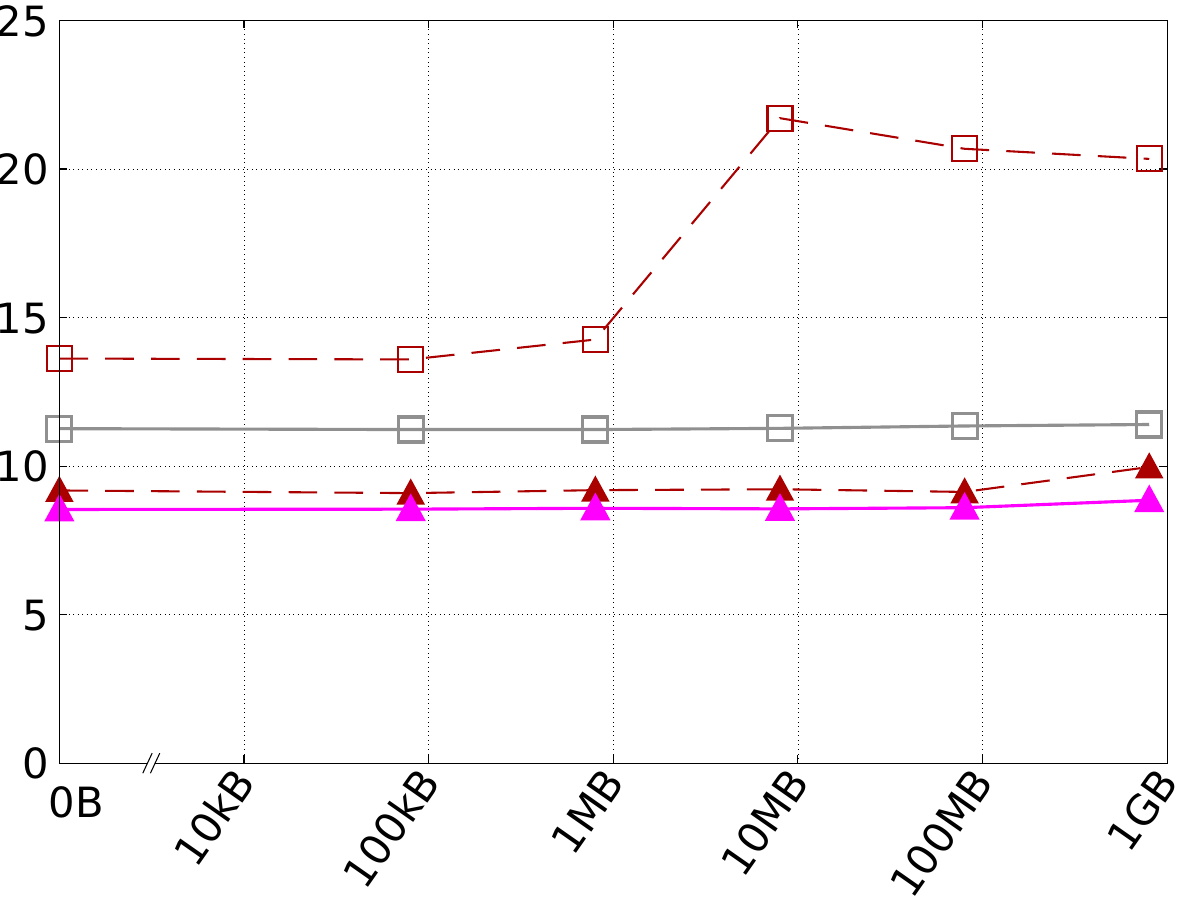}
  \end{subfigure}
  \begin{subfigure}{0.3\columnwidth}
    \centering
    \texttt{fibfp}
    \includegraphics[width=\columnwidth]{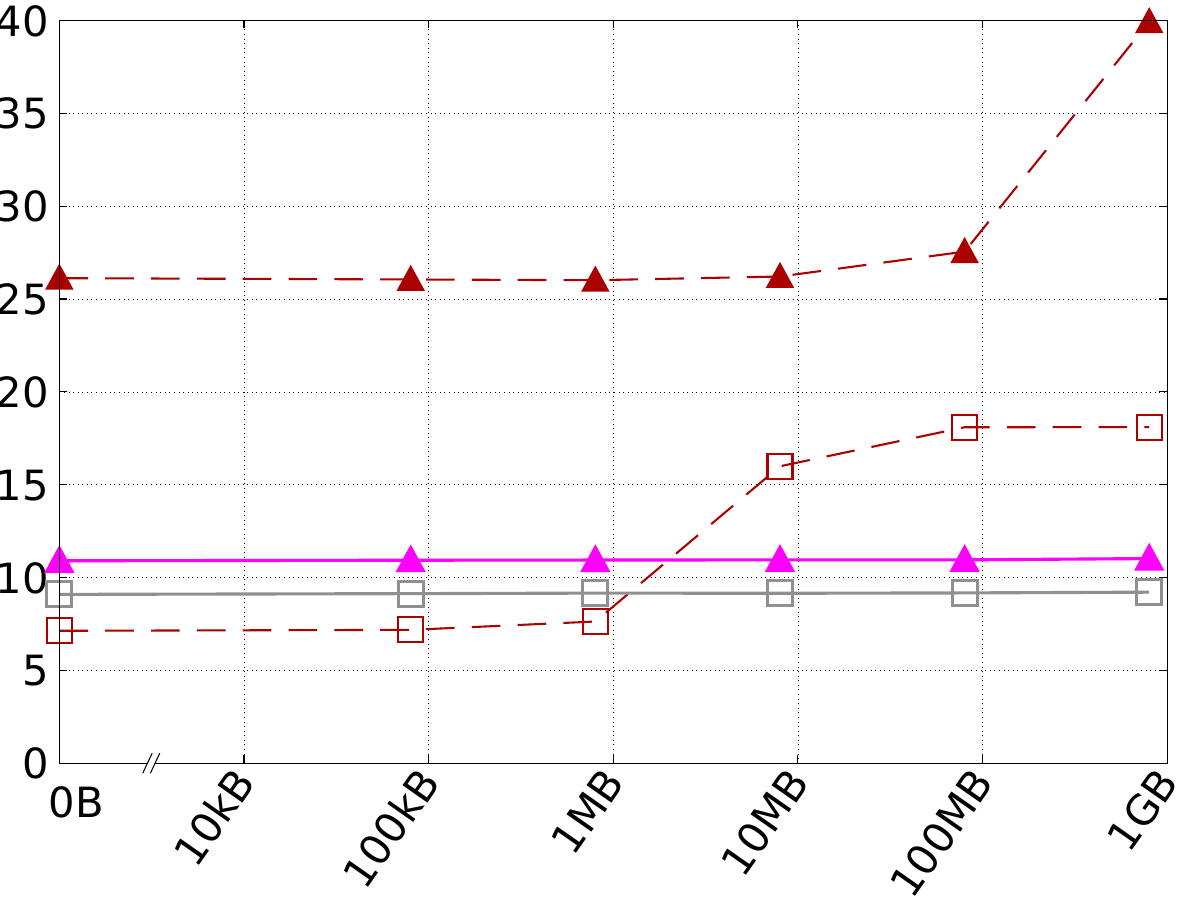}
  \end{subfigure}
  \begin{subfigure}{0.3\columnwidth}
    \centering
    \texttt{mbrot}
    \includegraphics[width=\columnwidth]{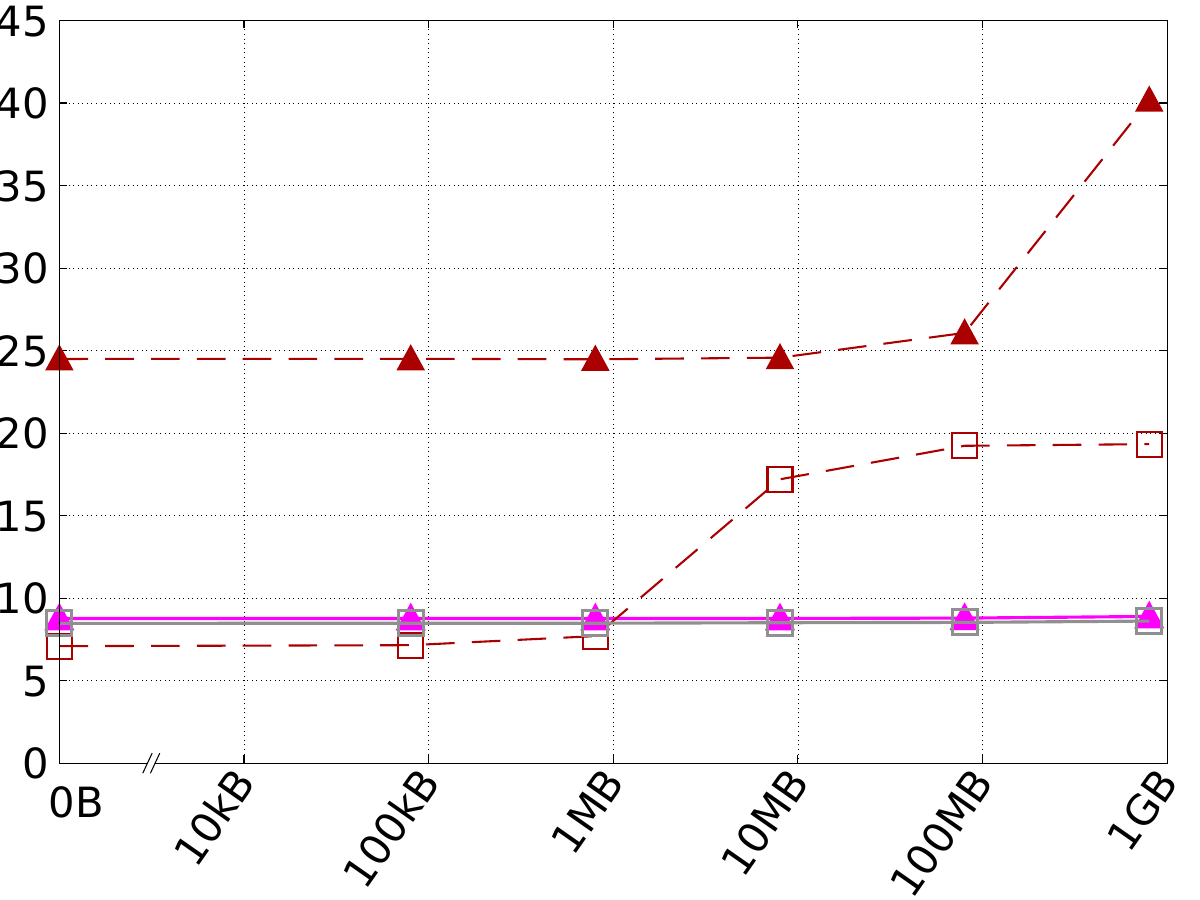}
  \end{subfigure}
  \\[0.5em]
  \begin{subfigure}{0.3\columnwidth}
    \centering
    \texttt{nucleic}
    \includegraphics[width=\columnwidth]{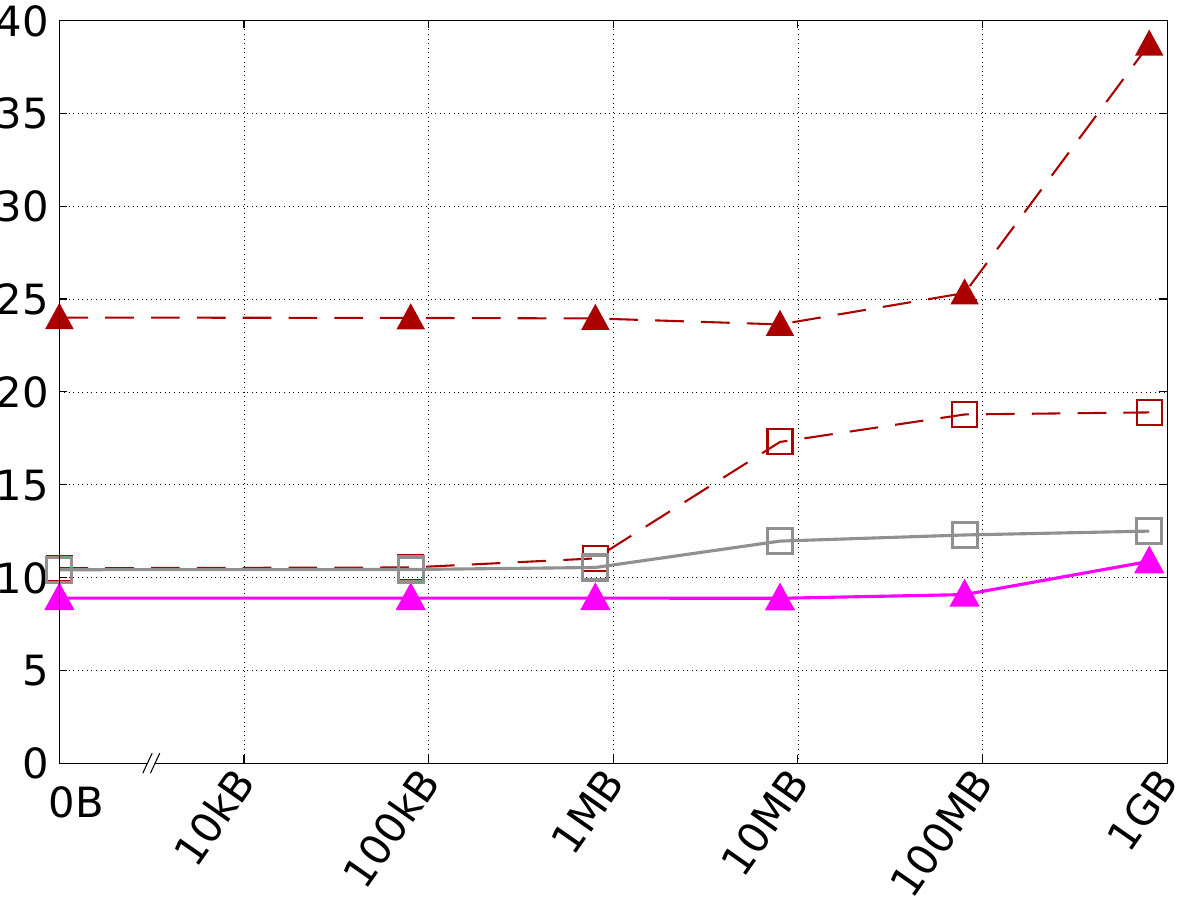}
  \end{subfigure}
  \begin{subfigure}{0.3\columnwidth}
    \centering
    \texttt{pnpoly}
    \includegraphics[width=\columnwidth]{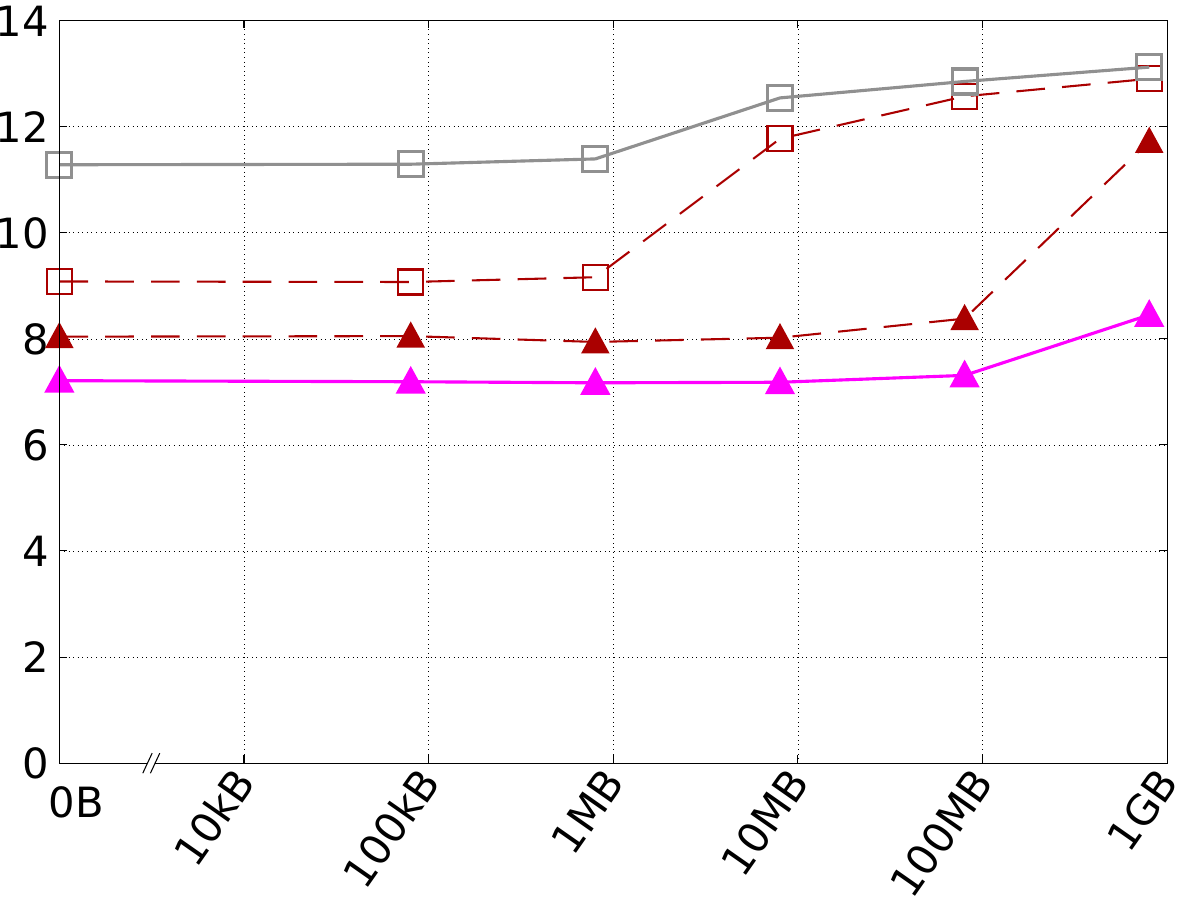}
  \end{subfigure}
  \begin{subfigure}{0.3\columnwidth}
    \centering
    \texttt{ray}
    \includegraphics[width=\columnwidth]{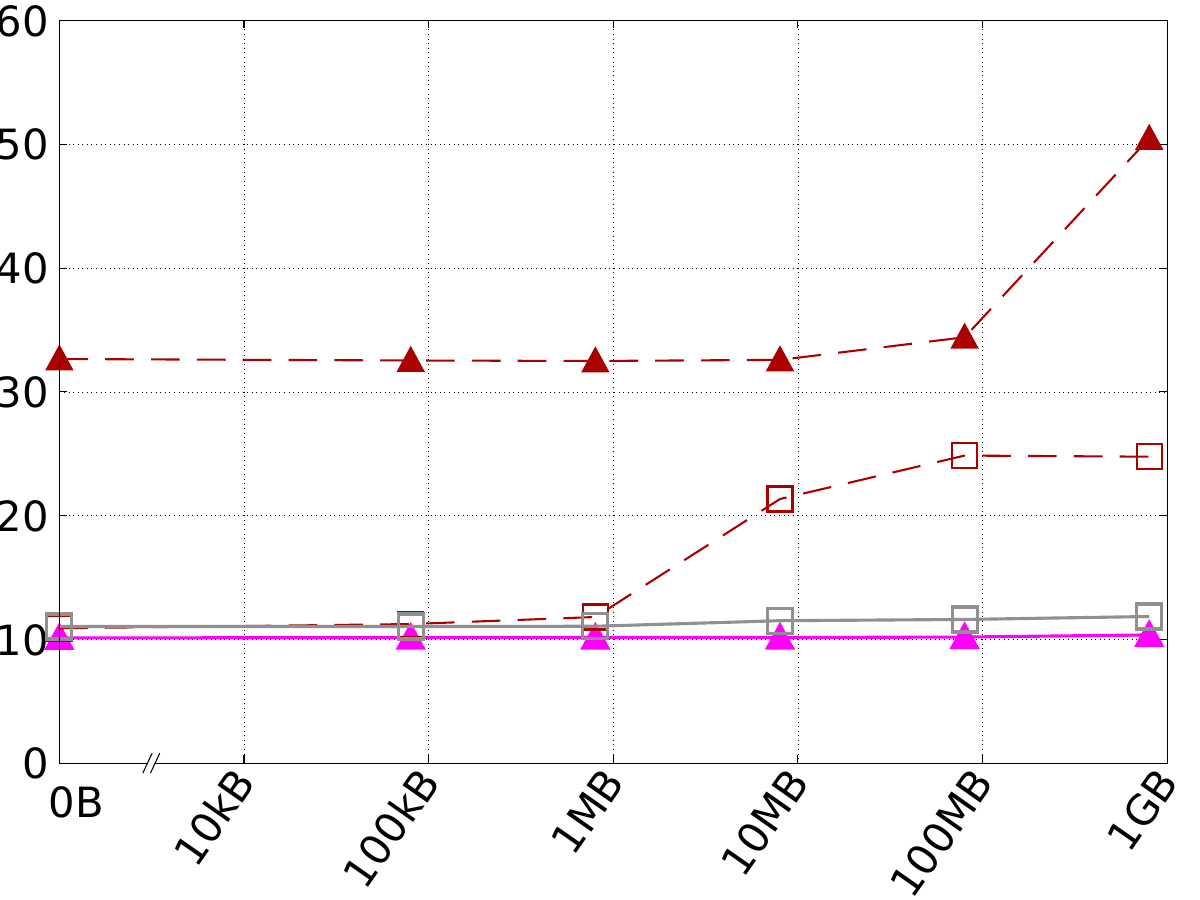}
  \end{subfigure}
  \\[0.5em]
  \begin{subfigure}{0.3\columnwidth}
    \centering
    \texttt{simplex}
    \includegraphics[width=\columnwidth]{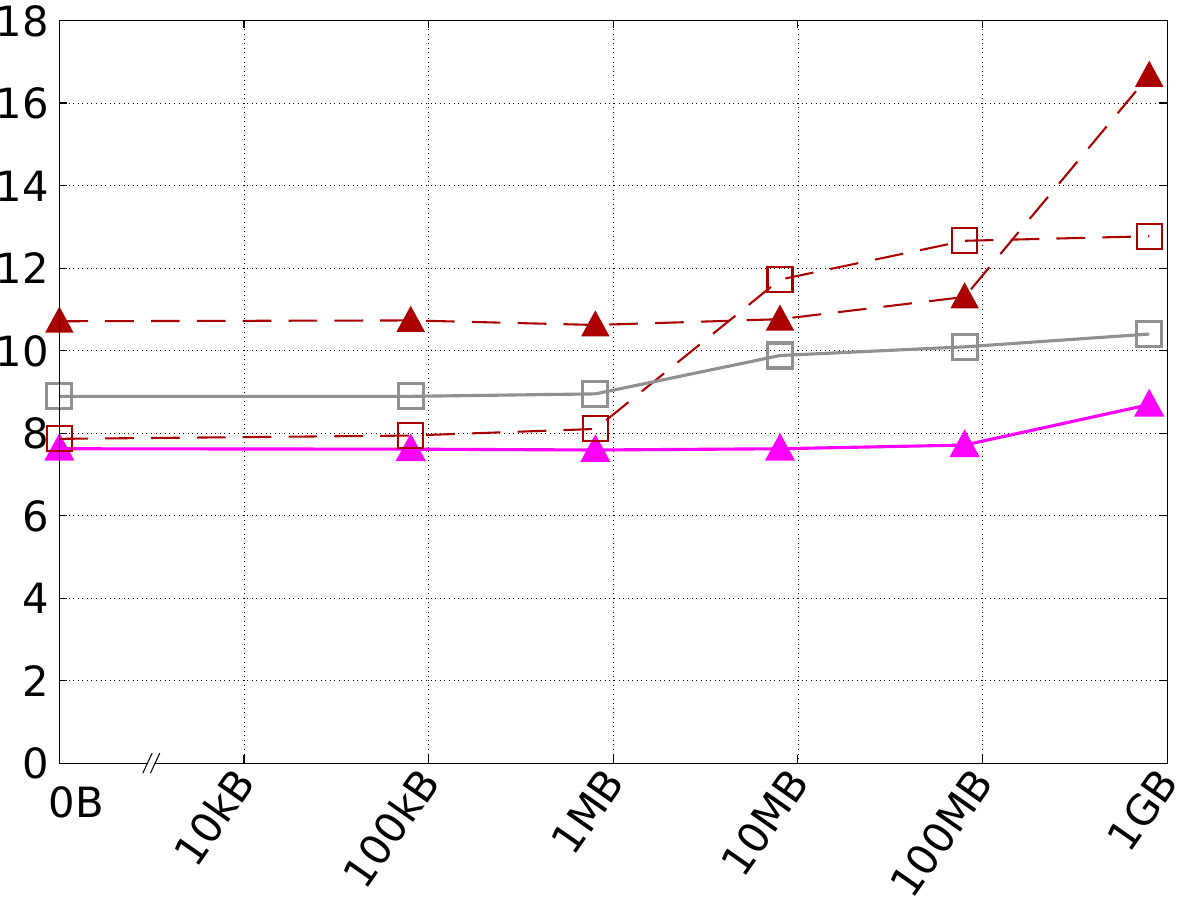}
  \end{subfigure}
  \begin{subfigure}{0.3\columnwidth}
    \centering
    \texttt{sum1}
    \includegraphics[width=\columnwidth]{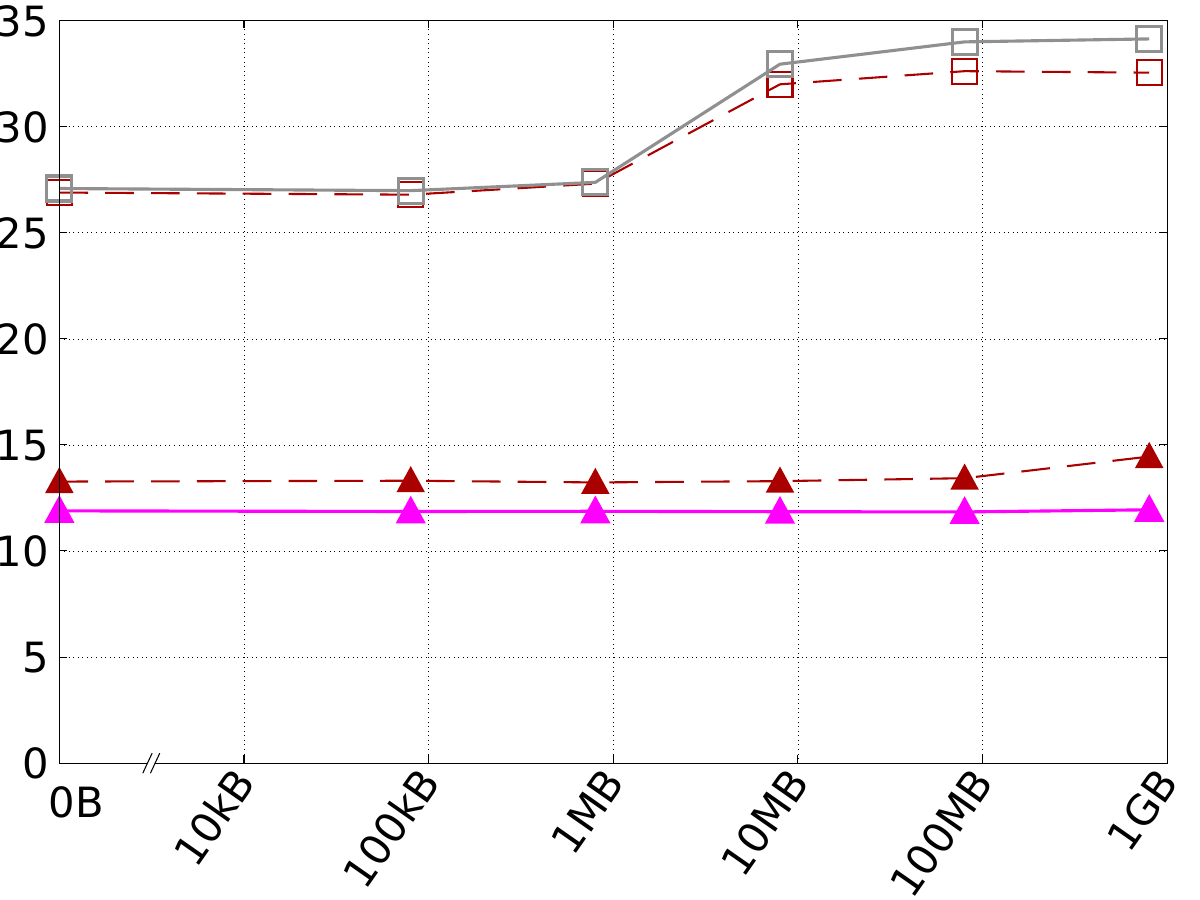}
  \end{subfigure}
  \begin{subfigure}{0.3\columnwidth}
    \centering
    \texttt{sumfp}
    \includegraphics[width=\columnwidth]{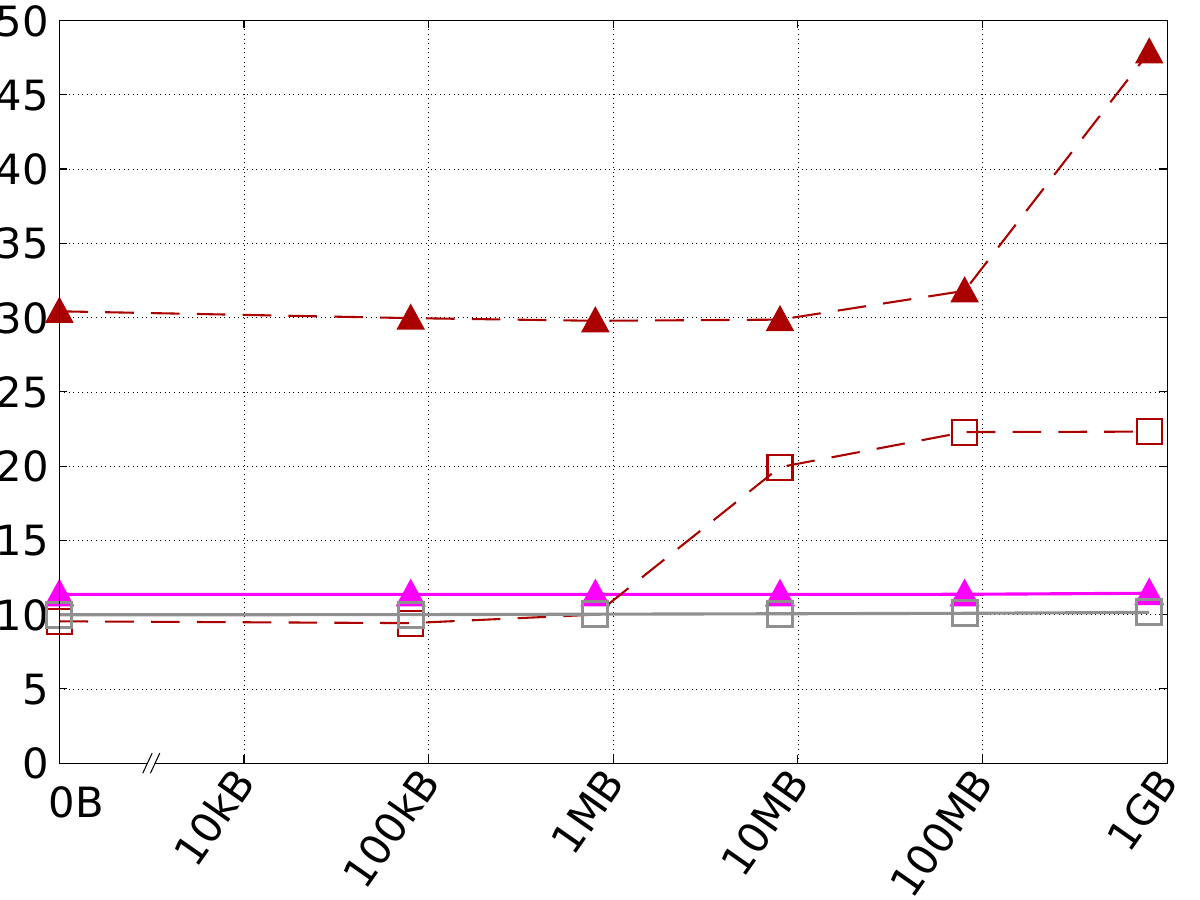}
  \end{subfigure}
  \\[-0.2em]
  \text{{\large Preallocated heap data size (bytes)}}
  \\[0.3em]
  \begin{subfigure}{0.8\columnwidth}
    \includegraphics[width=\columnwidth]{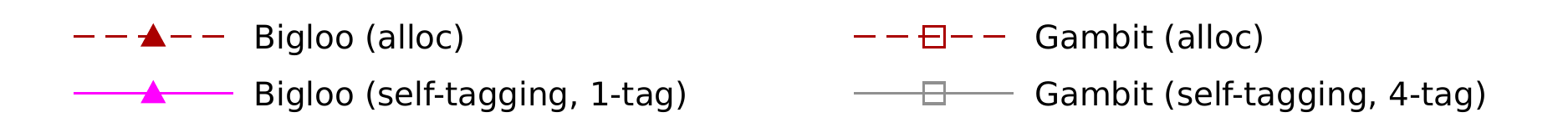}
  \end{subfigure}
  } \caption{ Execution time of float benchmarks with extra data allocated on
  heap before execution with Bigloo and Gambit on \textsc{Amd}. The x-axis shows
  the size of preallocated data in bytes on a logarithmic scale. The y-axis
  represents execution time in seconds (mean of {\experimentsrepetitions}
  repetitions). Execution time without preallocated data is added at $x=0$.
  Dashed lines correspond to times with allocated floats. Solid lines are with
  self-tagging. }
  \label{figure:gc-benchmarks}
\end{figure}

Figures~\ref{figure:bigloo-time-vs-alloc}~and~\ref{figure:gambit-time-vs-alloc}
show execution times of the best self-tagging variants on each implementation
(Bigloo 1-tag, and Gambit 4-tag respectively) relative to allocated floats. For
Bigloo, decreased memory allocations correlates with lower execution time on all
architectures (see Figure~\ref{figure:bigloo-memory}) with the exception of
\texttt{sum1}, an I/O bound benchmark.

Conversely, Gambit execution time slightly increases on many float benchmarks.
This difference is caused by its use of a stop-and-copy garbage collector with
bump allocation, while Bigloo uses the Boehm mark-and-sweep GC~\cite{BoehmW88}.
For programs with a nearly empty heap (default of R7RS benchmarks), the overhead
of self-tagging's encoding/decoding outweighs that of bump allocation.

However, since self-tagging improves performance by allocating fewer floats, its
impact is more pronounced for programs that spend more time in memory management
tasks. This section presents an experiment to measure execution time in a more
realistic setting where benchmarks are executed with live data on the heap, thus
putting more pressure on the garbage collector.

For this experiment, the R7RS benchmark suite is modified to allocate a vector
before the execution of each benchmark. The size of the allocated vector is
given as a parameter of the benchmark. Each float benchmark is then executed
with vectors of $10^4$ to $10^8$ fields, which correspond to about 80~kB and
800~MB respectively. An execution is also done with no preallocated vector.

Figure~\ref{figure:gc-benchmarks} compares execution times for Bigloo and Gambit
with self-tagging (1-tag and 4-tag respectively) and allocated floats with
increasing vector sizes. With a nearly empty heap, and thus little strain on the
garbage collector, the bump allocator of Gambit with allocated floats sometimes
outperforms self-tagging since it does not incur the cost of encoding and
decoding self-tagged floats. Once the heap is preloaded with about 1~MB (which
is the size of the L2 cache per core on the \textsc{Amd} machine) the strain of
the garbage collector starts to outweigh the cost of encoding and decoding
self-tagged floats, thus self-tagging hereafter outperforms allocated floats.

\subsection{Branch Prediction Considerations}
\label{section:branch-prediction}

This section discusses the impact of branch misprediction on the performance of
self-tagging by presenting another variant of self-tagging that was developed
based on intuitions that turned out to be wrong. This negative result is
detailed here to guide implementers when adapting self-tagging.

Self-tagging variants from previous sections use high exponent bits of floats.
This is convenient since it captures contiguous ranges of floats at the cost of
a small encoding/decoding overhead. Alternatively, low bits of a float's
mantissa can be used. For instance, by reserving the tag \texttt{000} for
self-tagged floats and heap allocating all floats whose low mantissa bits are not
\texttt{000}.

\newcommand\timemantissalinewidth{0.98}
\begin{figure}
  {\large Execution Time of Self-Tagging with Low Bits (Bigloo)}\\[0.5em]
  {\small \redrock}\\
  \begin{subfigure}{\timemantissalinewidth\linewidth}
    \center
    \includegraphics[width=\linewidth]{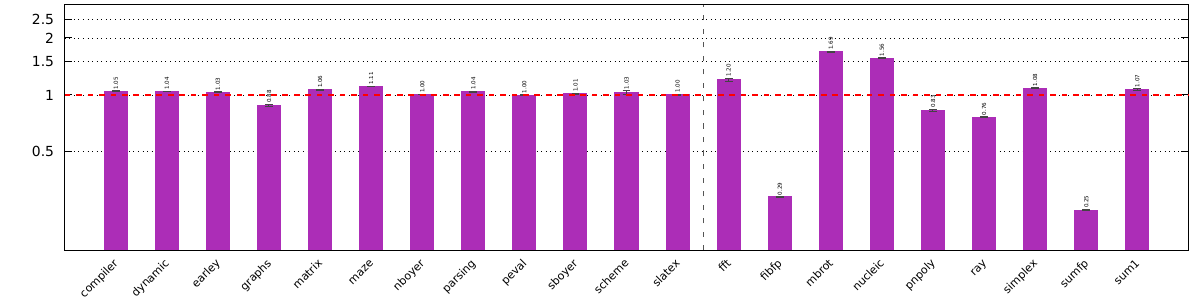}
  \end{subfigure}
  \legendbiglootimemantissa
  \caption{ Execution time of Bigloo with self-tagging using the mantissa low
    bits relative to allocated floats on the \textsc{Amd} microarchitecture. The y-axis
    shows execution time relative to the version that heap allocates
    all floats (dashed horizontal red line) on a logarithmic scale. $1.00\times$ indicates
    no change, lower means faster, and higher means slower execution than
    allocated floats. }
  \label{figure:bigloo-mantissa-vs-alloc}
\end{figure}

This encoding has no overhead since a self-tagged float corresponds exactly to
its IEEE754 value. It comes as the cost of no longer capturing contiguous ranges
of floats, but rather a set uniformly spread across all possible floats. As in
previous techniques, the proportion of captured floats can be increased by
assigning more tags to self-tagging. This variant is implemented in Bigloo with
tags \texttt{000} and \texttt{100}, thus self-tagging $1/4$ of all floats, in
particular all floats $n$ that are integers $|n| \leq 2^{51}$.

Figure~\ref{figure:bigloo-mantissa-vs-alloc} compares execution times of this
variant to allocated floats. As expected, benchmarks that extensively use floats
corresponding to integers (\texttt{fibfp} and \texttt{sumfp}) execute faster,
even faster than previous variants (see
Figure~\ref{figure:bigloo-time-vs-alloc}), due to most floats avoiding heap
allocation. However, other benchmarks (\texttt{mbrot} and \texttt{nucleic}) are
slower despite allocating about $1/4$ fewer floats.

This unpredictability is explainable by profiling the number of branch
mispredictions caused by this variant.
Figure~\ref{figure:bigloo-branch-vs-alloc} shows missed branch predictions of
float encodings, including self-tagging with mantissa bits. This encoding causes
a steep increase in missed branch predictions on some benchmarks, namely
\texttt{mbrot} and \texttt{nucleic}. This stems from the fact that when tags
occupy the low bits of the mantissa, even the smallest variations cause a
float's representation to switch between self-tagged and tagged pointer in a
hard to predict way. This is beyond the capabilities of the branch predictor and
execution suffers costly misprediction penalties~\cite{computersystemsbranch}.

\newcommand\branchlinewidth{0.98}
\begin{figure}
  {\large Branch Misprediction of Self-Tagging with Low Bits (Bigloo)}\\[0.5em]
  {\small \redrock}\\
  \begin{subfigure}{\branchlinewidth\linewidth}
    \center
    \includegraphics[width=\linewidth]{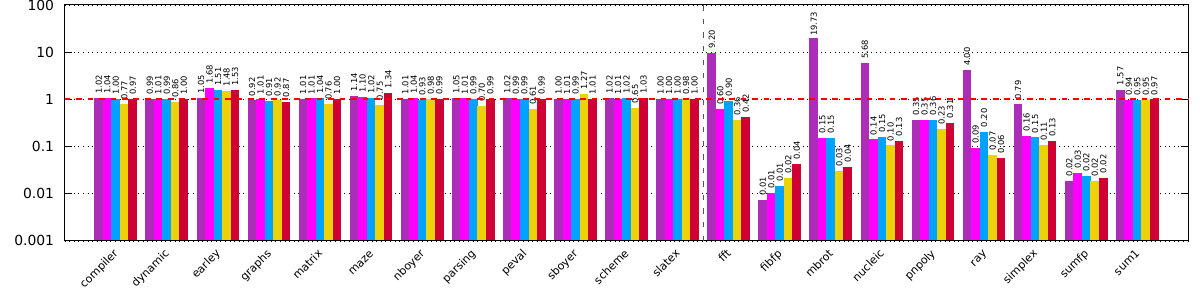}
  \end{subfigure}
  \legendbigloobranch
  \caption{ Branch mispredictions of Bigloo with each variant of self-tagging,
    NaN-boxing, and NuN-boxing relative to allocated floats by the \textsc{Amd}
    microarchitecture. The y-axis shows branch mispredictions relative to
    allocated floats (dashed horizontal red line) on a logarithmic scale. $1.00\times$
    indicates no change, lower means less, and higher means more mispredictions.
    }
  \label{figure:bigloo-branch-vs-alloc}
\end{figure}

This result refutes the intuition that self-tagging with mantissa bits may be a
good compromise due to its very cheap encoding/decoding. It also hints at the
necessity to avoid introducing tests for special values in encoding/decoding,
such as the test for $\pm0.0$ in the 2-tag self-tagging variant with
preallocated zeros, which has a higher occurrence of branch mispredictions than
1-tag self-tagging in Figure~\ref{figure:bigloo-branch-vs-alloc}.


\section{Self-Tagging on 32-bit Architectures}
\label{section:st-on-32-bits}

Although previous sections only considered double-precision floats on
64-bit architectures, self-tagging of single-precision floats is a
natural adaptation for 32-bit architectures.  Indeed, the core idea,
which is to superimpose a tag to a sequence of bits likely to appear
in practice, can be used for any machine word size.  A popular object
representation on 32-bit architectures is to use 2-bit tags (4
possible tags) and heap allocated objects aligned to 32-bit words.

Figure~\ref{figure:32-bit-tag-interval-table} shows the ranges
covered by each combination of the top 4 bits of the exponent,
which is 8 bits wide in the IEEE754 32-bit representation.

The 1-tag approach described previously can be adapted to the 
narrower exponent of the IEEE754 32-bit representation field: the self-tagged
encoding of a float whose IEEE754 32-bit representation is the integer
$n$ can be computed as $(n \oplus ((1 + 2 \times tag) << 27)) <<_{rot}
4$, where $\oplus$ is addition modulo $2^{32}$ and $<<_{rot}$ is the
left rotation operator. If the two lowest bits are equal to $tag$ then
the float is self-tagged. This will cover all float values in the
ranges: $0.0 .. 3.9 \times 10^{-34} \cup 3.1 \times 10^{-5} .. 1.3
\times 10^{5} \cup 1.0 \times 10^{34} .. Infinity/NaN$
(\onetagcolorname rows of Figure~\ref{figure:32-bit-tag-interval-table}
using $tag$=\texttt{00}).

The coverage can be doubled by using an adaptation of the
2-tag variant described in Section~\ref{section:2tags2}:
$0.0 .. 2.5 \times 10^{-29} \cup 4.7 \times 10^{-10} .. 8.6
\times 10^{9} \cup 1.6 \times 10^{29} .. Infinity/NaN$
(\twotagcolorname rows of Figure~\ref{figure:32-bit-tag-interval-table}
using $tag_1$=\texttt{00} and $tag_2$=\texttt{11}).
With this variant, it would be reasonable to reserve one tag
for small signed integers in the range $-2^{29} .. 2^{29}-1$
(-536870912 .. 536870911), and the remaining tag for heap allocated
objects, including the floats that can't be self-tagged.

\begin{figure}
\centering
\setlength\extrarowheight{2pt}
\begin{minipage}[t]{0.49\textwidth}
\center
\begin{tabular}{|c|c|c|}
  \multicolumn{1}{c}{\small \textbf{4 most signif.}} & \multicolumn{1}{c}{\small \textbf{4 most signif.}} & \multicolumn{1}{c}{}
    \\[-1ex]

  \multicolumn{1}{c}{\small \textbf{expo. bits}} & \multicolumn{1}{c}{\small \textbf{expo. bits + 1}} & \multicolumn{1}{c}{\small \textbf{Value}}
    \\
    \hline

    \cellcolor{2tag} & \cellcolor{1tag} & {\texttt{0.0}}
    \\

    \cline{3-3}
    \raisebox{0.8 em}[0 pt]{\cellcolor{2tag}0\underline{00}0} & \raisebox{0.8 em}[0 pt]{\cellcolor{1tag}0\underline{00}1} & {\footnotesize 1.4e-45 .. 3.9e-34}
    \\
    \hline

    {\cellcolor{2tag}0\underline{00}1} & {0010} & {\footnotesize 3.9e-34 .. 2.5e-29}
    \\
    \hline

    {0010} & {0011} & {\footnotesize 2.5e-29 .. 1.7e-24}
    \\
    \hline

    {0011} & {0100} & {\footnotesize 1.7e-24 .. 1.1e-19}
    \\
    \hline

    {0100} & {0101} & {\footnotesize 1.1e-19 .. 7.1e-15}
    \\
    \hline

    {0101} & {0110} & {\footnotesize 7.1e-15 .. 4.7e-10}
    \\
    \hline

    {\cellcolor{2tag}0\underline{11}0} & {0111} & {\footnotesize 4.7e-10 .. 3.1e-5}
    \\
    \hline

    {\cellcolor{2tag}0\underline{11}1} & {\cellcolor{1tag}1\underline{00}0} & {\footnotesize 3.1e-5 .. 2}
    \\
    \hline
\end{tabular}
\end{minipage}
\begin{minipage}[t]{0.49\textwidth}
\center
\begin{tabular}{|c|c|c|}
  \multicolumn{1}{c}{\small \textbf{4 most signif.}} & \multicolumn{1}{c}{\small \textbf{4 most signif.}} & \multicolumn{1}{c}{}
    \\[-1ex]

  \multicolumn{1}{c}{\small \textbf{expo. bits}} & \multicolumn{1}{c}{\small \textbf{expo. bits + 1}} & \multicolumn{1}{c}{\small \textbf{Value}}
    \\
    \hline

    {\cellcolor{2tag}1\underline{00}0} & {\cellcolor{1tag}1\underline{00}1} & {\footnotesize 2 .. 1.3e5}
    \\
    \hline

    {\cellcolor{2tag}1\underline{00}1} & {1010} & {\footnotesize 1.3e5 .. 8.6e9}
    \\
    \hline

    {1010} & {1011} & {\footnotesize 8.6e9 .. 5.6e14}
    \\
    \hline

    {1011} & {1100} & {\footnotesize 5.6e14 .. 3.7e19}
    \\
    \hline

    {1100} & {1101} & {\footnotesize 3.7e19 .. 2.4e24}
    \\
    \hline

    {1101} & {1110} & {\footnotesize 2.4e24 .. 1.6e29}
    \\
    \hline

    {\cellcolor{2tag}1\underline{11}0} & {1111} & {\footnotesize 1.6e29 .. 1.0e34}
    \\
    \hline

    \cellcolor{2tag} & \cellcolor{1tag} & {\footnotesize 1.0e34 .. 3.4e38}

    \\
    \cline{3-3}
    \raisebox{0.8 em}[0 pt]{\cellcolor{2tag}1\underline{11}1} & \raisebox{0.8 em}[0 pt]{\cellcolor{1tag}0\underline{00}0} & {\small \textit{Infinity/NaN}}
    \\
    \hline
\end{tabular}
\end{minipage}

\caption{
  The positive value ranges captured by each combination of the 4 most significant exponent bits of the IEEE754 32-bit representation.
  The rows coloured in \onetagcolorname are covered by the 1-tag variant (tag \texttt{00}) and
  the rows coloured in \twotagcolorname are covered by the 2-tag variant (tags \texttt{00} and \texttt{11}).
    }
\label{figure:32-bit-tag-interval-table}

\end{figure}
\section{Related Work}
\label{section:related-work}

Handling floats is a long-standing issue in dynamic language implementation. In
the last decades, little progress has been made improving the encoding of
floats, with most implementations either using heap allocated floats or
suffering from the overhead of NaN-boxing on pointers.  Thus, more general
strategies that use data flow analysis~\cite{DBLP:conf/lfp/Henglein92,
DBLP:conf/icfp/SerranoF96, 10.1007/978-3-642-28652-0_9, 10.1007/3-540-44854-3_9,
Kawai08} and partial evaluation~\cite{DBLP:conf/pldi/WurthingerWHWSS17} have
been developed to find locations where the type of a value is known across its
lifecycle.  This allows the generation of specialized code that safely handles
fully unboxed, untagged values. Such strategies are not specific to unboxing
floats. Rather, they tackle the more general problem of inferring types in
dynamic languages, which allows dropping type information in either tagged
pointers or NaN-boxed pointers. While more limited in scope, self-tagging solves
the problem of heap allocated floats with a new encoding that involves no
program analysis.

Self-tagging is straightforward to add to compilers that already use object
tagging.  Therefore, implementations that represent floats as tagged pointers
could benefit from it with minimal implementation effort. Such popular
implementations include CPython~\cite{cpythonpep703} and Google's
V8~\cite{v8_pointer_compression}. Compilers that use NaN/NuN-boxing, such as
Mozilla's SpiderMonkey~\cite{spidermonkey} and Apples's
JSC~\cite{javascriptcore} can also benefit from the lower impact that
self-tagging has on the performance of non-float types.

Due to the shortcomings of tagged pointers and NaN-boxing, additional strategies
are used to avoid boxing. A straightforward approach is to provide homogeneous
data structures such as \texttt{TypedArray} in JavaScript~\cite{ecmascript6} and
\texttt{numpy}'s \texttt{float64} arrays in Python~\cite{harris2020array}. Type
homogeneity then allows unboxing values within the collection. More generally,
compilers can detect data homogeneity at run time and use context-dependent
storage strategies to store data in a way that prevents pointer
chasing~\cite{DBLP:conf/oopsla/BolzDT13}. Since self-tagging avoids heap allocating most floats
in practice, it effectively prevents such pointer chasing in the case of floats
without static or run time analysis.

Even for general-purpose languages not specifically designed for numerical
computation, ensuring reasonably fast float operations is a major concern. All
\emph{production-quality} implementations surveyed for this paper devote
significant efforts to optimizing them. Unfortunately, the academic literature
is scarce when it comes to techniques used by these efficient implementations. 

In many cases, the source code itself has to be studied to understand the
implementation and compare it to self-tagging. The \text{NuN}-boxing
implementation used in experiments in this paper is based on that of
JSC\footnote{\url{https://github.com/WebKit/WebKit/commit/74c9b9bbc2c7527144ac366c3f62f84aaa1a8a4e}}.
NuN-boxing has the benefits over self-tagging that it encodes all floats
uniformly and does not reserve tags for them. Its drawbacks are that it
negatively impacts the type checking and accesses of all objects, it can
represent integers with at most 48 bits (usually 32 bits for practical reasons),
and it can't be used on 32-bit architectures.

NaN/NuN-boxing are trade-offs between the performance of floats and pointers
(NaN-boxing favors floats, NuN-boxing favors pointers). Self-tagging offers a
different trade-off between the performance of frequently used floats and other
floats. The fact that some ranges of floats are more commonly used has been
studied in scientific computing where real world data typically fall into a
narrow range of values~\cite{Brown07, Lindstrom06, Afroozeh23, Zurstrassen23}.
This is in accordance with the results presented in
Figure~\ref{figure:tag-interval-table}.

The techniques closest to self-tagging discovered among surveyed implementations
are those of
CRuby\footnote{\url{https://github.com/ruby/ruby/commit/b3b5e626ad69bf22be3228f847f94e1b68f40888}},
OpenSmallTalk\footnote{\url{https://clementbera.wordpress.com/2018/11/09/64-bits-immediate-floats}}
and the MonNom calculus~\cite{Muehlboeck21}. CRuby and OpenSmallTalk use a variant
close to the 1-tag encoding of Section~\ref{section:1tag}. The MonNom
implementation from~\cite{Muehlboeck21} uses a variant of the 2-tag encoding of
Section~\ref{section:2tags} with 2-bit tags instead of 3-bit. Bit patterns
captured by these implementations all exclude $\pm0.0$, which is treated as a
special case. This imposes complex sequences of instructions for
encoding/decoding that can hardly be inlined and risks increasing the number of
branch misses as discussed in Section~\ref{section:branch-prediction}.

To the best of the authors' knowledge, this is the first work
exposing the principles of self-tagging, carefully implementing it
in optimizing compilers, and evaluating its benefits, drawbacks,
and multiple variants.

In this paper, self-tagging was discussed in the context of dynamic
languages.  Yet, polymorphic languages such as OCaml and Haskell also attach
information to run time values and spend considerable effort to find efficient
encodings for abstract data types~\cite{DBLP:journals/pacmpl/BaudonRG23,
DBLP:journals/jfp/KennedyV12}.  Self-tagging could be applied in these
languages, either to floats or to the encoding of abstract data types.


\section{Conclusion}
\label{section:conclusion}

Avoiding heap allocation of floating point numbers has always been a major
concern for dynamic and polymorphic languages. This paper presents and evaluates
the popular techniques deployed to that end (NaN and NuN-boxing) as well as a
new approach, coined \emph{self-tagging}, that allows representing certain floats
without heap allocation.

The core idea is to apply a bitwise transformation that maps the most common
floats used in practice to bit patterns that contain the required type
information at the correct position. This allows encoding floats in specific
ranges as tagged values instead of heap allocated objects, reducing the strain
on the garbage collector and improving performance. Contrary to NaN-boxing and
NuN-boxing, which are designed to optimize the handling of floats, self-tagging
offers a different trade-off that adds no overhead to the handling of other
types. Since it does not rely on the specificity of the IEEE754 NaN encoding,
self-tagging is highly portable, including to 32-bit architectures.

It is also straightforward to retrofit in implementations that already use
object tagging. The only required changes are to reserve one or more tags for
float self-tagging and implement an encoding/decoding function for self-tagged
floats. The rest of the runtime stays unchanged.

As shown by the experimental evidence presented, float self-tagging
is a good candidate for general-purpose languages, such as JavaScript, whose
specification relies heavily on floats, but where the performance of other types
must not be sacrificed.

Given the purely bit-shuffling nature of float self-tagging boxing and
unboxing, it may also be a good candidate for an essentially zero-cost
implementation at the hardware level when floats are moved between
float and integer registers.


\section{Data Availability Statement}
\label{section:availability}

Data generated and analyzed for this paper is available in a companion
artifact~\citeartifact. This includes results for all microarchitectures listed
in Section~\ref{section:experiment}, some of which were omitted in this paper
for lack of space. The artifact also contains all source code used to run the
experiments.

\bibliographystyle{ACM-Reference-Format}
\bibliography{references}


\begin{thebibliography}{43}


\ifx \showCODEN    \undefined \def \showCODEN     #1{\unskip}     \fi
\ifx \showDOI      \undefined \def \showDOI       #1{#1}\fi
\ifx \showISBNx    \undefined \def \showISBNx     #1{\unskip}     \fi
\ifx \showISBNxiii \undefined \def \showISBNxiii  #1{\unskip}     \fi
\ifx \showISSN     \undefined \def \showISSN      #1{\unskip}     \fi
\ifx \showLCCN     \undefined \def \showLCCN      #1{\unskip}     \fi
\ifx \shownote     \undefined \def \shownote      #1{#1}          \fi
\ifx \showarticletitle \undefined \def \showarticletitle #1{#1}   \fi
\ifx \showURL      \undefined \def \showURL       {\relax}        \fi
\providecommand\bibfield[2]{#2}
\providecommand\bibinfo[2]{#2}
\providecommand\natexlab[1]{#1}
\providecommand\showeprint[2][]{arXiv:#2}

\bibitem[iee(2019)]%
        {ieee754}
 \bibinfo{year}{2019}\natexlab{}.
\newblock \showarticletitle{IEEE Standard for Floating-Point Arithmetic}.
\newblock \bibinfo{journal}{\emph{IEEE Std 754-2019 (Revision of IEEE
  754-2008)}} (\bibinfo{year}{2019}), \bibinfo{pages}{1--84}.
\newblock
\urldef\tempurl%
\url{https://doi.org/10.1109/IEEESTD.2019.8766229}
\showDOI{\tempurl}


\bibitem[r7r(2024)]%
        {r7rs-benchmarks}
 \bibinfo{year}{2024}\natexlab{}.
\newblock \bibinfo{title}{{R7RS} Benchmarks}.
\newblock
  \bibinfo{howpublished}{\url{https://github.com/ecraven/r7rs-benchmarks}}.
\newblock


\bibitem[Afroozeh et~al\mbox{.}(2023)]%
        {Afroozeh23}
\bibfield{author}{\bibinfo{person}{Azim Afroozeh}, \bibinfo{person}{Leonardo~X.
  Kuffo}, {and} \bibinfo{person}{Peter~A. Boncz}.}
  \bibinfo{year}{2023}\natexlab{}.
\newblock \showarticletitle{{ALP:} Adaptive Lossless floating-Point
  Compression}.
\newblock \bibinfo{journal}{\emph{Proc. {ACM} Manag. Data}}
  \bibinfo{volume}{1}, \bibinfo{number}{4} (\bibinfo{year}{2023}),
  \bibinfo{pages}{230:1--230:26}.
\newblock
\urldef\tempurl%
\url{https://doi.org/10.1145/3626717}
\showDOI{\tempurl}


\bibitem[Appel(1998)]%
        {moderncompilerinctagging}
\bibfield{author}{\bibinfo{person}{Andrew~W. Appel}.}
  \bibinfo{year}{1998}\natexlab{}.
\newblock \bibinfo{booktitle}{\emph{Modern Compiler Implementation in C}}.
\newblock \bibinfo{publisher}{Cambridge University Press},
  \bibinfo{address}{Cambridge, UK}, Chapter~16, \bibinfo{pages}{372--373}.
\newblock


\bibitem[Baudon et~al\mbox{.}(2023)]%
        {DBLP:journals/pacmpl/BaudonRG23}
\bibfield{author}{\bibinfo{person}{Tha{\"{\i}}s Baudon},
  \bibinfo{person}{Gabriel Radanne}, {and} \bibinfo{person}{Laure Gonnord}.}
  \bibinfo{year}{2023}\natexlab{}.
\newblock \showarticletitle{Bit-Stealing Made Legal: Compilation for Custom
  Memory Representations of Algebraic Data Types}.
\newblock \bibinfo{journal}{\emph{Proc. {ACM} Program. Lang.}}
  \bibinfo{volume}{7}, \bibinfo{number}{{ICFP}} (\bibinfo{year}{2023}),
  \bibinfo{pages}{813--846}.
\newblock
\urldef\tempurl%
\url{https://doi.org/10.1145/3607858}
\showDOI{\tempurl}


\bibitem[Bellard and Gordon(2024)]%
        {quickjs}
\bibfield{author}{\bibinfo{person}{Fabrice Bellard} {and}
  \bibinfo{person}{Charlie Gordon}.} \bibinfo{year}{2024}\natexlab{}.
\newblock \bibinfo{title}{QuickJS JavaScript Engine}.
\newblock \bibinfo{howpublished}{\url{https://bellard.org/quickjs/}}.
\newblock


\bibitem[Boehm and Weiser(1988)]%
        {BoehmW88}
\bibfield{author}{\bibinfo{person}{Hans{-}Juergen Boehm} {and}
  \bibinfo{person}{Mark~D. Weiser}.} \bibinfo{year}{1988}\natexlab{}.
\newblock \showarticletitle{Garbage Collection in an Uncooperative
  Environment}.
\newblock \bibinfo{journal}{\emph{Softw. Pract. Exp.}} \bibinfo{volume}{18},
  \bibinfo{number}{9} (\bibinfo{year}{1988}), \bibinfo{pages}{807--820}.
\newblock
\urldef\tempurl%
\url{https://doi.org/10.1002/SPE.4380180902}
\showDOI{\tempurl}


\bibitem[Bolz et~al\mbox{.}(2013)]%
        {DBLP:conf/oopsla/BolzDT13}
\bibfield{author}{\bibinfo{person}{Carl~Friedrich Bolz}, \bibinfo{person}{Lukas
  Diekmann}, {and} \bibinfo{person}{Laurence Tratt}.}
  \bibinfo{year}{2013}\natexlab{}.
\newblock \showarticletitle{Storage strategies for collections in dynamically
  typed languages}. In \bibinfo{booktitle}{\emph{Proceedings of the 2013 {ACM}
  {SIGPLAN} International Conference on Object Oriented Programming Systems
  Languages {\&} Applications, {OOPSLA} 2013, part of {SPLASH} 2013,
  Indianapolis, IN, USA, October 26-31, 2013}},
  \bibfield{editor}{\bibinfo{person}{Antony~L. Hosking},
  \bibinfo{person}{Patrick~Th. Eugster}, {and} \bibinfo{person}{Cristina~V.
  Lopes}} (Eds.). \bibinfo{publisher}{{ACM}}, \bibinfo{pages}{167--182}.
\newblock
\urldef\tempurl%
\url{https://doi.org/10.1145/2509136.2509531}
\showDOI{\tempurl}


\bibitem[Brown et~al\mbox{.}(2007)]%
        {Brown07}
\bibfield{author}{\bibinfo{person}{Ashley~W. Brown}, \bibinfo{person}{Paul
  H.~J. Kelly}, {and} \bibinfo{person}{Wayne Luk}.}
  \bibinfo{year}{2007}\natexlab{}.
\newblock \showarticletitle{Profiling floating point value ranges for
  reconfigurable implementation}. In \bibinfo{booktitle}{\emph{Proceedings of
  the 1st HiPEAC Workshop on Reconfigurable Computing}}.
  \bibinfo{pages}{6--16}.
\newblock


\bibitem[Bryant and O'Hallaron(2016)]%
        {computersystemsbranch}
\bibfield{author}{\bibinfo{person}{Randal~E. Bryant} {and}
  \bibinfo{person}{David~R. O'Hallaron}.} \bibinfo{year}{2016}\natexlab{}.
\newblock \bibinfo{booktitle}{\emph{Computer Systems - A Programmer's
  Perspective} (\bibinfo{edition}{{T}hird {G}lobal} ed.)}.
\newblock \bibinfo{publisher}{Pearson}, Chapter~5, \bibinfo{pages}{577--579}.
\newblock


\bibitem[Corporation(2024)]%
        {intel_instruction_set}
\bibfield{author}{\bibinfo{person}{Intel Corporation}.}
  \bibinfo{year}{2024}\natexlab{}.
\newblock \bibinfo{booktitle}{\emph{Intel® 64 and IA-32 Architectures Software
  Developer's Manual - Instruction Set Reference}}.
\newblock
\urldef\tempurl%
\url{https://www.intel.com/content/www/us/en/developer/articles/technical/intel-sdm.html}
\showURL{%
\tempurl}


\bibitem[Engelen(2024)]%
        {tinylisp}
\bibfield{author}{\bibinfo{person}{Robert~Van Engelen}.}
  \bibinfo{year}{2024}\natexlab{}.
\newblock \bibinfo{title}{Lisp in 99 lines of C and how to write one yourself}.
\newblock
  \bibinfo{howpublished}{\url{https://github.com/Robert-van-Engelen/tinylisp}}.
\newblock


\bibitem[Foundation(2024)]%
        {spidermonkey}
\bibfield{author}{\bibinfo{person}{Mozilla Foundation}.}
  \bibinfo{year}{2024}\natexlab{}.
\newblock \bibinfo{title}{SpiderMonkey 131.0 - Mozilla's JavaScript and
  WebAssembly Engine}.
\newblock \bibinfo{howpublished}{\url{https://spidermonkey.dev/}}.
\newblock


\bibitem[Gross(2023)]%
        {cpythonpep703}
\bibfield{author}{\bibinfo{person}{Sam Gross}.}
  \bibinfo{year}{2023}\natexlab{}.
\newblock \bibinfo{title}{PEP 703 - Making the Global Interpreter Lock Optional
  in CPython}.
\newblock \bibinfo{howpublished}{\url{https://peps.python.org/pep-0703}}.
\newblock


\bibitem[Gudeman(1993)]%
        {Gudeman1993RepresentingTI}
\bibfield{author}{\bibinfo{person}{David~A. Gudeman}.}
  \bibinfo{year}{1993}\natexlab{}.
\newblock \bibinfo{booktitle}{\emph{Representing Type Information in
  Dynamically Typed Languages}}.
\newblock \bibinfo{type}{{T}echnical {R}eport}.
  \bibinfo{institution}{University of Arizona}.
\newblock
\urldef\tempurl%
\url{https://www.cs.arizona.edu/sites/default/files/TR93-27.pdf}
\showURL{%
\tempurl}


\bibitem[Harris et~al\mbox{.}(2020)]%
        {harris2020array}
\bibfield{author}{\bibinfo{person}{Charles~R. Harris},
  \bibinfo{person}{K.~Jarrod Millman}, \bibinfo{person}{St{\'{e}}fan~J. van~der
  Walt}, \bibinfo{person}{Ralf Gommers}, \bibinfo{person}{Pauli Virtanen},
  \bibinfo{person}{David Cournapeau}, \bibinfo{person}{Eric Wieser},
  \bibinfo{person}{Julian Taylor}, \bibinfo{person}{Sebastian Berg},
  \bibinfo{person}{Nathaniel~J. Smith}, \bibinfo{person}{Robert Kern},
  \bibinfo{person}{Matti Picus}, \bibinfo{person}{Stephan Hoyer},
  \bibinfo{person}{Marten~H. van Kerkwijk}, \bibinfo{person}{Matthew Brett},
  \bibinfo{person}{Allan Haldane}, \bibinfo{person}{Jaime~Fern{\'{a}}ndez del
  R{\'{i}}o}, \bibinfo{person}{Mark Wiebe}, \bibinfo{person}{Pearu Peterson},
  \bibinfo{person}{Pierre G{\'{e}}rard-Marchant}, \bibinfo{person}{Kevin
  Sheppard}, \bibinfo{person}{Tyler Reddy}, \bibinfo{person}{Warren Weckesser},
  \bibinfo{person}{Hameer Abbasi}, \bibinfo{person}{Christoph Gohlke}, {and}
  \bibinfo{person}{Travis~E. Oliphant}.} \bibinfo{year}{2020}\natexlab{}.
\newblock \showarticletitle{Array programming with {NumPy}}.
\newblock \bibinfo{journal}{\emph{Nature}} \bibinfo{volume}{585},
  \bibinfo{number}{7825} (\bibinfo{date}{Sept.} \bibinfo{year}{2020}),
  \bibinfo{pages}{357--362}.
\newblock
\urldef\tempurl%
\url{https://doi.org/10.1038/s41586-020-2649-2}
\showDOI{\tempurl}


\bibitem[Henglein(1992)]%
        {DBLP:conf/lfp/Henglein92}
\bibfield{author}{\bibinfo{person}{Fritz Henglein}.}
  \bibinfo{year}{1992}\natexlab{}.
\newblock \showarticletitle{Global Tagging Optimization by Type Inference}. In
  \bibinfo{booktitle}{\emph{Proceedings of the Conference on Lisp and
  Functional Programming, {LFP} 1992, San Francisco, California, USA, 22-24
  June 1992}}, \bibfield{editor}{\bibinfo{person}{Jon~L. White}} (Ed.).
  \bibinfo{publisher}{{ACM}}, \bibinfo{pages}{205--215}.
\newblock
\urldef\tempurl%
\url{https://doi.org/10.1145/141471.141542}
\showDOI{\tempurl}


\bibitem[Inc.(2024)]%
        {javascriptcore}
\bibfield{author}{\bibinfo{person}{Apple Inc.}}
  \bibinfo{year}{2024}\natexlab{}.
\newblock \bibinfo{title}{JavaScriptCore 2.46.3 - JavaScript Engine for
  WebKit}.
\newblock \bibinfo{howpublished}{\url{https://webkit.org}}.
\newblock


\bibitem[International(2015)]%
        {ecmascript6}
\bibfield{author}{\bibinfo{person}{{ECMA} International}.}
  \bibinfo{year}{2015}\natexlab{}.
\newblock \bibinfo{booktitle}{\emph{Standard {ECMA}-262 - {ECMAS}cript Language
  Specification} (\bibinfo{edition}{6th} ed.)}.
\newblock
\urldef\tempurl%
\url{https://www.ecma-international.org/ecma-262/6.0/}
\showURL{%
\tempurl}


\bibitem[Jones et~al\mbox{.}(2012a)]%
        {gchandbooktags}
\bibfield{author}{\bibinfo{person}{Richard Jones}, \bibinfo{person}{Antony
  Hosking}, {and} \bibinfo{person}{Eliot Moss}.}
  \bibinfo{year}{2012}\natexlab{a}.
\newblock \bibinfo{booktitle}{\emph{The Garbage Collection Handbook: The Art of
  Automatic Memory Management}}.
\newblock \bibinfo{publisher}{Chapman and Hall/CRC}, \bibinfo{address}{Florida,
  USA}, Chapter~11, \bibinfo{pages}{168--171}.
\newblock


\bibitem[Jones et~al\mbox{.}(2012b)]%
        {gchandbookgenerations}
\bibfield{author}{\bibinfo{person}{Richard Jones}, \bibinfo{person}{Antony
  Hosking}, {and} \bibinfo{person}{Eliot Moss}.}
  \bibinfo{year}{2012}\natexlab{b}.
\newblock \bibinfo{booktitle}{\emph{The Garbage Collection Handbook: The Art of
  Automatic Memory Management}}.
\newblock \bibinfo{publisher}{Chapman and Hall/CRC}, \bibinfo{address}{Florida,
  USA}, Chapter~9, \bibinfo{pages}{111--114}.
\newblock


\bibitem[Kawai(2008)]%
        {Kawai08}
\bibfield{author}{\bibinfo{person}{Shiro Kawai}.}
  \bibinfo{year}{2008}\natexlab{}.
\newblock \showarticletitle{Efficient floating-point number handling for
  dynamically typed scripting languages}. In
  \bibinfo{booktitle}{\emph{Proceedings of the 2008 Symposium on Dynamic
  Languages, {DLS} 2008, July 8, 2008, Paphos, Cyprus}},
  \bibfield{editor}{\bibinfo{person}{Johan Brichau}} (Ed.).
  \bibinfo{publisher}{{ACM}}, \bibinfo{pages}{6}.
\newblock
\urldef\tempurl%
\url{https://doi.org/10.1145/1408681.1408687}
\showDOI{\tempurl}


\bibitem[Kennedy and Vytiniotis(2012)]%
        {DBLP:journals/jfp/KennedyV12}
\bibfield{author}{\bibinfo{person}{Andrew~J. Kennedy} {and}
  \bibinfo{person}{Dimitrios Vytiniotis}.} \bibinfo{year}{2012}\natexlab{}.
\newblock \showarticletitle{Every bit counts: The binary representation of
  typed data and programs}.
\newblock \bibinfo{journal}{\emph{J. Funct. Program.}} \bibinfo{volume}{22},
  \bibinfo{number}{4-5} (\bibinfo{year}{2012}), \bibinfo{pages}{529--573}.
\newblock
\urldef\tempurl%
\url{https://doi.org/10.1017/S0956796812000263}
\showDOI{\tempurl}


\bibitem[Leijen(2022)]%
        {Leijen2022}
\bibfield{author}{\bibinfo{person}{Daan Leijen}.}
  \bibinfo{year}{2022}\natexlab{}.
\newblock \bibinfo{booktitle}{\emph{What About the Integer Numbers? Fast
  Arithmetic with Tagged Integers - A Plea for Hardware Support}}.
\newblock \bibinfo{type}{{T}echnical {R}eport} MSR-TR-2022-17.
  \bibinfo{institution}{Microsoft Research}.
\newblock
\urldef\tempurl%
\url{https://www.microsoft.com/en-us/research/publication/what-about-the-integer-numbers-fast-arithmetic-with-tagged-integers-a-plea-for-hardware-support/}
\showURL{%
\tempurl}


\bibitem[Limited(2009)]%
        {arm_instruction_set}
\bibfield{author}{\bibinfo{person}{Arm Limited}.}
  \bibinfo{year}{2009}\natexlab{}.
\newblock \bibinfo{booktitle}{\emph{Instruction Set Assembly Guide for Armv7
  and earlier Arm® architectures Reference Guide}}.
\newblock
\newblock
\shownote{\url{https://developer.arm.com/documentation/100076/0200}}.


\bibitem[Lindahl and Sagonas(2003)]%
        {10.1007/3-540-44854-3_9}
\bibfield{author}{\bibinfo{person}{Tobias Lindahl} {and}
  \bibinfo{person}{Konstantinos Sagonas}.} \bibinfo{year}{2003}\natexlab{}.
\newblock \showarticletitle{Unboxed Compilation of Floating Point Arithmetic in
  a Dynamically Typed Language Environment}. In
  \bibinfo{booktitle}{\emph{Implementation of Functional Languages}},
  \bibfield{editor}{\bibinfo{person}{Ricardo Pe{\~{n}}a} {and}
  \bibinfo{person}{Thomas Arts}} (Eds.). \bibinfo{publisher}{Springer Berlin
  Heidelberg}, \bibinfo{address}{Berlin, Heidelberg},
  \bibinfo{pages}{134--149}.
\newblock
\showISBNx{978-3-540-44854-9}
\urldef\tempurl%
\url{https://dx.doi.org/10.1007/3-540-44854-3_9}
\showURL{%
\tempurl}


\bibitem[Lindstrom and Isenburg(2006)]%
        {Lindstrom06}
\bibfield{author}{\bibinfo{person}{Peter Lindstrom} {and}
  \bibinfo{person}{Martin Isenburg}.} \bibinfo{year}{2006}\natexlab{}.
\newblock \showarticletitle{Fast and Efficient Compression of Floating-Point
  Data}.
\newblock \bibinfo{journal}{\emph{IEEE Transactions on Visualization and
  Computer Graphics}} \bibinfo{volume}{12}, \bibinfo{number}{5}
  (\bibinfo{year}{2006}), \bibinfo{pages}{1245--1250}.
\newblock
\urldef\tempurl%
\url{https://doi.org/10.1109/TVCG.2006.143}
\showDOI{\tempurl}


\bibitem[{Manuel Serrano}(2024)]%
        {bigloo}
\bibfield{author}{\bibinfo{person}{{Manuel Serrano}}.}
  \bibinfo{year}{2024}\natexlab{}.
\newblock \bibinfo{title}{{B}igloo}.
\newblock
  \bibinfo{howpublished}{\url{https://www-sop.inria.fr/indes/fp/Bigloo}}.
\newblock


\bibitem[{Marc Feeley}(2025)]%
        {gambit}
\bibfield{author}{\bibinfo{person}{{Marc Feeley}}.}
  \bibinfo{year}{2025}\natexlab{}.
\newblock \bibinfo{title}{{G}ambit}.
\newblock \bibinfo{howpublished}{\url{https://gambitscheme.org}}.
\newblock


\bibitem[Mason(2022)]%
        {DBLP:conf/iwst/Mason22}
\bibfield{author}{\bibinfo{person}{Dave Mason}.}
  \bibinfo{year}{2022}\natexlab{}.
\newblock \showarticletitle{Design Principles for a High-Performance
  Smalltalk}. In \bibinfo{booktitle}{\emph{Proceedings of the International
  Workshop on Smalltalk Technologies 2022 co-located with the 28th European
  Smalltalk User Group Conference {(ESUG} 2022), Novi Sad, Serbia, August
  24th-26th, 2022}} \emph{(\bibinfo{series}{{CEUR} Workshop Proceedings},
  Vol.~\bibinfo{volume}{3325})}, \bibfield{editor}{\bibinfo{person}{Lo{\"{\i}}c
  Lagadec} {and} \bibinfo{person}{Vincent Aranega}} (Eds.).
  \bibinfo{publisher}{CEUR-WS.org}.
\newblock
\urldef\tempurl%
\url{https://ceur-ws.org/Vol-3325/regular2.pdf}
\showURL{%
\tempurl}


\bibitem[Melan\c{c}on et~al\mbox{.}(2025)]%
        {artifact}
\bibfield{author}{\bibinfo{person}{Olivier Melan\c{c}on}, \bibinfo{person}{Marc
  Feeley}, {and} \bibinfo{person}{Manuel Serrano}.}
  \bibinfo{year}{2025}\natexlab{}.
\newblock \bibinfo{booktitle}{\emph{Float Self-Tagging Artifact}}.
\newblock
\urldef\tempurl%
\url{https://doi.org/10.5281/zenodo.16356364}
\showDOI{\tempurl}


\bibitem[Muehlboeck and Tate(2021)]%
        {Muehlboeck21}
\bibfield{author}{\bibinfo{person}{Fabian Muehlboeck} {and}
  \bibinfo{person}{Ross Tate}.} \bibinfo{year}{2021}\natexlab{}.
\newblock \showarticletitle{Transitioning from structural to nominal code with
  efficient gradual typing}.
\newblock \bibinfo{journal}{\emph{Proc. {ACM} Program. Lang.}}
  \bibinfo{volume}{5}, \bibinfo{number}{{OOPSLA}} (\bibinfo{year}{2021}),
  \bibinfo{pages}{1--29}.
\newblock
\urldef\tempurl%
\url{https://doi.org/10.1145/3485504}
\showDOI{\tempurl}


\bibitem[Newman et~al\mbox{.}(2024)]%
        {sbcl}
\bibfield{author}{\bibinfo{person}{William~Harold Newman},
  \bibinfo{person}{Christophe Rhodes}, \bibinfo{person}{Nikodemus Siivola},
  \bibinfo{person}{Juho Snellman}, \bibinfo{person}{Paul Khuong},
  \bibinfo{person}{Jan Moringen}, {and} \bibinfo{person}{Douglas Katzman}.}
  \bibinfo{year}{2024}\natexlab{}.
\newblock \bibinfo{title}{Steel Bank Common Lisp 2.4}.
\newblock \bibinfo{howpublished}{\url{https://www.sbcl.org}}.
\newblock


\bibitem[Pall(2024)]%
        {luajit}
\bibfield{author}{\bibinfo{person}{Mike Pall}.}
  \bibinfo{year}{2024}\natexlab{}.
\newblock \bibinfo{title}{LuaJIT 2.1 - Just-In-Time Compiler for Lua}.
\newblock \bibinfo{howpublished}{\url{https://luajit.org}}.
\newblock


\bibitem[Petersen and Glew(2012)]%
        {10.1007/978-3-642-28652-0_9}
\bibfield{author}{\bibinfo{person}{Leaf Petersen} {and} \bibinfo{person}{Neal
  Glew}.} \bibinfo{year}{2012}\natexlab{}.
\newblock \showarticletitle{GC-Safe interprocedural unboxing}. In
  \bibinfo{booktitle}{\emph{Proceedings of the 21st International Conference on
  Compiler Construction}} (Tallinn, Estonia) \emph{(\bibinfo{series}{CC'12})}.
  \bibinfo{publisher}{Springer-Verlag}, \bibinfo{address}{Berlin, Heidelberg},
  \bibinfo{pages}{165--184}.
\newblock
\showISBNx{9783642286513}
\urldef\tempurl%
\url{https://doi.org/10.1007/978-3-642-28652-0_9}
\showDOI{\tempurl}


\bibitem[{Roberto Ierusalimschy}(2024)]%
        {luainterpreter}
\bibfield{author}{\bibinfo{person}{{Roberto Ierusalimschy}}.}
  \bibinfo{year}{2024}\natexlab{}.
\newblock \bibinfo{title}{Lua 5.4}.
\newblock \bibinfo{howpublished}{\url{https://www.lua.org}}.
\newblock


\bibitem[Serrano(2021)]%
        {DBLP:journals/pacmpl/Serrano21}
\bibfield{author}{\bibinfo{person}{Manuel Serrano}.}
  \bibinfo{year}{2021}\natexlab{}.
\newblock \showarticletitle{Of JavaScript {AOT} compilation performance}.
\newblock \bibinfo{journal}{\emph{Proc. {ACM} Program. Lang.}}
  \bibinfo{volume}{5}, \bibinfo{number}{{ICFP}} (\bibinfo{year}{2021}),
  \bibinfo{pages}{1--30}.
\newblock
\urldef\tempurl%
\url{https://doi.org/10.1145/3473575}
\showDOI{\tempurl}


\bibitem[Serrano and Feeley(1996)]%
        {DBLP:conf/icfp/SerranoF96}
\bibfield{author}{\bibinfo{person}{Manuel Serrano} {and} \bibinfo{person}{Marc
  Feeley}.} \bibinfo{year}{1996}\natexlab{}.
\newblock \showarticletitle{Storage Use Analysis and its Applications}. In
  \bibinfo{booktitle}{\emph{Proceedings of the 1996 {ACM} {SIGPLAN}
  International Conference on Functional Programming, {ICFP} 1996,
  Philadelphia, Pennsylvania, USA, May 24-26, 1996}},
  \bibfield{editor}{\bibinfo{person}{Robert Harper} {and}
  \bibinfo{person}{Richard~L. Wexelblat}} (Eds.). \bibinfo{publisher}{{ACM}},
  \bibinfo{pages}{50--61}.
\newblock
\urldef\tempurl%
\url{https://doi.org/10.1145/232627.232635}
\showDOI{\tempurl}


\bibitem[Shaughnessy(2013)]%
        {ruby_under_microscope}
\bibfield{author}{\bibinfo{person}{Pat Shaughnessy}.}
  \bibinfo{year}{2013}\natexlab{}.
\newblock \bibinfo{booktitle}{\emph{Ruby Under a Microscope: An Illustrated
  Guide to Ruby Internals}}.
\newblock \bibinfo{publisher}{No Starch Press}, \bibinfo{address}{San
  Francisco, USA}.
\newblock
\showISBNx{1593275277}


\bibitem[Sheludko and Solanes(2020)]%
        {v8_pointer_compression}
\bibfield{author}{\bibinfo{person}{Igor Sheludko} {and}
  \bibinfo{person}{Santiago~Aboy Solanes}.} \bibinfo{year}{2020}\natexlab{}.
\newblock \bibinfo{title}{Pointer Compression in V8}.
\newblock
  \bibinfo{howpublished}{\url{https://v8.dev/blog/pointer-compression}}.
\newblock


\bibitem[Tobin-Hochstadt and Felleisen(2010)]%
        {occurrence-typing}
\bibfield{author}{\bibinfo{person}{Sam Tobin-Hochstadt} {and}
  \bibinfo{person}{Matthias Felleisen}.} \bibinfo{year}{2010}\natexlab{}.
\newblock \showarticletitle{Logical types for untyped languages}. In
  \bibinfo{booktitle}{\emph{Proceedings of the 15th ACM SIGPLAN International
  Conference on Functional Programming}} (Baltimore, Maryland, USA)
  \emph{(\bibinfo{series}{ICFP '10})}. \bibinfo{publisher}{Association for
  Computing Machinery}, \bibinfo{address}{New York, NY, USA},
  \bibinfo{pages}{117--128}.
\newblock
\showISBNx{9781605587943}
\urldef\tempurl%
\url{https://doi.org/10.1145/1863543.1863561}
\showDOI{\tempurl}


\bibitem[W{\"{u}}rthinger et~al\mbox{.}(2017)]%
        {DBLP:conf/pldi/WurthingerWHWSS17}
\bibfield{author}{\bibinfo{person}{Thomas W{\"{u}}rthinger},
  \bibinfo{person}{Christian Wimmer}, \bibinfo{person}{Christian Humer},
  \bibinfo{person}{Andreas W{\"{o}}{\ss}}, \bibinfo{person}{Lukas Stadler},
  \bibinfo{person}{Chris Seaton}, \bibinfo{person}{Gilles Duboscq},
  \bibinfo{person}{Doug Simon}, {and} \bibinfo{person}{Matthias Grimmer}.}
  \bibinfo{year}{2017}\natexlab{}.
\newblock \showarticletitle{Practical partial evaluation for high-performance
  dynamic language runtimes}. In \bibinfo{booktitle}{\emph{Proceedings of the
  38th {ACM} {SIGPLAN} Conference on Programming Language Design and
  Implementation, {PLDI} 2017, Barcelona, Spain, June 18-23, 2017}},
  \bibfield{editor}{\bibinfo{person}{Albert Cohen} {and}
  \bibinfo{person}{Martin~T. Vechev}} (Eds.). \bibinfo{publisher}{{ACM}},
  \bibinfo{pages}{662--676}.
\newblock
\urldef\tempurl%
\url{https://doi.org/10.1145/3062341.3062381}
\showDOI{\tempurl}


\bibitem[Zurstrassen et~al\mbox{.}(2023)]%
        {Zurstrassen23}
\bibfield{author}{\bibinfo{person}{Niko Zurstrassen},
  \bibinfo{person}{Lennart~M. Reimann}, \bibinfo{person}{Nils Bosbach},
  \bibinfo{person}{Lukas Juenger}, {and} \bibinfo{person}{Rainer Leupers}.}
  \bibinfo{year}{2023}\natexlab{}.
\newblock \showarticletitle{Evaluation of the RISC-V Floating Point
  Extensions}. In \bibinfo{booktitle}{\emph{DVCon Europe 2023; Design and
  Verification Conference and Exhibition Europe}}. \bibinfo{pages}{57--64}.
\newblock
\urldef\tempurl%
\url{https://ieeexplore.ieee.org/abstract/document/10461383}
\showURL{%
\tempurl}


\end{thebibliography}


\end{document}